%% file: main.tex
\documentclass[12pt]{article}
\usepackage[utf8]{inputenc}
\usepackage[DIV=13]{typearea}\usepackage{ragged2e}
\usepackage{calligra,amsmath,amsfonts,bbm,mathrsfs,amssymb}
\usepackage{slashed,cancel}
\usepackage{cite}
\usepackage{bm}
\usepackage{tablefootnote}

\usepackage[table, dvipsnames]{xcolor}
\usepackage{imakeidx}
\makeindex[intoc, title=Index of commands]
\usepackage{hyperref}
\hypersetup{colorlinks=true,urlcolor=magenta,anchorcolor=blue,citecolor=blue,filecolor=blue,linkcolor=magenta,menucolor=blue,linktocpage=true,pdfproducer=medialab,
pdftitle={ALPaca: The ALP Automatic Computing Algorithm},
pdfauthor={Jorge Alda, Marta Fuentes Zamoro, Luca Merlo, Xavier Ponce Díaz, Stefano Rigolin}}
\usepackage[english]{babel}
\usepackage{catchfile}
\usepackage{indentfirst}
\usepackage{graphicx}
\usepackage{float}
\usepackage{enumerate}
\usepackage{caption}
\usepackage{subcaption}
\usepackage{multirow}
\usepackage[normalem]{ulem} 
\usepackage{booktabs}
\usepackage{physics}
\usepackage{url}
\usepackage{soul}
\usepackage{verbatim}
\usepackage{lscape}
\usepackage{multicol}
\usepackage{bbold}
\usepackage{tabularx}
\usepackage{fontawesome}
\usepackage{listings}
\usepackage{tcolorbox}

\newcommand{\email}[1]{\href{mailto:#1}{\tt #1}}

\numberwithin{equation}{section}

\newcommand{\blue}[1]{\color{blue} #1 \color{black}}
\newcommand{\magenta}[1]{\color{magenta} #1 \color{black}}

\newcommand{\alpaca}{{\fontfamily{cmss}\selectfont ALPaca }}

\newcommand{\nn}{\nonumber}
\newcommand{\be}{\begin{equation}}
\newcommand{\ee}{\end{equation}}

\newcommand{\TeV}{\ \text{TeV}}
\newcommand{\GeV}{\ \text{GeV}}

\usepackage[frozencache,cachedir=minted-cache]{minted}
\usepackage{verbatim}

\definecolor{codegreen}{rgb}{0,0.6,0}
\definecolor{codegray}{rgb}{0.5,0.5,0.5}
\definecolor{codepurple}{rgb}{0.58,0,0.82}
\definecolor{backcolour}{rgb}{0.95,0.95,0.92}
\definecolor{codeblack}{rgb}{0.05,0.05,0.05}
\definecolor{codewhite}{rgb}{0.95,0.95,0.95}

\setminted[pwsh-session]{
bgcolor=codeblack,
style=github-dark,
escapeinside=||
}

\setminted[python]{
bgcolor=codewhite,
}

\usepackage{etoolbox}
\makeatletter
\patchcmd{\minted@colorbg}{\medskip}{}{}{}
\patchcmd{\endminted@colorbg}{\medskip}{}{}{}
\makeatother

\begin{document} 
\renewcommand*{\thefootnote}{\fnsymbol{footnote}}

\begin{titlepage}

\vspace*{-1cm}
\flushleft{\magenta{IFT-UAM/CSIC-25-82}} 
\\[1cm]
\vskip 1cm

\begin{center}
\blue{\bf \Large \alpaca\!\!: the ALP Automatic Computing Algorithm}
\centering
\vskip .3cm
\end{center}
\vskip 0.5  cm
\begin{center}
{\large\bf Jorge Alda}$^{a,b}$~\footnote{\email{jorge.alda@pd.infn.it}},
{\large\bf Marta Fuentes Zamoro}$^{c}$~\footnote{\email{marta.zamoro@uam.es}},
{\large\bf Luca Merlo}$^{c}$~\footnote{\email{luca.merlo@uam.es}},\vskip 0.5cm
{\large\bf Xavier Ponce D\'iaz}$^{d}$~\footnote{\email{xavier.poncediaz@unibas.ch}},
and 
{\large\bf Stefano Rigolin}$^{a}$~\footnote{\email{stefano.rigolin@pd.infn.it}},
\vskip .7cm
{\footnotesize
$^a$~Dipartamento di Fisica e Astronomia ``G.~Galilei" and Istituto Nazionale di Fisica Nucleare,\\
Sezione di Padova, Universit\`a degli Studi di Padova, I-35131 Padova, Italy{\par\centering \vskip 0.25 cm\par}
$^b$~Centro de Astropart\'iculas y F\'isica de Altas Energ\'ias (CAPA)\\ Pedro Cerbuna 12, E-50009 Zaragoza, Spain
{\par\centering \vskip 0.25 cm\par}
$^c$~Departamento de F\'isica Te\'orica and Instituto de F\'isica Te\'orica UAM/CSIC,\\
Universidad Aut\'onoma de Madrid, Cantoblanco, 28049, Madrid, Spain \\
{\par\centering \vskip 0.25 cm\par}
$^d$~Department of Physics, University of Basel,  Klingelbergstrasse 82, \\ CH-4056 Basel, 
Switzerland
}

\end{center}
\vskip 2cm
\begin{abstract}
\justify
The ALP Automatic Computing Algorithm, \alpaca\!\!, is an open source Python library devoted to studying the 
phenomenology of Axion-Like Particles (ALPs) with masses in the ranges $m_a \in [0.01 - 10]$  GeV. \alpaca 
provides a flexible and comprehensive framework to define ALP couplings at arbitrary energy scales, perform 
Renormalisation Group evolution and matching down to the desired low energy scale, and compute a large variety 
of ALP observables, with particular care to the meson decay sector. The package includes support for UV completions, 
experimental constraints, and visualisation tools, enabling both detailed analyses and broad parameter space 
exploration. The program is publicly available at \href{https://github.com/alp-aca/alp-aca}{\alpaca  \faicon{github}}.
\end{abstract}
\end{titlepage}
\setcounter{footnote}{0}

\pdfbookmark[1]{Table of Contents}{tableofcontents}
\tableofcontents
\renewcommand*{\thefootnote}{\arabic{footnote}}

\bigskip

\section{Introduction}

The study of Axion-Like Particles (ALPs)~\cite{Georgi:1986df} is a rapidly evolving field at the interface of particle physics, cosmology and astrophysics. ALPs emerge naturally in various extensions of the Standard Model (SM). They appear in string theory contexts~\cite{Witten:1984dg,Choi:2006qj,Svrcek:2006yi,Arvanitaki:2009fg,Cicoli:2012sz}, in composite Higgs models~\cite{Merlo:2017sun,Brivio:2017sdm,
Alonso-Gonzalez:2018vpc,Alonso-Gonzalez:2020wst}, and in supersymmetric scenarios~\cite{Bellazzini:2017neg}.  They may have an impact on cosmological observables~\cite{Ferreira:2018vjj,DEramo:2018vss,Escudero:2019gvw,Arias-Aragon:2020qtn,Arias-Aragon:2020qip,
Arias-Aragon:2020shv,Ferreira:2020bpb,Escudero:2021rfi,Araki:2021xdk,DEramo:2021psx,DEramo:2021lgb,
DEramo:2022nvb}, or play the role of a Dark Matter candidate~\cite{Gelmini:1984pe,Berezinsky:1993fm,Lattanzi:2007ux,Bazzocchi:2008fh,Lattanzi:2013uza,Queiroz:2014yna}. In many cases, they have been studied in association with flavour model dynamics~\cite{Davidson:1981zd,Wilczek:1982rv,Ema:2016ops,
Calibbi:2016hwq,Arias-Aragon:2017eww,Arias-Aragon:2022ats,DiLuzio:2023ndz,Greljo:2024evt} and neutrino mass generation~\cite{Chikashige:1980qk,Chikashige:1980ui,Gelmini:1980re,deGiorgi:2023tvn,Liang:2024vnd,Greljo:2025suh}.

Following the seminal paper in Ref.~\cite{Georgi:1986df}, complementary studies focussed on the effective description (EFT) of ALPs~\cite{Choi:1986zw,Salvio:2013iaa,
Brivio:2017ije,Alonso-Alvarez:2018irt,Gavela:2019wzg,Chala:2020wvs,Bauer:2020jbp,DiLuzio:2020oah,Bonilla:2021ufe,Arias-Aragon:2022byr,
Arias-Aragon:2022iwl,Song:2023lxf,DiLuzio:2023cuk}, that has then been used to analyse their possible signals both at  colliders~\cite{Jaeckel:2012yz,Mimasu:2014nea,
Jaeckel:2015jla,Alves:2016koo,Knapen:2016moh,Brivio:2017ije,Bauer:2017nlg,Mariotti:2017vtv,Bauer:2017ris,
Baldenegro:2018hng,Craig:2018kne,Bauer:2018uxu,Gavela:2019cmq,Haghighat:2020nuh,Wang:2021uyb,Liu:2021lan,deGiorgi:2022oks,
Bonilla:2022pxu,Ghebretinsaea:2022djg,Vileta:2022jou,Marcos:2024yfm,Arias-Aragon:2024gpm} and low-energy facilities~\cite{Izaguirre:2016dfi,Merlo:2019anv,Aloni:2019ruo,Bauer:2019gfk,Bauer:2020jbp,Bauer:2021mvw,Carmona:2021seb,
Guerrera:2021yss,Gallo:2021ame,Bertholet:2021hjl,Cheng:2021kjg,Bonilla:2022qgm,Bonilla:2022vtn,deGiorgi:2022vup,Guerrera:2022ykl,
Bonilla:2023dtf,Arias-Aragon:2023ehh,DiLuzio:2024jip,deGiorgi:2024str,Alda:2024cxn,Alda:2024xxa,
Arias-Aragon:2024qji,Arias-Aragon:2024gdz,Bisht:2024hbs,Gao:2025ohi,Alda:2025uwo}.
Their rich phenomenology offers a unique opportunity to probe physics beyond the Standard Model (BSM) across a broad range of energy scales.

The ALP Automatic Computing Algorithm (\alpaca\!\!) is a Python package designed to systematically handle the complexities of ALP phenomenology in the meson sector. The code provides an integrated framework to define ALP effective interactions at arbitrary energy scales, perform Renormalisation Group running and matching, and compute a wide range of observables. The implementation of ALP interactions, including their Renormalisation Group evolution and the relevant formulae for phenomenological analyses, is based on results from recent theoretical developments in the literature (see Ref.~\cite{Alda:2025uwo} and references therein). Additionally, \alpaca also incorporates a database of experimental data and includes routines for numerical analysis and plotting.

\alpaca is structured around modular building blocks that reflect the underlying theoretical framework. These include multiple operator bases for ALP couplings, support for user-defined and pre-implemented UV completions, and the computation of both tree-level and loop-induced processes. Moreover, the package implements a database of the latest experimental constraints and provides tools for parameter space scans and exclusion plotting.

The aim of \alpaca is to assist in automating and organising the many technical steps that arise in realistic ALP analyses. By making complex calculations and data handling accessible and reproducible, \alpaca enables both rapid prototyping and in-depth studies of ALP scenarios, from theory to phenomenological predictions. In a companion paper~\cite{Alda:2025uwo}, we provide the latest phenomenological analysis on an ALP with mass $m_a \in [0.01 - 10]\GeV$, using \alpaca\!\!.

Summarising, \alpaca provides a versatile tool for ongoing and future analyses of the physics involving ALPs:
\begin{itemize}
\item[-] It allows one to test specific UV models as well as generic ALP-EFT Lagrangians. 
\item[-] It permits one to define ALP couplings at arbitrary energy scales.
\item[-] It supports multiple operator bases, handling both flavour-universal and flavour-specific ALP couplings with fermions.
\item[-] It encodes the full 1-loop RGE of the shift-symmetric ALP couplings.
\item[-] It performs the matching and the running across various thresholds, from the UV scale $\Lambda$ down to meson-mass energies. This can be done with several integrator methods to adapt to the user requirements.
\item[-] It includes the routines to compute ALP decays into photons, hadrons and fermions. Particular attention is devoted to the hadronic decay treatment including the state-of-art $\chi$PT description, as well as VMD contributions. Additionally, it is possible to provide a user-defined branching ratio to a dark sector.
\item[-] It accounts for various ALP production mechanisms, such as quarkonia radiative decays, FNCN meson decays, LFV decays and non-resonant production.
\item[-] It is equipped with the latest available experimental measurements.
\item[-] It handles prompt, displaced vertices and invisible signatures. 
\item[-] It includes statistical tools to compute likelihoods and confidence intervals.
\item[-] It scans over parameter spaces and allows to produce exclusion plots and signal regions.
\item[-] It integrates with the iNSPIRE-HEP REST API for automatic citation generation. 
\end{itemize}

This manual is organised as follows: in Section~\ref{sec:installation} we show how \alpaca can be easily installed along with its dependencies. In Section~\ref{sec:ALPbases} we describe the different bases implemented in the program and how the matching and running is implemented internally. Section~\ref{sec:UVmodels} shows how the user can define different ALP UV-models. Sections~\ref{sec:alp_prod} and ~\ref{sec:exp_searches} describe the different ALP-production and ALP-detection relevant processes, which are implemented in \alpaca\!\!. Finally, the last sections explain different additional functionalities of the program such as statistics in Section~\ref{sec:statistics}, plotting of exclusion limits in Section~\ref{sec:plotting}, while in Section~\ref{sec:citations} how one can easily access the references associated to experimental searches or theoretical computations.
 
\alpaca is an open-source project under MIT license, publicly available at \href{https://github.com/alp-aca/alp-aca}{\alpaca  \faicon{github}} including useful tutorials and use cases in \href{https://github.com/alp-aca/examples}{\alpaca\!\!/examples  \faicon{github}} and its functionality is fully documented in this manual.

\section{Installation and first steps}
\input{installation}
\section{Defining the ALP couplings, running and matching}
\label{sec:ALPbases}
\input{alpcouplings}
\section{UV-complete models}
\label{sec:UVmodels}
\input{uv_model}

\section{ALP processes}
\label{sec:alp_prod}
\input{alp_process}
\section{Experimental ALP searches}
\label{sec:exp_searches}
\input{exp_signatures}
\input{exp}
\section{Statistics}\label{sec:statistics}
\input{stats} 
\section{Plotting}\label{sec:plotting}
\input{plotting.tex}

\section{Managing bibliography and references}\label{sec:citations}
\input{references.tex}

\section{Conclusions}
This manual gives an overview of the features of \alpaca at the time corresponding to the release of version 1.0, as well as several examples on how to use the functions presented here. The full content of the package can be found in \href{https://github.com/alp-aca/alp-aca}{\alpaca  \faicon{github}}, while some examples, associated to the phenomenological study of Ref.~\cite{Alda:2025uwo} and to this manual, are implemented in \href{https://github.com/alp-aca/examples/tree/main/manual}{\alpaca\!\!/examples  \faicon{github}}.

Despite, the robustness of the package presented in this paper, there are several ways in which \alpaca may be improved in the following versions. 
To highlight some of them, long-lived particle searches in beam-dump experiments could also be considered. These types of searches require Monte Carlo simulations to compute ALP production across all possible channels. Such computations could be integrated into \alpaca by linking it with the existing software ALPINIST ~\cite{Jerhot:2022chi}, enabling a more comprehensive treatment of ALP phenomenology. Another viable improvement is to incorporate astrophysical bounds and searches, though it is technically more challenging, as it would require a careful review of the assumptions underlying each constraint. Moreover, it may be useful to introduce chirality breaking bases: as in this case, not all the ALP--fermion couplings can be moved to the derivative basis, it will be necessary to provide \alpaca with the RGE analysis in the chirality breaking description. Finally, it may be a promising extension to incorporate the matching of specific UV completions to ALP EFTs, using for example tools such as Matchete~\cite{Fuentes-Martin:2020udw,Fuentes-Martin:2022jrf} or MatchMakerEFT~\cite{Carmona:2021xtq}.  

All in all, we expect \alpaca to play a central role in the global effort to test the properties of ALPs.

\section*{Acknowledgements}
We thank Pilar Coloma, Enrique Fern\'andez Mart\'inez, Maksym Ovchynnikov, Maria Ramos for useful discussions. JA kindly thanks the hospitality of Universidad Aut\'onoma de Madrid/IFT during the completion of this work. MFZ kindly thanks the University of Padova for the hospitality and the COST Action COSMIC WISPers (CA21106) for financial support during the early stages of this work.

JA has received funding from the Fundaci\'on Ram\'on Areces ``Beca para ampliaci\'on de estudios en el extranjero en el campo de las Ciencias de la Vida y de la Materia'', and acknowledges support by the grants PID2021-126078NB-C21 funded by MCIN/\hspace{0pt}AEI/\hspace{0pt}10.13039/\hspace{0pt}501100011033 and ``ERDF A way of making Europe''.  The work of MFZ is supported by the Spanish MIU through the National Program FPU (grant number FPU22/\hspace{0pt}03625). XPD has partially received funding from the Swiss National Science Foundation (SNF) through the Eccellenza Professorial Fellowship ``Flavor Physics at the High Energy Frontier,'' project number 186866.
We acknowledge partial financial support by the European Union's Horizon 2020 research and innovation programme under the Marie Sk\l odowska-Curie grant agreement No.~101086085-ASYMMETRY and by the Spanish Research Agency (Agencia Estatal de Investigaci\'on) through the grant IFT Centro de Excelencia Severo Ochoa No CEX2020-001007-S and by the grant PID2022-137127NB-I00 funded by MCIN/\hspace{0pt}AEI/\hspace{0pt}10.13039/501100011033.
This article is based upon work from COST Action COSMIC WISPers CA21106, supported by COST (European Cooperation in Science and Technology).

\appendix
\section{Database of experimental measurements}
\input{database_experiments}
\printindex

\bibliographystyle{utphys}
\bibliography{biblio}

\end{document}

%% file: installation.tex
\label{sec:installation}

\alpaca can be installed directly from the standard python repository \texttt{PyPI}, by executing in the terminal of the operating system the following command: 
\begin{minted}{pwsh-session}
$ |\textcolor{white}{pip3 install alpaca-ALPs}|
\end{minted}

This command installs \alpaca as well as the following depenences
\begin{itemize}
    \item \texttt{sympy}: Symbolic algebra~\cite{sympy}.
    \item \texttt{vegas}: Numerical integration~\cite{Lepage:1977sw,vegas}.
    \item \texttt{flavio}: Calculates hadronic form factors and running of the SM parameters~\cite{Straub:2018kue,flavio}.
    \item \texttt{distinctipy}: Generates palettes for plots~\cite{distinctipy}.
    \item \texttt{requests}: Connects to the iNSPIRE-HEP API~\cite{Moskovic:2021zjs} to generate bibliographical records~\cite{requests}. See Section~\ref{sec:citations} for more information.
\end{itemize}

Additionally, a working installation of \texttt{matplotlib}~\cite{Hunter:2007} is required for the plotting capabilities described in Section~\ref{sec:plotting}, and \texttt{IPython}~\cite{PER-GRA:2007} provides enhanced output when working on notebook environments.

It is \emph{strongly recommended} to install \alpaca inside a virtual environment (venv), in order to avoid clashes with conflicting versions of the dependencies. In order to create a venv, execute the following command
\begin{minted}{pwsh-session}
$ |\textcolor{white}{python3 -m venv pathToVenv}|
\end{minted}
where \verb|pathToVenv| is the location where the files of the venv will be stored. In order to activate the venv, for Linux or MacOS using \texttt{bash} or \texttt{zsh}
\begin{minted}{pwsh-session}
$ |\textcolor{white}{source pathToVenv/bin/activate}|
\end{minted}
for Windows using \texttt{cmd.exe}
\begin{minted}{pwsh-session}
C:\> |\textcolor{white}{pathToVenv{\textbackslash}Scripts{\textbackslash}Activate.bat}|
\end{minted}
and for Windows using \texttt{PowerShell}
\begin{minted}{pwsh-session}
PS C:\> path_to_venv\Scripts\Activate.ps1
\end{minted}
Once the venv is activated, \alpaca and the packages listed above can be normally installed and used.\\

\alpaca can be imported in a Python script with
\begin{minted}{python}
import alpaca
\end{minted}

%% file: alpcouplings.tex
There exist several possibilities to define the ALP-EFT Lagrangian at a given scale $\Lambda$. For this reason, the logic of \alpaca is to give the user the flexibility of defining the couplings in different bases and at various scales, following mainly Ref.~\cite{Bauer:2020jbp}. In the package, the collection of ALP couplings at a given energy scale $\Lambda$ and in a specific basis is represented using an \texttt{ALPcouplings}\index{\texttt{ALPcouplings}} object which stores all the information related to the couplings present in the Lagrangian. As the form of the Lagrangian depends on the energy scale, there are different possible bases implemented in \alpaca\!\!.

\subsection{Bases for ALP couplings}

\begin{figure}
    \centering
    \includegraphics[trim={0 3cm 0 35mm},clip,width=0.9\textwidth]{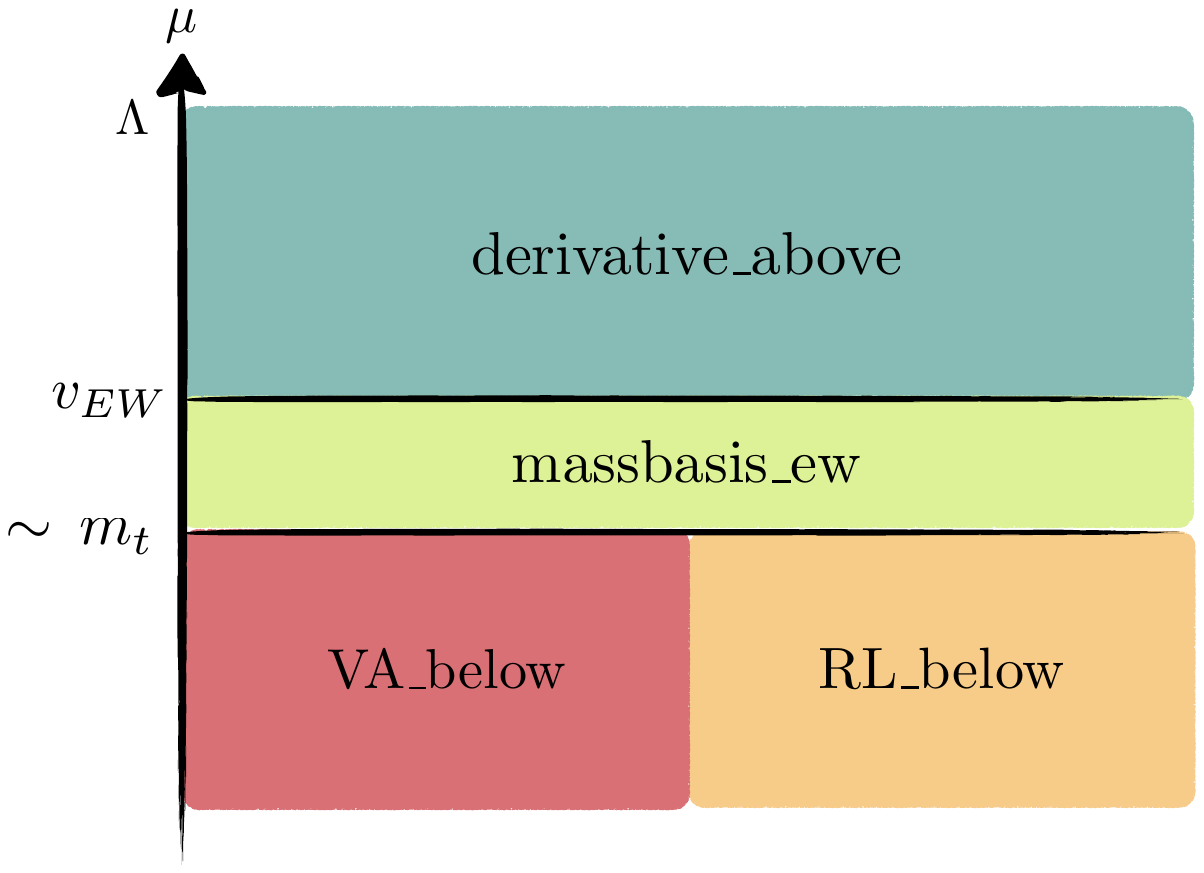}
    \caption{\em Summary of the different bases depending on the energy scale that are available to the user.}
    \label{fig:bases-diagram}
\end{figure}

In this section we define the various bases available, providing the explicit analytical expression and the list of parameters that enter each description. Each basis holds in a specific energy interval. A summarising scheme is reported in Fig.~\ref{fig:bases-diagram}. 

It is relevant to mention that all the bases included in \alpaca are ``chirality preserving'', that is with derivative couplings to fermions. The advantage of this choice is that the shift symmetry of the ALP is manifest in the fermionic couplings. 

The alternative basis is the ``Yukawa-like'' or ``chirally breaking'' one, that is the common basis where the ALP originates as an orbital degree of freedom after the spontaneous breaking of a continuous global symmetry. The switch between the two basis can be performed via appropriate chiral rotations of the fermions, but only if the ALP--fermion couplings are shift symmetry invariant. We will be back to this aspect in the following.

\subsubsection{Derivative basis above the EW scale}
This basis is defined above the electroweak scale (EW), $\mu_\mathrm{EW}$, with three generations of fermions, that is $n_{q_L} = n_{u_R} = n_{d_R} = n_{\ell_L} = n_{e_R} = 3$. As usual, $q_L$ and $\ell_L$ refer to the left-handed (LH) quark and lepton $\mathrm{SU}(2)_L$ doublets, while $u_R$, $d_R$ and $e_R$ are the right-handed (RH) up- and down-quark and charged lepton $\mathrm{SU}(2)_L$ singlets. 

This is a commonly used basis in ALP-EFT descriptions, as couplings with fermions include a derivative of the ALP, $\partial_\mu a$, that are manifestly shift symmetry invariant. The only shift symmetry breaking terms are the ALP--gauge boson couplings, also dubbed as ``anomalous'' terms, and the ALP mass term. No ALP--Higgs coupling is introduced as it is not independent from the other operators~\cite{Georgi:1986df}. Moreover, this basis is valid up to the scale $\Lambda$ that represents the mass scale of heavier degrees of freedom, integrated out of the spectrum. This is the case, for example, of RH neutrinos, which are commonly present in Majoron models. Another example can be found in all the models where the ALP arises as a Goldstone boson of a singlet scalar field that breaks the Peccei-Quinn (PQ)~\cite{Peccei:1977hh} symmetry by its vacuum expectation value (vev): the radial model is typically very heavy. 

The label of the derivative basis, its explicit definition and the associated parameters are listed as follows:
\begin{itemize}
\item \textbf{Name:} \verb|derivative_above|

\item \textbf{Lagrangian:}
Besides the SM Lagrangian and the canonical kinetic terms of the ALP, the lowest dimensional operators describing the ALP mass and couplings are
\begin{align}
\mathcal{L} \supset& -\dfrac12m^2_a a^2+\frac{a}{f_a}\left(\frac{g_B^2}{16\pi^2} c_{B}B_{\mu\nu}\tilde{B}^{\mu\nu} + \frac{g_W^2}{16\pi^2}c_{W} W^i_{\mu\nu}\tilde{W}^{i,\mu\nu}+\frac{g_s^2}{16\pi} c_{G}G^a_{\mu\nu}\tilde{G}^{a,\mu\nu}\right)+\nn \\ &+ \frac{\partial^\mu a}{f_a} \left(\bar{q}_L c_{q_L} \gamma_\mu q_L  + \bar{u}_R c_{u_R} \gamma_\mu u_R + \bar{d}_R c_{d_R} \gamma_\mu d_R + \bar{\ell}_L c_{\ell_L} \gamma_\mu \ell_L + \bar{e}_R c_{e_R} \gamma_\mu e_R \right)\,.
\label{eq:derivative_above}
\end{align}
In the anomalous couplings $aV\tilde{V}$, $\tilde{V}$ is the dual field defined with the convention \mbox{$\epsilon^{0123}=1$}.

\item \textbf{Parameters:}
\begin{itemize}
    \item \texttt{cB}: Coupling to $B$ bosons $c_B$.
   \item \texttt{cW}: Coupling to $W$ bosons $c_W$.
   \item \texttt{cG}: Coupling to gluons $c_G$.
  \item \texttt{cqL}: Coupling matrix to left-handed quarks $c_{q_L}$.
  \item \texttt{cuR}: Coupling matrix to right-handed up-type quarks $c_{u_R}$.
  \item \texttt{cdR}: Coupling matrix to right-handed down-type quarks $c_{d_R}$.
  \item \texttt{clL}: Coupling matrix to left-handed leptons $c_{\ell_L}$.
  \item \texttt{ceR}: Coupling matrix to right-handed charged leptons $c_{e_R}$.
\end{itemize}
\end{itemize}

\subsubsection{Mass basis at the EW scale}
This basis is defined at the electroweak scale in the EW broken phase and with three generations of fermions, that is $n_{u_L} = n_{d_L}=n_{u_R} = n_{d_R} = n_{e_L} = n_{\nu_L} = n_{e_R} = 3$, in the mass basis.

\begin{itemize}
    \item \textbf{Name: }\verb|massbasis_ew|
    \item \textbf{Lagrangian:}
    \be
    \begin{split}
        \mathcal{L} \supset&-\dfrac12m_a^2a^2+ \dfrac{a}{f_a}\left(\frac{\alpha_\mathrm{em}}{4\pi}c_{\gamma} F_{\mu\nu}\tilde{F}^{\mu\nu} + \frac{\alpha_\mathrm{em}}{2\pi s_w c_w}c_{\gamma Z} F_{\mu\nu}\tilde{Z}^{\mu\nu} + \right.  \\
        &+ \left.\frac{\alpha_\mathrm{em}}{4\pi s_w^2 c_w^2}c_{Z} Z_{\mu\nu}\tilde{Z}^{\mu\nu}+\frac{\alpha_\mathrm{em}}{2\pi s_w^2}c_{W} W^+_{\mu\nu}\tilde{W}^{-\mu\nu} + \frac{\alpha_s}{4\pi}c_{G} G^a_{\mu\nu}\tilde{G}^{a,\mu\nu}\right)+ \\
        &+ \frac{\partial^\mu a}{f_a} \left(\bar{u}_L c'_{u_L} \gamma_\mu u_L  + \bar{u}_R c'_{u_R} \gamma_\mu u_R + \bar{d}_L c'_{d_L} \gamma_\mu d_L + \bar{d}_R c'_{d_R} \gamma_\mu d_R\right. \\ 
        &+\left. \bar{e}_L c'_{e_L} \gamma_\mu e_L + \bar{e}_R c'_{e_R} \gamma_\mu e_R + \bar{\nu}_L c'_{\nu_L} \gamma_\mu \nu_L \right)\,.
    \end{split}
    \ee
\item \textbf{Parameters:}
\begin{itemize}
    \item  \texttt{cgamma}: Coupling to photons $c_\gamma$.
    \item \texttt{cgammaZ}: Coupling to photon and $Z$ boson $c_{\gamma Z}$.
    \item \texttt{cZ}: Coupling to $Z$ bosons $c_Z$.
    \item \texttt{cW}: Coupling to $W$ bosons $c_W$.
    \item \texttt{cG}: Coupling to gluons $c_G$.
    \item \texttt{cuL}: Coupling matrix to left-handed up-type quarks $c'_{u_L}$.
    \item \texttt{cuR}: Coupling matrix to right-handed up-type quarks $c'_{u_R}$.
    \item \texttt{cdL}: Coupling matrix to left-handed down-type quarks $c'_{d_L}$.
    \item  \texttt{cdR}: Coupling matrix to right-handed down-type quarks $c'_{d_R}$.
    \item  \texttt{ceL}: Coupling matrix to left-handed charged leptons $c'_{e_L}$.
    \item \texttt{ceR}: Coupling matrix to right-handed charged leptons $c'_{e_R}$.
    \item \texttt{cnuL}: Coupling matrix to left-handed neutrinos $c'_{\nu_L}$.
\end{itemize}
\end{itemize}

The user can use this basis as an starting input, or can be obtained after matching from the bases above. In the former case, the user can introduce the numerical inputs, while in the latter the couplings in the basis \texttt{massbasis\_ew} are calculated from the couplings in the basis \texttt{derivative\_above} according to
\begin{align}\label{eq:massbasis}
    c_\gamma &= c_B + c_W\,, \nonumber\\
    c_Z &= c_W \cos^2\theta_W -c_B \sin^2\theta_W\,,\nonumber\\
    c_{\gamma Z} &=  c_W \cos^4\theta_W + c_B \sin^4\theta_W\,,\nonumber\\
    c'_{u_L} &= V_{u_L}^\dagger c_{q_L} V_{u_L}\,,\nonumber\\
    c'_{d_L} &= V_{d_L}^\dagger c_{q_L} V_{d_L}\,,\nonumber\\
    c'_{u_R} &= V_{u_R}^\dagger c_{u_R} V_{u_R}\,,\nonumber\\
    c'_{d_R} &= V_{d_R}^\dagger c_{d_R} V_{d_R}\,,\nonumber\\
    c'_{e_L} &= c_{\nu_L} = c_{\ell_L}\,,\nonumber\\
    c'_{e_R} &= c_{e_R}\,,
\end{align}
where $\theta_W$ is the Weinberg angle and $V_i$ are the unitary transformations necessary to move to the fermion mass basis. 

If the user introduces the inputs in this basis directly, it is possible to break $\mathrm{SU}(2)_L$, e.g.~$c'_{u_L}\neq V_\mathrm{CKM} c'_{d_L} V_\mathrm{CKM}^\dagger$, however, then it is not possible to return to a $\mathrm{SU}(2)_L$ unbroken basis. Hence, in \alpaca it is not allowed to move from this basis to \texttt{derivative\_above}.

\subsubsection{Right/left basis below the EW scale}
This basis is defined below the EW scale with fermions in the mass basis, where the top-quark as well as the massive gauge bosons and the physical Higgs have been removed from the theory. In terms of number of generations, $n_{u_L} =n_{u_R} =2$ and $n_{d_L}=n_{d_R} = n_{e_L} = n_{e_R} = n_{\nu_L}  = 3$. This basis is typically used to compute amplitudes in low-energy physics.

\begin{itemize}
    \item \textbf{Name: }\verb|RL_below|
    \item \textbf{Lagrangian:}
    \be
    \begin{split}
        \mathcal{L} \supset& -\dfrac12m_a^2 a^2+\frac{a}{f_a}\left(\frac{\alpha_\mathrm{em}}{4\pi}c_{\gamma} F_{\mu\nu}\tilde{F}^{\mu\nu} + \frac{\alpha_s}{4\pi}c_{G} G^a_{\mu\nu}\tilde{G}^{a,\mu\nu}\right)+\\
        &+ \frac{\partial^\mu a}{f_a} \left(\bar{u}_L c_{u_L} \gamma_\mu u_L  + \bar{u}_R c_{u_R} \gamma_\mu u_R + \bar{d}_L c_{d_L} \gamma_\mu d_L + \bar{d}_R c_{d_R} \gamma_\mu d_R+\right. \\\ 
        &\hspace{1.5cm}+\left. \bar{e}_L c_{e_L} \gamma_\mu e_L + \bar{e}_R c_{e_R} \gamma_\mu e_R + \bar{\nu}_L c_{\nu_L} \gamma_\mu \nu_L \right)\,.
    \end{split}
    \ee
\item \textbf{Parameters:}
\begin{itemize}
    \item  \texttt{cgamma}: Coupling to photons $c_\gamma$.
    \item \texttt{cG}: Coupling to gluons $c_G$.
    \item \texttt{cuL}: Coupling matrix to left-handed up-type quarks $c_{u_L}$.
    \item \texttt{cuR}: Coupling matrix to right-handed up-type quarks $c_{u_R}$.
    \item \texttt{cdL}: Coupling matrix to left-handed down-type quarks $c_{d_L}$.
    \item  \texttt{cdR}: Coupling matrix to right-handed down-type quarks $c_{d_R}$.
    \item  \texttt{ceL}: Coupling matrix to left-handed charged leptons $c_{e_L}$.
    \item \texttt{ceR}: Coupling matrix to right-handed charged leptons $c_{e_R}$.
    \item \texttt{cnuL}: Coupling matrix to left-handed neutrinos $c_{\nu_L}$.
\end{itemize}
\end{itemize}

\subsubsection{Vector/axial basis below EW scale}
This basis undergoes the same conditions as the previous one, that is defined below the EW scale with $n_{u_L} = n_{u_R} =2$ and $n_{d_L}=n_{d_R}= n_{e_L}=n_{e_R}=n_{\nu_L} = 3$. The vector and axial couplings are defined as function of the LH and RH couplings as $c^{V/A}_f= c^{L}_{f} \pm  c^{R}_{f}$. This basis is particularly useful for low-energy phenomenology.
\begin{itemize}
    \item \textbf{Name: }\verb|VA_below|
    \item \textbf{Lagrangian:}
    \begin{align}
        \mathcal{L} \supset& -\dfrac12m_a^2a^2+\frac{a}{f_a}\left(\frac{\alpha_\mathrm{em}}{4\pi}c_{\gamma} F_{\mu\nu}\tilde{F}^{\mu\nu} + \frac{\alpha_s}{4\pi}c_{G} G^a_{\mu\nu}\tilde{G}^{a,\mu\nu}\right)+\\ 
        &+ \frac{\partial^\mu a}{2f_a} \left(\bar{u} \gamma_\mu (c_u^V + c_u^A \gamma_5) u + \bar{d} \gamma_\mu (c_d^V + c_d^A \gamma_5) d + \bar{e} \gamma_\mu (c_e^V + c_e^A \gamma_5) e + 2\bar{\nu}_L c_\nu \gamma_\mu \nu_L \right)\,.\nn
    \end{align}
    \item \textbf{Parameters:}
    \begin{itemize}
        \item \texttt{cgamma}: Coupling to photons $c_\gamma$.
        \item \texttt{cG}: Coupling to gluons $c_G$.
        \item \texttt{cuV}: Vectorial coupling matrix to up-type quarks $c_u^V$.
        \item \texttt{cuA}: Axial coupling matrix to up-type quarks $c_u^A$.
        \item \texttt{cdV}: Vectorial coupling matrix to down-type quarks $c_d^V$.
        \item \texttt{cdA}: Axial coupling matrix to down-type quarks $c_d^A$.
        \item \texttt{ceV}: Vectorial coupling matrix to charged leptons $c_e^V$.
        \item \texttt{ceA}: Axial coupling matrix to charged leptons $c_e^A$.
        \item \texttt{cnu}: Coupling to matrix left-handed neutrinos $c_\nu$.
    \end{itemize}
\end{itemize}
\subsubsection{Implementation}
The first argument to \texttt{ALPcouplings} is a \texttt{dict} containing the non-zero couplings. For couplings to fermions, if only a real number (\texttt{float}) is provided, \alpaca assumes flavour-universal couplings. If instead a $n_f \times n_f$ matrix (with $n_f$ equal to the number of fermions $f$ active at the scale $\Lambda$), \texttt{numpy.array} or \texttt{numpy.matrix}, is provided, the couplings will be non-universal. Couplings to gauge bosons are always scalars. Finally, \texttt{ALPcouplings} accept the optional argument \verb|ew_scale| indicating at which energy scale $\mu_\mathrm{ew}$ (in GeV) the matching between theories ``above/at the EW scale'' and ``below the EW scale'' is performed; the default value is $100\,\mathrm{GeV}$ (for more details on the value choice, see Sect.~\ref{sec:running_matching}). If \texttt{scale} $<(>)$ \verb|ew_scale|, the basis must end in \verb|_below| (\verb|_above|). 

Finally, if the \texttt{ALPcouplings} object is defined above/at the EW scale, the user can also provide the arguments \texttt{VuL}, \texttt{VdL}, \texttt{VuR} and \texttt{VdR}, corresponding to the unitary matrices that  diagonalize the SM Yukawa couplings in the quark sector,
\begin{equation}
    Y_u^\mathrm{diag} = V_{u_L}^\dagger Y_u V_{u_R}\,,\qquad Y_d^\mathrm{diag} = V_{d_L}^\dagger Y_d V_{d_R}\,.
\end{equation}
\alpaca automatically imposes consistency with the CKM matrix, so if \texttt{VuL} is provided by the user, \texttt{VdL} is calculated as $V_{d_L} = V_{u_L} V_\mathrm{CKM}$, while if \texttt{VdL} is provided, \texttt{VuL} is calculated as $V_{u_L} = V_{d_L}V_\mathrm{CKM}^\dagger$. If neither is provided, \alpaca uses $V_{u_L} = \mathbb{1}$ and $V_{d_L} = V_\mathrm{CKM}$, and providing both results in an error. If \texttt{VuR} or \texttt{VdR} are not provided, they are set equal to the identity. No unitary matrices are considered in the leptonic sector as neutrinos are treated as massless at the Lagrangian level. Notice that none of the observables calculable with \alpaca depends on the neutrino masses. 

As an example, the following code defines the ALP coupling to LH quarks equal to $1$, while the rest are vanishing, in the \verb|derivative_above| basis at the scale $\Lambda=1\TeV$. Additionally, in this case $Y_d$ is assumed to be diagonal and the electroweak scale is fixed to $91\GeV$:

\begin{minted}{python}
import alpaca
import numpy as np

couplings = alpaca.ALPcouplings(
    {'cqL': 1.0},
    scale=1000,
    basis='derivative_above',
    ew_scale=91,
    VdL = np.eye(3)
)
\end{minted}

Alternatively, another example consists in selecting a non-universal texture on RH leptons and introducing and ALP coupling to W bosons:

\begin{minted}{python}
import alpaca
import numpy as np

couplings = alpaca.ALPcouplings(
    {'ceR':np.array([[0.5, 0.1, 0],[-0.2, 0.4, 0],[0, 0, 0]]), 'cW':1.0},
    scale=1000,
    basis='derivative_above'
)
\end{minted}
In this case, as no unitary matrix $V_i$ as been defined, the corresponding assumption is $V_{u_L}=\mathbb{1}=V_{u_R}=V_{d_R}$ and $V_{d_L} = V_\mathrm{CKM}$, leading to a diagonal $Y_u$, but $Y_d^\mathrm{diag} = V_{d_L}^\dagger Y_d$. Moreover, the high-energy scale $\Lambda=1\,\mathrm{TeV}$, while the electroweak scale is fixed at the default value.

\subsection{Matching and running}
\label{sec:running_matching}
Effective field theories allow to focus on the low-energy phenomenology, integrating out the degrees of freedom heavier than a given energy scale $\Lambda$. Below this scale, the Wilson coefficients of the effective description encode the contributions from the heavier physics. They need to be ``run'' down to the characteristic energy scale of the observables that one is interested in computing. This procedure is known in the literature~\cite{MartinCamalich:2020dfe,Bauer:2020jbp,Bonilla:2021ufe,DasBakshi:2023lca,Bresciani:2024shu} and it is implemented in \alpaca\!\!.

\subsubsection{Theory}

The running of the ALP couplings and SM parameters is obtained by solving the Renormalisation Group equations, either numerically or relying on approximations:
\begin{equation}
    \frac{d}{d\log\mu} c_X(\mu) = \frac{\beta_X(\mu)}{16\pi^2}\;.
\end{equation}

\alpaca implements the 1-loop Renormalisation Group equations for the ALP couplings from Ref.~\cite{Bauer:2020jbp}. In the unbroken phase, the beta functions\footnote{To obtain these exact expressions, we have set the $\beta_i$ variables defined in Ref.~\cite{Bauer:2020jbp} to be $\beta_u=-1$, $\beta_d=\beta_e=1$ and $\beta_Q=\beta_L=0$. This corresponds to eliminating the operator $O_\phi$ in favour of RH quark operators.} are
\begin{align}
    \beta_{q_L} &= \frac{1}{2}\{Y_uY_u^\dagger + Y_dY_d^\dagger,\,c_{q_L}\}-Y_u c_{u_R} Y_u^\dagger - Y_d c_{d_R} Y_d^\dagger - \left(16\alpha_s^2 \tilde{c}_G + 9\alpha_2^2 \tilde{c}_W+\frac{1}{3}\alpha_1^2 \tilde{c}_B\right) \mathbb{1}\,,\nonumber\\
    \beta_{u_R} &= \{Y_u^\dagger Y_u,\,c_{u_R}\} -2 Y_u^\dagger c_{q_L} Y_u +\left(-2X + 16\alpha_s^2 \tilde{c}_G+\frac{16}{3}\alpha_1^2 \tilde{c}_B\right)\mathbb{1}\,, \nonumber\\
    \beta_{d_R} &= \{Y_d^\dagger Y_d,\,c_{d_R}\} -2 Y_d^\dagger c_{q_L} Y_d +\left(2X + 16\alpha_s^2 \tilde{c}_G+\frac{4}{3}\alpha_1^2 \tilde{c}_B\right)\mathbb{1}\,, \nonumber\\
    \beta_{\ell_L} &= \frac{1}{2}\{Y_eY_e^\dagger,\,c_{\ell_L}\} - Y_e c_{e_R} Y_e^\dagger - \left(9\alpha_2^2\tilde{c}_W+3\alpha_1^2 \tilde{c}_B\right)\mathbb{1}\,,\nonumber\\
    \beta_{e_R} &= \{Y_e^\dagger Y_e,\,c_{e_R}\}- 2Y_e^\dagger c_{\ell_L} Y_e + \left(2X+12\alpha_1^2\tilde{c}_B\right)\mathbb{1}\,,\nonumber \\
    \beta_G &=\beta_W = \beta_B = 0\,,
\end{align}
with the definitions
\begin{align}\label{eq:tilde_couplings}
    X &= \Tr[3 c_{q_L}(Y_u Y_u^\dagger-Y_dY_d^\dagger)-3c_{u_R}Y_u^\dagger Y_u +3c_{d_R}Y_d^\dagger Y_d -c_{\ell_L} Y_eY_e^\dagger +c_{e_R}Y_e^\dagger Y_e ]\,,\nonumber\\
    \tilde{c}_G &= c_G + \frac{1}{2} \Tr(c_{u_R}+c_{d_R}-2c_{q_L})\,,\nonumber\\
    \tilde{c}_W &= c_W -\frac{1}{2}\Tr(3c_{q_L}+c_{\ell_L})\,,\nonumber\\
    \tilde{c}_B &= c_B +\Tr\left(\frac{4}{3}c_{u_R}+\frac{1}{3}c_{d_R}-\frac{1}{6}c_{q_L}+c_{e_R}-\frac{1}{2}c_{\ell_L}\right)\,.
\end{align}

Additionally, \alpaca also implements the beta functions for the Yukawa couplings of the quarks,
\begin{align}
    \beta_{Y_u} &=\left(\frac{3}{2}Y_u Y_u^\dagger-\frac{3}{2}Y_dY_d^\dagger +\Tr(3 Y_uY_u^\dagger+3Y_dY_d^\dagger+Y_eY_e^\dagger)-\frac{9}{4}g_2^2-\frac{17}{12}g_1^2-8g_s^2\right)Y_u\,,\nonumber\\
    \beta_{Y_d} &=\left(\frac{3}{2}Y_d Y_d^\dagger-\frac{3}{2}Y_uY_u^\dagger +\Tr(3 Y_uY_u^\dagger+3Y_dY_d^\dagger+Y_eY_e^\dagger)-\frac{9}{4}g_2^2-\frac{5}{12}g_1^2-8g_s^2\right)Y_d\,.
\end{align}

The evolution of $Y_e$ and the gauge couplings is, on the other hand, extracted from the package \texttt{wilson}. In principle, the beta functions of the gauge running get a one-loop contribution from diagrams involving ALPs of order $m_a^2/f_a^2$, which \alpaca does not include. The Yukawa coupling do not receive ALP contributions at one loop.

The matching between the high-energy and low-energy regimes, obtained by integrating out $t$, $h$, $W^\pm$ and $Z^0$, is performed in the mass basis. The rotation to the mass basis is detailed in Eq.~\eqref{eq:massbasis}, with the Yukawa matrices and ALP couplings evaluated at the matching scale. This process is typically implemented integrating out at the same time the massive gauge bosons, the physical Higgs and the top-quark at a scale $\mu_\mathrm{EW}$, which is a good approximation. By default in the program this scale is set to $\mu_\mathrm{EW}=100 \,$GeV. These matching contributions to the diagonal fermion couplings have been computed in Ref.~\cite{Bauer:2020jbp}. Here we just report the results implemented in the code for completeness:

\begin{align}\label{eq:fermion_matching_1}
   c_{f_L}'(\mu_\mathrm{EW}) 
   &= \frac{3 y_t^2}{8\pi^2}\,(c_{u_R}-c_{q_L})_{33}\,\big( T_3^f - Q_f\,  s_w^2 \big)\,
    \ln\frac{\mu_\mathrm{EW}^2}{m_t^2}\,\mathbbm{1}+ \nonumber\\
   &\quad\mbox{}+ \frac{3\alpha^2}{8\pi^2}\,\bigg[    
    \frac{c_{WW}}{2s_w^4} \left( \ln\frac{\mu_\mathrm{EW}^2}{m_W^2} -\frac{13}{6} \right) 
    + \frac{2c_{\gamma Z}}{s_w^2\, c_w^2}\,Q_f\,\big( T_3^f - Q_f\, s_w^2 \big)
    \left( \ln\frac{\mu_\mathrm{EW}^2}{m_Z^2}  -\frac{13}{6} \right)+ \nonumber \\
   & \quad + \frac{c_{ZZ}}{s_w^4\, c_w^4}\,\big( T_3^f - Q_f\, s_w^2 \big)^2
    \left( \ln\frac{\mu_\mathrm{EW}^2}{m_Z^2} -\frac{19}{6} \right) \bigg]\,\mathbbm{1} 
    + \delta_{fd} \Delta c_{d_L}'(\mu_\mathrm{EW}) \,, \\
    \label{eq:fermion_matching_2}
   c_{f_R}'(\mu_\mathrm{EW}) 
   &= \frac{3 y_t^2}{8\pi^2}\,(c_{u_R}-c_{q_L})_{33}\,\big( - Q_f\, s_w^2 \big)+\,
    \ln\frac{\mu_\mathrm{EW}^2}{m_t^2}\,\mathbbm{1} \nonumber\\
   &\quad\mbox{}+ \frac{3\alpha^2}{8\pi^2}\,Q_f^2 \left[
    \frac{2c_{\gamma Z}}{c_w^2} \left( \ln\frac{\mu_\mathrm{EW}^2}{m_Z^2} -\frac{13}{6} \right) 
    - \frac{c_{ZZ}}{c_w^4} \left( \ln\frac{\mu_\mathrm{EW}^2}{m_Z^2} -\frac{19}{6} \right) \right] 
    \mathbbm{1} \, .
\end{align}
The off-diagonal matching contributions are only relevant for the down-type LH quarks
\begin{align}
    (\Delta c_{d_L}')_{ij}&=\frac{y_t^2}{16\pi^2}\,\bigg\{
    V_{mi}^* V_{nj} \left[ c_{u_L}'(\mu_\mathrm{EW}) \right]_{mn} \left( \delta_{m3} + \delta_{n3} \right)
    \left[ - \frac14\ln\frac{\mu_\mathrm{EW}^2}{m_t^2} - \frac38
    + \frac34\,\frac{1-x_t+\ln x_t}{\left(1-x_t\right)^2} \right]+ \notag\\
   &+ V_{3i}^* V_{3j} \left[ c_{u_L}'(\mu_\mathrm{EW}) \right]_{33} 
    + V_{3i}^* V_{3j} \left[ c_{u_R}'(\mu_\mathrm{EW}) \right]_{33} \left[ 
    \frac12\ln\frac{\mu_\mathrm{EW}^2}{m_t^2} - \frac14 
    - \frac32\,\frac{1-x_t+\ln x_t}{\left(1-x_t\right)^2} \right]+\notag\\
   &\hspace{1.7cm}\mbox{}- \frac{3\alpha}{2\pi s_w^2}\,c_{WW}\,V_{3i}^* V_{3j}\,
    \frac{1-x_t+x_t\ln x_t}{\left(1-x_t\right)^2} \bigg\} \,, 
\end{align}
where for shortness of notation $V\equiv V_\mathrm{CKM}$ here. Technically, at lower energies, one should integrate out the $b-$ and $c-$quarks: as this would lead to negligible $\sim 1\%$ corrections, this procedure is not included in \alpaca.

The gauge boson couplings to the ALP also get the contribution of integrating out the top, which in this case leads to 
\be
\begin{aligned}
       \tilde c_{G}(\mu\lesssim\mu_\mathrm{EW}) &= c_{G} + \frac12\,\sum_{q\ne t}\, c_{qq}(\mu) \,, \\
   \tilde c_{\gamma}(\mu\lesssim\mu_\mathrm{EW}) &=  c_{\gamma} + \sum_{f\ne t}\, N_c^f\, Q_f^2\,c_{ff}(\mu) \,.
\end{aligned}
\ee
 
All in all, the beta functions in the low-energy regime are
\be
\begin{aligned}
    \beta_{c_{q_R}} &=-\beta_{c_{q_L}} = 16\alpha_s^2 \tilde{c}_G + 12\alpha_\mathrm{em}^2 \tilde{c}_\gamma\,,\\
    \beta_{c_{e_R}} &= -\beta_{c_{e_L}} =  12\alpha_\mathrm{em}^2 \tilde{c}_\gamma\,,\\
    \beta_G &= \beta_\gamma = 0\,,
\end{aligned}
\ee
with
\be
\begin{aligned}
    \tilde{c}_G(\mu) =& c_G +\frac{1}{2}\sum_i (c_{u_R}^{ii}-c_{u_L}^{ii})\,\Theta(\mu-m_{u_i})  +\frac{1}{2}\sum_i (c_{d_R}^{ii}-c_{d_L}^{ii})\,\Theta(\mu-m_{d_i})\,,\\
    \tilde{c}_\gamma(\mu) =& c_\gamma +\frac{4}{3}\sum_i (c_{u_R}^{ii}-c_{u_L}^{ii})\,\Theta(\mu-m_{u_i})  +\frac{1}{3}\sum_i (c_{d_R}^{ii}-c_{d_L}^{ii})\,\Theta(\mu-m_{d_i})+\\
    &+\sum_i (c_{e_R}^{ii}-c_{e_L}^{ii})\,\Theta(\mu-m_{e_i})\,.
\end{aligned}
\ee

One must also consider the running of the SM parameters, which in \alpaca is handled internally using \texttt{flavio}\cite{flavio}.\\

All in all, \alpaca performs the matching and running of the Wilson coefficients down to the value of $0.04\GeV$, correctly taking into consideration the integration out of top quark, massive gauge bosons and physical Higgs. The integration out of the $b$-quark and any lighter particle is not implemented and the corresponding theoretical error is estimated to be at most of $1\%$.

\subsubsection{Implementation}\label{sec:running_impl}
Running and matching of the \texttt{ALPcouplings} object is implemented through the \verb|match_run|\index{\texttt{ALPcouplings}!\texttt{.match\_run}} method. Notice that in \alpaca the running is performed only towards a lower scale, otherwise an error appears. An example is:

\begin{minted}{python}
import alpaca

couplings_1000 = alpaca.ALPcouplings(
    {'cqL': 1.0},
    scale=1000,
    basis='derivative_above'
)

couplings_10 = couplings_1000.match_run(
    scale_out=10,
    basis='VA_below',
    integrator='scipy',
    beta='full',
    match_tildecouplings=True,
    scipy_method='DOP853',
    scipy_atol=1e-6,
    scipy_rtol=1e-8
)
\end{minted}
The arguments of \verb|match_run| are:
\begin{itemize}
    \item \verb|scale_out|: Scale of the ALP couplings after running and matching, in GeV.
    \item \texttt{basis}: Basis of the ALP couplings, as described in Section~\ref{sec:ALPbases}.
    \item \texttt{integrator}: Method used to integrate the Renormalisation Group equations. The available options are:
    \begin{itemize}
        \item \texttt{scipy} (default): Uses the function \verb|scipy.integrate.solve_ivp()|. This method is generally slower, but produces more precise solutions. 
        \item \texttt{leadinglog}: Uses the leading logarithm approximation. It is much faster, but less precise, specially when the initial and final scales are separated by several orders of magnitude.
        \item \texttt{symbolic}: Uses the leading logarithm approximation with symbolic algebra.
        \item \verb|no_rge|: Changes the scale, translates the basis and performs the matching without running the RG equations.
    \end{itemize}
    \item \texttt{beta}: If \texttt{"full"} (default), it uses the full expression of the $\beta$ functions; if \texttt{"ytop"}, it only retains the dependence on the Yukawa coupling $y_t$ and the gauge couplings.
    \item \verb|match_tildecouplings|: If \texttt{True}, in the matching at the electroweak scale, the ALP couplings to gauge vectors are replaced by their ``tilde'' versions defined in Eq.~\eqref{eq:tilde_couplings}
    which captures some of the 2-loops matching contributions. The default is \texttt{True}. 
    \item \verb|scipy_method|: One of the available methods for \verb|scipy.integrate.solve_ivp| compatible with complex arguments: \texttt{"RK23"}, \texttt{"RK45"} and \texttt{"DOP853"} are explicit Runge-Kutta methods,
    
    and \texttt{"BDF"} uses a backward differentiation formula. The default is \texttt{"RK45"}, while \texttt{"DOP853"} is recommended when using small tolerances.
    \item \verb|scipy_atol| and \verb|scipy_rtol| are the absolute and relative tolerances, respectively, used by \verb|scipy.integrate.solve_ivp|. The integrator stops when the estimate of the error is smaller than \verb|scipy_atol + scipy_rtol * abs(c)|, where \texttt{c} is the corresponding ALP coupling.
\end{itemize}
Most of the functions described in the following sections involve the running of ALP couplings, and consequently will accept the optional arguments \texttt{integrator}, \texttt{beta},\linebreak
\verb|match_tildecouplings|, \verb|scipy_method|, \verb|scipy_atol| and \verb|scipy_rtol|.

Internally, the running above the EW scale is implemented in the\linebreak 
\verb|derivative_above| basis, and below the EW scale in the \verb|RL_below| basis, while the matching is performed between \verb|massbasis_ew| and \verb|RL_below|. 

It follows that if the input parameters are given at the EW scale in the \verb|massbasis_ew| basis or below the EW scale in the \verb|VA_below| one, then the couplings are translated to the \verb|RL_below| basis before performing the running. On the other hand, if the user requests the results to be expressed in terms of the \verb|massbasis_ew| or \verb|VA_below| bases, then the couplings are translated to these bases after the running.

\subsection{Other operations with ALP couplings}
The \texttt{ALPcouplings} objects are subscriptable\index{\texttt{ALPcouplings}!subscripting}, so individual couplings can be accessed and modified as if they were elements of a \texttt{dict}:

\begin{minted}{python}
couplings = alpaca.ALPcouplings(
    {'cqL': 1.0},
    scale=1000,
    basis='derivative_above'
)

couplings['cW'] = 0.5
print(f"c_qL^33 = {couplings['cqL'][2,2]}\n c_W = {couplings['cW']}")
\end{minted}
\vspace{-8mm}
\begin{minted}{pwsh-session}
c_qL^33 = 1.0
c_W = 0.5
\end{minted}

Two \texttt{ALPcouplings} objects defined at the same scale and basis can be added or subtracted, and an \texttt{ALPcouplings} object can be multiplied or divided by a number. \texttt{ALPcouplings} can also be multiplied by \texttt{numpy} arrays, but only if the array is in the left-hand side of the multiplication~\index{\texttt{ALPcouplings}!arithmetic operations}. For example, the following code creates a grid of logarithmically generated couplings in the $c_\gamma-c_W$ plane in the range $10^{-8}$ to $1$ by using addition and multiplication:

\begin{minted}{python}
import numpy as np

coupling_gamma = alpaca.ALPcouplings(
    {'cgamma': 1.0},
    scale=1000,
    basis='massbasis_ew'
)
coupling_W = alpaca.ALPcouplings(
    {'cW': 1.0},
    scale=1000,
    basis='massbasis_ew'
)

cgamma = np.logspace(-8, 0, 100)
cW = np.logspace(-8, 0, 100)
x_cgamma, y_cW = np.meshgrid(cgamma, cW)
couplings_grid = x_cgamma * coupling_gamma + y_cW * coupling_W
\end{minted}

\texttt{ALPcouplings} objects, as well as any Python container (\texttt{list}, \texttt{tuple}, \texttt{dict}) that stores them, can be saved as JSON files. For individual objects, the methods \texttt{save}\index{\texttt{ALPcouplings}!\texttt{.save}} and \texttt{load}\index{\texttt{ALPcouplings}!\texttt{.load}} can be used
\begin{minted}{python}
couplings_grid[13,15].save('ALP_coupling.json')
coupling_fromfile = ALPcouplings.load('ALP_coupling.json')
\end{minted}
Alternatively, the argument to \texttt{save} and \texttt{load} can also be a text stream, for example a file object
\begin{minted}{python}
with open('ALP_coupling.json', 'wt') as f:
    couplings_grid[13,15].save(f)
\end{minted}

For containers, instead, the \texttt{ALPcouplingsEncoder}\index{\texttt{ALPcouplingsEncoder}} and \texttt{ALPcouplingsDecoder}\index{\texttt{ALPcouplingsDecoder}} classes must be used in conjunction with the \texttt{json} module of the Python standard library:
\begin{minted}{python}
from alpaca import ALPcouplingsEncoder, ALPcouplingsDecoder
import json

with open('ALP_grid.json', 'wt') as f:
    # Numpy arrays can not be stored directly as JSON
    # We have to convert it to list before
    json.dump(couplings_grid.tolist(), f, cls=ALPcouplingsEncoder)

with open('ALP_grid.json', 'rt') as f:
    couplings_grid2 = json.load(f, cls=ALPcouplingsDecoder)
\end{minted}

%% file: uv_model.tex
In \alpaca it is possible to work with UV completions. To do so, the user can either work with preexisting models already implemented in the program, or create their custom model by defining the relations that the EFT couplings must satisfy according to the conditions established by the UV completion. Let us start with the first case.

\subsection{Predefined classes of models}
The preexisting models present in \alpaca are the traditional invisible QCD axions constructions, the DFSZ \cite{Zhitnitsky:1980tq,Dine:1981rt} and KSVZ \cite{Kim:1979if,Shifman:1979if}, and the more recent Flaxion model~\cite{Ema:2016ops,Calibbi:2016hwq}. The user can modify the charge assignment of the PQ symmetry within these constructions to adapt to the specific case considered (see the accompanying paper~\cite{Alda:2025uwo} for examples). 

For models where the SM fermions are charged under $\mathrm{U}(1)_\mathrm{PQ}$ (e.g. DFSZ-like, flaxion, etc.) with charge $\mathcal{X}_{f_i}$, where $f= q_L, u_R, d_R, \ell_L, e_R$ and $i=1,2,3$, the matching to the ALP Lagrangian in the derivative basis is
\begin{align}
    c_f^{ij} &= -\delta_{ij}\mathcal{X}_{f_i}\,,\nonumber\\
    N = \frac{N_\mathrm{DW}}{2}= c_G &= -\frac{1}{2}\sum_i \Big(2\mathcal{X}_{q_L i} - \mathcal{X}_{u_Ri}-\mathcal{X}_{d_Ri}\Big)\,,\nonumber\\
    c_W  &= -\frac{1}{2}\sum_i\Big(3\mathcal{X}_{q_L i}+\mathcal{X}_{\ell_L i}\Big)\,,\nonumber\\
    c_B &= -\sum_i \left(\frac{1}{6}\mathcal{X}_{q_L i} -\frac{4}{3}\mathcal{X}_{u_R i}-\frac{1}{3}\mathcal{X}_{d_R i} +\frac{1}{2}\mathcal{X}_{\ell_L i} - \mathcal{X}_{e_R i}\right)\,,\nonumber\\
    E = c_\gamma = c_B + c_W &= -\sum_i \left(\frac{5}{3}\mathcal{X}_{q_L i} -\frac{4}{3}\mathcal{X}_{u_R i}-\frac{1}{3}\mathcal{X}_{d_R i} +\mathcal{X}_{\ell_L i} - \mathcal{X}_{e_R i}\right)\,.
\end{align}
Notice that it is possible to express the ALP--fermion couplings in the chirality preserving basis, as no shift symmetry breaking terms are present: it is straightforward to check that the shift symmetry invariance conditions reported in Ref.~\cite{Chala:2020wvs} are satisfied. 

On the contrary, for KSVZ-like models with heavy fermions $\mathcal{F}_i$, taken to be LH and each fermion charged under $\mathrm{SU}(3)_C\times \mathrm{SU}(2)_L\times \mathrm{U}(1)_Y\times \mathrm{U}(1)_\mathrm{PQ}$ with charges $(\mathcal{C}_i,\,\mathcal{I}_i,\,\mathcal{Y}_i,\,\mathcal{X}_i)$, the generated couplings to gauge bosons are
\begin{align}\label{eq:match_KSVZ}
    c_G &= -\sum_i \mathcal{X}_i\,d(\mathcal{I}_i)\,T(\mathcal{C}_i)\,,\nonumber\\
    c_W &= -\sum_i \mathcal{X}_i\,d(\mathcal{C}_i)\,T(\mathcal{I}_i)\,,\nonumber\\
    c_B &= -\sum_i \mathcal{X}_i\,d(\mathcal{C}_i)\,d(\mathcal{I}_i)\,\mathcal{Y}_i^2\,,
\end{align}
where $d$ is the dimension of the representation, and $T$ its Dynkin index. For representations of $\mathrm{SU}(2)$, if $d(\mathcal{I}_i)=2j_i+1$ where $j_i$ is the isospin, then $T(\mathcal{I}_i) = j_i(j_i+1)(2j_i+1)/3$. For $\mathrm{SU}(3)_c$, each representation can be identified by two Dynkin labels $\mathcal{C}_i \equiv (\alpha_i\,,\beta_i)$ with $\alpha,\,\beta\in\mathbb{N}$,\footnote{Physically, a composite system formed by $\alpha$ quarks and $\beta$ anti-quarks will be in the reducible representation $\underbrace{\boldsymbol{3}\otimes\cdots\otimes\boldsymbol{3}}_\alpha\otimes\underbrace{\boldsymbol{\bar{3}}\otimes\cdots\otimes\boldsymbol{\bar{3}}}_\beta$, and its largest irreducible component is the representation $(\alpha,\,\beta)$.} then its dimension and Dynkin index are~\cite{Slansky:1981yr}
\begin{align}
    d(\mathcal{C}_i) &= \frac{1}{2}(\alpha_i+1)(\beta_i+1)(\alpha_i+\beta_i+2)\,,\nonumber\\
    T(\mathcal{C}_i) &= \frac{1}{24}d(\mathcal{C}_i)(\alpha_i^2+3\alpha_i+\alpha_i\beta_i+3\beta_i+\beta_i^2)\,.
\end{align}
For example, for the singlet representation $\boldsymbol{1}\equiv(0,\,0)$, $T(\boldsymbol{1})=0$, while for the colour triplet one $\boldsymbol{3}\equiv(1,\,0)$, $T(\boldsymbol{3}) = 1/2$, and for the colour octet $\boldsymbol{8}\equiv(1,\,1)$, $T(\boldsymbol{8}) = 3$. 
Also note that the conjugate representation of $\mathcal{C}\equiv(\alpha,\,\beta)$ is $\bar{\mathcal{C}}\equiv (\beta,\,\alpha)$, and consequently $T(\mathcal{C})=T(\bar{\mathcal{C}})$.

A final comment is in order in the case of the ``classic'' DFSZ model: loop contributions have been computed in Refs.~\cite{Freytsis:2009ct,Gao:2025ohi} and we estimate that they provide a contribution of at most $\mathcal{O}(10\%)$. We do not include these contributions in the present version of the code.

\subsubsection{Implementation}
The various UV-complete models are included in the module \texttt{alpaca.uvmodels}\index{\texttt{uvmodels}}. The python class \texttt{uvmodels.PQChargedModel}\index{\texttt{uvmodels}!\texttt{.PQChargedModel}} describes models in which the SM fermions are charged under the PQ symmetry.
The models are implemented with a \texttt{dict} with the PQ charges for each type of fermion, \texttt{"qL"}, \texttt{"uR"}, \texttt{"dR"}, \texttt{"lL"} and \texttt{"eR"}; if any of the charges is not specified, then it is assigned the default value $\mathcal{X}=0$. If the value is a number, flavour universality is assumed, while 3-elements lists are used for non-universal charges.

As an example, let us consider the DFSZ-IV model with maximal $E/N$ introduced in Ref.~\cite{DiLuzio:2017pfr}, that exhibits the non-universal PQ charges
\begin{equation}
    \mathcal{X}_{u_R} = (2,\ 6,\ 12)\mathcal{X}_\Phi\,,\qquad\mathcal{X}_{d_R} = -(0,\ 4,\ 14)\mathcal{X}_\Phi\,,\qquad\mathcal{X}_{e_R} = (14,\ 42,\ 98)\mathcal{X}_\Phi\,.
\end{equation}
It can be implemented with the following code:
\begin{minted}{python}
from alpaca.uvmodels import PQChargedModel
import sympy as sp

X = sp.Symbol(r'\mathcal{X}_\Phi')
maximal_model = PQChargedModel("DFSZ-IV with maximal E/N", {
    'uR': [2*X, 6*X, 12*X],
    'dR': [0, -4*X, -14*X],
    'eR': [14*X, 42*X, 98*X]
})

couplings = maximal_model.get_couplings({X: 1.0}, scale=1000.0)
print(maximal_model.couplings_latex())
\end{minted}
The result is given by the couplings
\begin{align}
c_{u_R} &= \left(\begin{matrix}- 2 \mathcal{X}_\Phi & 0 & 0\\0 & - 6 \mathcal{X}_\Phi & 0\\0 & 0 & - 12 \mathcal{X}_\Phi\end{matrix}\right) \nonumber \\
c_{d_R} &= \left(\begin{matrix}0 & 0 & 0\\0 & 4 \mathcal{X}_\Phi & 0\\0 & 0 & 14 \mathcal{X}_\Phi\end{matrix}\right) \nonumber \\
c_{e_R} &= \left(\begin{matrix}- 14 \mathcal{X}_\Phi & 0 & 0\\0 & - 42 \mathcal{X}_\Phi & 0\\0 & 0 & - 98 \mathcal{X}_\Phi\end{matrix}\right) \nonumber \\
c_G &= \mathcal{X}_\Phi \nonumber \\
c_B &= \frac{524 \mathcal{X}_\Phi}{3} \,.
\end{align}

The flaxion is a further type of models in which the SM fermions are charged under $\mathrm{U}(1)_\mathrm{PQ}$. In addition to the ALP couplings, the PQ charges also determine the texture of the Yukawa matrices. In \alpaca they are implemented with the class \texttt{uvmodels.Flaxion}\index{\texttt{uvmodels}!\texttt{.Flaxion}}. The following code defines a flaxion model with the assignement of PQ charges $\mathcal{X}_{q_L} = (4, 3, 0)$, $\mathcal{X}_{u_R} = (-4, -1, 0)$, $\mathcal{X}_{d_R} = (-4, -3, -3)$, $\mathcal{X}_{\ell_L} = (1, 0, 0)$, $\mathcal{X}_{e_R} = (-5, -3, -2)$ and $\epsilon = \langle \phi\rangle/\Lambda = 0.2$:

\begin{minted}{python}
from alpaca.uvmodels import Flaxion
my_flaxion = Flaxion("Flaxion", {
    'qL': [4, 3, 0],
    'uR': [-4, -1, 0],
    'dR': [-4, -3, -3],
    'lL': [1, 0, 0],
    'eR': [-5, -3, -2]
})

couplings_myflaxion = my_flaxion.get_couplings(eps=0.2, scale=1000.0)
\end{minted}

In order to define KSVZ-like models, we first need to specify the heavy fermions that are added to the theory. They are described by the class \texttt{alpaca.uvmodels.HeavyFermion}\index{\texttt{uvmodels}!\texttt{.HeavyFermion}}. The user is requested to provide the representation of each fermion under $\mathrm{SU}(3)_c\times \mathrm{SU}(2)_L\times \mathrm{U}(1)_Y\times \mathrm{U}(1)_\mathrm{PQ}$. For $\mathrm{SU}(3)_c$, the argument must be a \texttt{int} or \texttt{str} indicating the dimension of the representation (so $\boldsymbol{3}$ is \texttt{3} or \texttt{"3"}), while the conjugate representation is obtained by appending \texttt{\_bar} at the end (so $\boldsymbol{\bar{3}}$ becomes \texttt{"3\_bar"}), and primes are used when there is more than one representation of the same dimension (e.g. $\boldsymbol{15}$ and $\boldsymbol{15'}$). Alternatively, the $\mathrm{SU}(3)_c$ representation can be specified with a 2-element tuple for the Dynkin labels. The $\mathrm{SU}(2)_L$ representation is specified by its dimension, and $\mathrm{U}(1)_Y$ and $\mathrm{U}(1)_\mathrm{PQ}$ by their respective charges.

The KSVZ-like models are then created with the class \texttt{alpaca.uvmodels.KSVZ\_model}\index{\texttt{uvmodels}!\texttt{.KSVZ\_model}}, passing as an argument the list of the \texttt{alpaca.uvmodels.HeavyFermion} objects representing the new heavy fermions. For example, the following model including a heavy lepton $\mathcal{E}\sim(\boldsymbol{1},\,\boldsymbol{1},\,\mathcal{Y},\,\mathcal{X})$ only generates an anomaly for the $B$ field, but not for the gluon or the $W^\pm$ bosons:
\begin{minted}{python}
from alpaca.uvmodels import HeavyFermion, KSVZ_model
import scipy as sp

X = sp.Symbol(r"\mathcal{X}")
Y = sp.Symbol(r"\mathcal{Y}")
E = HeavyFermion(1, 1, Y, X)
b_ksvz = KSVZ_model("B-KSVZ", [E])

couplings = b_ksvz.get_couplings({X: 1, Y: 1}, scale=1000.0)
print(b_ksvz.couplings)
\end{minted}
which produces $c_B = \mathcal{Y}^2\mathcal{X}$ and $c_G=c_W=0$.

Table~\ref{tab:uv_models} summarises the predefined models implemented in \alpaca\!\!. The DFSZ-like models use the angle $\beta$, given by \texttt{alpaca.uvmodels.beta}\index{\texttt{uvmodels}!\texttt{.beta}}, and the KSVZ-like models the PQ charge $\mathcal{X}$ given by \texttt{alpaca.uvmodels.KSVZ\_charge}\index{\texttt{uvmodels}!\texttt{.KSVZ\_charge}}. An example of how to obtain the couplings from these predefined models is given below:
\begin{minted}{python}
import alpaca
import numpy as np
from alpaca.uvmodels import QED_DFSZ, beta
from alpaca.uvmodels import Q_KSVZ, KSVZ_charge

fa = 1000

couplings_QED_DFSZ = QED_DFSZ.get_couplings({beta: np.pi/4}, 4*np.pi*fa)
couplings_Q_KSVZ = Q_KSVZ.get_couplings({KSVZ_charge: 1}, 4*np.pi*fa)
\end{minted}

\begin{table}[tb]
    \centering
    \begin{tabular}{c|c|c}
        Model & Type & Definition \\\hline
        \rowcolor{lightgray!40}\texttt{uvmodels.QED\_DFSZ}\index{\texttt{uvmodels}!\texttt{.QED\_DFSZ}} & DFSZ-like & $\mathcal{X}_{e_R} = -2\cos^2\beta$, $\mathcal{X}_{d_R} = -\mathcal{X}_{u_R} = 2\sin^2\beta$ \\
        \texttt{uvmodels.u\_DFSZ}\index{\texttt{uvmodels}!\texttt{.u\_DFSZ}} & DFSZ-like & $\mathcal{X}_{e_R} = 1$, $ \mathcal{X}_{d_R} = -2$ \\
        \rowcolor{lightgray!40}\texttt{uvmodels.e\_DFSZ}\index{\texttt{uvmodels}!\texttt{.e\_DFSZ}} & DFSZ-like & $\mathcal{X}_{u_R} = 1$, $\mathcal{X}_{d_R}  = 1$ \\
        \texttt{uvmodels.Q\_KSVZ}\index{\texttt{uvmodels}!\texttt{.Q\_KSVZ}} & KSVZ-like & $\mathcal{Q} \sim (\boldsymbol{3},\,\boldsymbol{1},\,0,\,\mathcal{X})$ \\
        \rowcolor{lightgray!40}\texttt{uvmodels.L\_KSVZ}\index{\texttt{uvmodels}!\texttt{.L\_KSVZ}} & KSVZ-like & $\mathcal{E} \sim (\boldsymbol{1},\,\boldsymbol{2},\,0,\,\mathcal{X})$\\
        \texttt{uvmodels.Y\_KSVZ}\index{\texttt{uvmodels}!\texttt{.Y\_KSVZ}} & KSVZ-like & $\mathcal{Y} \sim (\boldsymbol{1},\,\boldsymbol{1},\,\frac{1}{2},\,\mathcal{X})$ \\
        \rowcolor{lightgray!40} &  & $\mathcal{X}_{q_L} = (3, 2, 0)$, $\mathcal{X}_{u_R} =(-5,-1,0)$,  \\
        \rowcolor{lightgray!40} &  & $\mathcal{X}_{d_R} = (-4,-3,-3)$ , $\mathcal{X}_{\ell_L} = (1,0,0)$\\
        \rowcolor{lightgray!40}\multirow{-3}{*}{\texttt{uvmodels.flaxion\_benchmark}}\index{\texttt{uvmodels}!\texttt{.flaxion\_benchmark}} & \multirow{-3}{*}{Flaxion} & $\mathcal{X}_{e_R} = (-8, -5,-3)$ \\
         && $\mathcal{X}_{d_R} = \mathcal{X}_{e_R} = -\cos^2\beta$, $\mathcal{X}_{u_R} = -\sin^2\beta$ \\
        \multirow{-2}{*}{\texttt{uvmodels.nonuniversal\_DFSZ}}\index{\texttt{uvmodels}!\texttt{.nonuniversal\_DFSZ}} & \multirow{-2}{*}{ DFSZ-like } & $\mathcal{X}_{q_L} = -\mathcal{X}_{\ell_L} = (0,0,1)$
    \end{tabular}
    \caption{\em Models predefined in \texttt{alpaca.uvmodels}.}
    \label{tab:uv_models}
\end{table}

\subsection{User-defined classes of models}
It is also possible to work with novel classes of UV completions, outside DFSZ or KSVZ setups. The only requisite is that the user provides \alpaca with the relations among the couplings in the ``derivative'' basis above EW scale (see Eq.~\eqref{eq:derivative_above}). Notice that the only supported basis is \verb|derivative_above|.

\subsubsection{Implementation}
All the UV models implemented in \alpaca inherit from the class \texttt{uvmodels}\texttt{.ModelBase}\index{\texttt{uvmodels}!\texttt{.ModelBase}}. The following code shows how users can create their own models by subclassing and implementing the \verb|__init__| method. The couplings are defined in \texttt{self.couplings}, a \texttt{dict} with the same keys as the \texttt{derivative\_above} basis and values formed with \texttt{sympy} expressions. The code also creates an instance of the newly defined model, and obtains its couplings in \LaTeX{} format and as an \texttt{ALPcouplings} object.

\begin{minted}{python}
from alpaca.uvmodels import ModelBase
import sympy as sp

fermion_couplings = ['cuR', 'cdR', 'cqL', 'ceR', 'clL']
class UniversalModel(ModelBase):
    """Model with universal couplings to fermions."""
    def __init__(self, name: str, cf: sp.Expr):
        super().__init__(name)
        self.couplings = {c: cf for c in fermion_couplings}
        
cf = sp.Symbol("c_f")
univ = UniversalModel("Universal", cf)

print(univ.couplings_latex())
couplings_universal = univ.get_couplings({cf: 1.0}, scale=1000)
\end{minted}

\vspace{-8mm}
\begin{minted}{pwsh-session}
\begin{align}
c_{u_R} &= c_{f} \nonumber \\
c_{d_R} &= c_{f} \nonumber \\
c_{q_L} &= c_{f} \nonumber \\
c_{e_R} &= c_{f} \nonumber \\
c_{\ell_L} &= c_{f} \nonumber 
\end{align}
\end{minted}

\texttt{ModelBase} and its subclasses define the following methods:
\begin{itemize}
    \item \texttt{get\_couplings}: Returns an \texttt{ALPcouplings} object obtained by replacing the arithmetic values of the couplings by numerical values.
    \item \texttt{couplings\_latex}: Returns the couplings of the model, in \LaTeX{} format. The non-zero couplings are displayed in an \texttt{align} environment, and equation numbers can be added with the argument \texttt{eqnumber=True}.
    \item \texttt{E\_over\_N}: Returns the algebraic expression for $E/N=c_\gamma/c_G$.
\end{itemize}

%% file: alp_process.tex
In this section we focus on the low-energy phenomenology of ALPs, that is, their production and decays. In this first version of \alpaca\!\!, only production in mesonic searches has been implemented, with the exception of Lepton--Flavour--Violating (LFV) decays such as $\mu \to e a$.
In terms of ALP decays, the program includes the state-of-the-art hadronic decays that has developed quite rapidly in the last year, see Refs.~\cite{Bai:2024lpq,Ovchynnikov:2025gpx,Bai:2025fvl,Balkin:2025enj}.
The goal of this section is to include all relevant formulae implemented in \alpaca\!\!, without detailing the computations, and to reference the original works from which they were obtained.

\subsection{ALP decays}\label{sec:ALPdecays}
Depending on its mass, the ALP can decay into different particles. A summary of all the decay channels implemented in the program can be seen in Table \ref{tab:decay-channel}. In the remainder of this subsection, we will delve into the details regarding the calculation of each process.

\begin{table}[h!]
    \centering
    \begin{tabular}{c|c}
     Decay channel & Final particles  \\\hline
     \rowcolor{lightgray!40} Bosonic & $\gamma\gamma$, $gg$\\
     Fermionic & $e^+e^-$, $\mu^+\mu^-$, $\tau^+\tau^-$, $c\bar{c}$, $b\bar{b}$, $\mu^\pm e^\mp$, $\tau^\pm e^\mp$, $\tau^\pm\mu^\mp$\\
     \rowcolor{lightgray!40} Hadronic & $\pi^0\pi^0\pi^0$, $\pi^0\pi^+\pi^-$, $\gamma\pi\pi$, $\eta^{(\prime)} \pi^0\pi^0$, $\eta^{(\prime)} \pi^+\pi^-$, $\omega\omega$ \\
     Dark sector & $\chi\bar{\chi}$
    \end{tabular}
    \caption{\em ALP decays implemented in \alpaca\!\!.}
    \label{tab:decay-channel}
\end{table}

\subsubsection{Decays to photons}
In the case of the decay of ALP to photons, it is important to consider the mixing between the ALP and the lighter neutral pseudoscalar mesons, namely $\pi$, $\eta$ and $\eta^\prime$. According to Ref.~\cite{Aloni:2018vki}, it is possible to write the decay width as
\begin{equation}
\label{eq:decay-photon}
\Gamma \left(a\rightarrow \gamma\gamma\right)=\frac{\alpha_\text{EM}m_a^3}{(4\pi)^3f_a^2}\left|c_\gamma +\mathcal{C}_\gamma^\chi+\mathcal{C}_\gamma^\text{VMD}+ \mathcal{C}_\gamma^\ell+\mathcal{C}_\gamma^{q}+\mathcal{C}_\gamma^W\right|^2\,,
\end{equation}
where in addition to the direct contribution from $c_\gamma$, we have included the following effective contributions.
\\
\noindent\textbf{Fermion contributions:}
Leptons and quarks (in the perturbative regime) contribute to the photon couplings via one-loop diagrams~\cite{Bauer:2020jbp} as
    \begin{align}
        \mathcal{C}_\gamma^\ell &= \sum_{\ell=e,\mu,\tau} c_\ell^A B_1(4m_\ell^2/m_a^2)\,, \\
        \mathcal{C}_\gamma^q &= \sum_q N_c Q_q^2 c^A_q B_1(4m_q^2/m_a^2)\,
    \end{align}
    where $B_1$ is the loop function, defined as
    \begin{equation}
        B_1(x)=1-xf^2(x),\quad \text{with}\; f(x)=\begin{cases} \arcsin{\frac{1}{\sqrt{x}}} & x\geq 1 \\
                     \frac{\pi}{2}+\frac{i}{2}\log{\frac{1+\sqrt{1-x}}{1-\sqrt{1-x}}} &  x<1\,.
       \end{cases}
    \end{equation}
    
Above $m_a>2.1\,\mathrm{GeV}$, there is a two-loop effect in which the coupling to gluons generates an effective coupling to quarks, which in turn generates an effective coupling to photons. This contribution can be estimated as~\cite{Bauer:2020jbp}
    \begin{equation}
        \mathcal{C}_\gamma^\mathrm{pQCD(2)} = -\frac{N_c \alpha_s^2(m_a)}{2\pi^2}c_G \sum_q Q_q^2 B_1(4m_q^2/m_a^2) \log\frac{\Lambda^2}{\tilde{m}_q^2}\,,
    \end{equation}
The contribution of the top quark is only taken into account in the regime where it has not been integrated out. In the 2-loop contribution, the logarithms are regulated by the mass of the quarks, $\tilde{m}_q=m_q$.\\
\noindent\textbf{Chiral Perturbation Theory:} 
Contribution from the chiral rotation, which is present even when $c_\gamma=0$ and is a result of the chiral rotation to quark fields
    \begin{equation}
        \mathcal{C}_\gamma^\chi = -2c_G N_c \mathrm{Tr}(\boldsymbol{\kappa}\mathbf{Q}^2) = -1.92(4)c_G\,,
    \end{equation}
where this result is computed at NLO in $\chi$PT.\\
\noindent\textbf{Vector meson dominance:}
The photon $\gamma$ interacts with hadrons by first mixing into vector mesons, which will then interact with other mesons. Then, ALP interactions come from its mixing with pseudoscalar meson. This type of contribution can be extracted by using the framework of Vector Meson Dominance, see Ref.~\cite{Aloni:2018vki}, that allows us to write these effective interactions as
\begin{equation}
    \mathcal{C}_\gamma^\mathrm{VMD} = \mathcal{F}(m_a) \frac{f_a}{f_\pi}\left[3\mathrm{Tr}(\boldsymbol{a} \boldsymbol{\rho^0}\boldsymbol{\rho^0}) + \frac{1}{3}\mathrm{Tr}(\boldsymbol{a} \boldsymbol{\omega}\boldsymbol{\omega}) + \frac{2}{3} \mathrm{Tr}(\boldsymbol{a} \boldsymbol{\phi}\boldsymbol{\phi}) +2 \mathrm{Tr}(\boldsymbol{a} \boldsymbol{\rho^0}\boldsymbol{\omega}) \right]\,.
\end{equation}
Here, the trace takes into account the $\chi$PT group factors, while $\mathcal{F}(m_a)$ is obtained by fitting to experimental data (see Fig.2 in Ref.~\cite{Aloni:2018vki}). \alpaca follows the procedure outlined in Ref.~\cite{Ovchynnikov:2025gpx} in order to compute explicitly this contribution $\mathcal{C}_\gamma^\mathrm{VMD}$.\\
\noindent\textbf{Electroweak contributions:}
     Finally, if the $W^\pm$ bosons have not been integrated out, they also generate a one-loop contribution,
    \begin{equation}
        \mathcal{C}_\gamma^W = \frac{2\alpha_\mathrm{em}}{\pi s^2_w}c_W B_2(4m_W^2/m_a^2)\,,
    \end{equation}
which for light ALPs is strongly suppressed, since $B_2(4m_W^2/m_a^2) \approx m_a^2/(6m_W^2)$.

\subsubsection{Decays to hadrons}
Depending on its mass, the ALP can decay to different hadrons. In \alpaca\!\!, the most relevant hadronic decays have been included: $a\rightarrow 3\pi$, $a\rightarrow \omega\omega$, $a\rightarrow \pi\pi\gamma$, $a\rightarrow \eta \pi\pi$, and $a\to\eta^\prime \pi\pi$. 
The expressions of the decay widths are extracted from Ref.~\cite{Ovchynnikov:2025gpx}, including the most general expressions accounting for the contributions of all the couplings. In this case, for $m_a\lesssim 1\,$GeV the contribution is basically the tree-level $\chi$PT computation~\cite{Bauer:2021mvw,DiLuzio:2022tbb}, while for heavier masses $\chi$PT is modified by the resonant contributions extracted from experimental data. 

For the regime of heavier masses we use the quark-hadron duality to estimate the total decay of the ALP into hadrons. This decay is estimated as~\cite{Bauer:2021mvw}
\begin{equation}
    \Gamma(a\to\mathrm{light\ hadrons})\Big|_{m_a\gtrsim1.8\GeV} \approx \Gamma(a\to gg) = \frac{\alpha_s^2}{8\pi^3}\frac{m_a^3}{f_a^2}|c_G^\mathrm{eff}|^2 \left(1 +\frac{\alpha_s}{4\pi}\frac{291-14n_q}{12}\right)\, ,
\end{equation}
where the coupling $c_G^\mathrm{eff}$ accounts for the one loop quark effects
\begin{equation}\label{eq:cG_eff}
    c_G^\mathrm{eff} = c_G + \frac{1}{2} \sum_{q\neq t} c_{q} B_1\left(\frac{4m_q^2}{m_a^2}\right)\,. 
\end{equation}

The crossover from one regime to the other is dynamically implemented in \alpaca\!\!, such that the total decay width is ``differentiable''. Concretely, the regime--change happens when the perturbative estimation of the decays is larger than the sum of the exclusive hadronic channels,
\begin{equation}
    \Gamma(a\to gg)+\Gamma(a\to \bar{b}b)+\Gamma(a\to \bar{c}c) \geq \sum_i \Gamma(a\to \mathrm{had}_i),
\end{equation}
where here $\Gamma(a\to \mathrm{had}_i)$ are the explicit individual decays listed above.

\subsubsection{Decays to fermions}
Contrary to the previous sections, the decay of the ALP to fermions is relatively simple. Depending on the ALP mass, different fermionic channels will be allowed, as $m_a\geq 2m_f$ must be satisfied. The decay rate of $a\rightarrow f \bar{f}$ is given by:
\begin{equation}
\Gamma\left(a\rightarrow f \bar{f}\right)=N_f\frac{m_a m_f^2}{8\pi f_a^2}\left|c_f^{\text{eff}}\right|^2\sqrt{1-\frac{4 m_f^2}{m_a^2}}\,,
\end{equation}
where $c_f^\text{eff}$ stands for the effective coupling to fermions, which includes the one-loop contributions of the matching and running of Eqs.~\eqref{eq:fermion_matching_1} and \eqref{eq:fermion_matching_2}, and the loop contribution by photons and gluons~\cite{Bauer:2017ris}
\begin{equation}
    c_{f}^\textrm{eff}=(c_{f_R}-c_{f_L})(\mu)-12 \left(Q_f^2 \alpha_\mathrm{em}^2\, c_\gamma+\delta_{fq}\alpha_\mathrm{S}^2 \, c_G \right) \left[\log\frac{\mu^2}{m_f^2}-\frac{11}{3}+g\left(\frac{4m_f^2}{m_a^2}\right)\right] \, ,
\end{equation}
with
\begin{equation}
    g(\tau) = 5 + \frac{4}{3} \int_0^1 dx \, \frac{1 - 4\tau(1 - x)^2 - 2x + 4x^2}{\sqrt{\tau(1 - x)^2 - x^2}} \arctan\left( \frac{x}{\sqrt{\tau(1 - x)^2 - x^2}} \right),
\end{equation}
and where the gluon contribution is only applied to quarks.
\subsubsection{Decay into dark sector}\label{sec:darksector}
\alpaca can also accommodate models where the ALP acts as a portal to a dark sector through the decay $a\to\chi\bar{\chi}$. The calculation of $\mathrm{BR}(a\to\chi\bar{\chi})$ is out of the scope of the program, and instead should be provided by the user. The dark sector increases the total decay width of the ALP
\begin{equation}
    \Gamma_a = \frac{1}{1-\mathrm{BR}(a\to\chi\bar{\chi})} \sum_{Y\in\mathrm{SM}}\Gamma(a\to Y)\,,
\end{equation}
while decreasing the branching ratio to all SM channels,
\begin{equation}
    \mathrm{BR}(a\to X) = \Big[1-\mathrm{BR}(a\to\chi\bar{\chi})\Big]\frac{\Gamma(a\to X)}{\sum_{Y\in \mathrm{SM}} \Gamma(a\to Y)}\,.
\end{equation}
This branching ratio is set to zero in \alpaca by default, but it can be included following the details in Sect.~\ref{sec:structure_process}.

\subsection{ALP production}\label{sec:ALPproduction}
The ALP can be produced in different processes, depending on its couplings and mass. In Table \ref{tab:production-channel} we show a summary of the production processes included in the program.

\begin{table}[h!]
    \centering
    \begin{tabular}{c|c}
        Type & Processes \\\hline
        \rowcolor{lightgray!40} Quarkonia decays & $J/\psi\to \gamma a$, $\Upsilon(1S)\to\gamma a$, $\Upsilon(3S)\to \gamma a$\\
        Non-resonant production & $e^+e^-\to \gamma a$ \\
        \rowcolor{lightgray!40} & $B^\pm\to K^{(*)\pm} a$, $B^0\to K^{(*)0} a$, $B_s^0\to \phi a$,\\
        \rowcolor{lightgray!40}& $B^\pm\to\pi^\pm a$, $B^0\to\pi^0 a$, $B^\pm\to\rho^\pm a$, $B^0\to\rho^0 a$,\\
        \rowcolor{lightgray!40}& $K^\pm\to\pi^\pm a$, $K_{L,S}^0\to\pi^0 a$,\\
        \rowcolor{lightgray!40}& $D^0\to\pi^0 a$, $D^0\to\eta^{(\prime)}a$, $D^0\to\rho^0a$, \\
        \rowcolor{lightgray!40}\multirow{-5}{*}{FCNC meson decays}& $D^\pm\to\pi^\pm a$, $D^\pm\to\rho^\pm a$, $D_s^\pm\to K^{(*)\pm}a$\\
        LFV lepton decays & $\mu^\pm \to e^\pm a$, $\tau^\pm\to e^\pm a$, $\tau^\pm \to \mu^\pm a$
    \end{tabular}
    \caption{\em ALP production channels, classified depending on the initial particles.}
    \label{tab:production-channel}
\end{table}

\subsubsection{Quarkonia radiative decays and non-resonant production}

A well-known process that allows to test ALPs is the production associated to Quarkonia radiative decays. Here we will outline just the most important formulae and concepts associated to these processes. However, one can find all the relevant details in Refs.~\cite{Merlo:2019anv,DiLuzio:2024jip}. 

Searches of ALPs in quarkonia follow different production channels depending on how the final state has been produced. Here we can distinguish tagged searches or direct production of the final quarkonia state (or untagged searches). In the former process, the quarkonia state that is studied is tagged by the decay products of an excited resonance, e.g.~$J/\Psi(2S)\to J/\Psi\,\pi \pi$. Then, ALPs are produced via the radiative decay $V\to \gamma \, a$, which is given by 
\begin{equation}
\label{eq:quarkonia_rad_dec}
    \mathcal{B}(V\to\gamma a)\simeq\frac{\alpha_\mathrm{em} Q_Q^2}{24}\frac{m_V\,f_V^2}{\Gamma_V\,f_a^2}
    \left(1-\frac{m_a^2}{m_V^2}\right)
    \left|c_\gamma\,\frac{\alpha_\mathrm{em}}{\pi}\left(1-\frac{m_a^2}{m_V^2}\right)-\tilde{F}_V\frac{m_V}{f_{V}}\,c^A_{qq}\right|^2 \, ,
\end{equation}
where $m_V$ is the mass, $\Gamma_V$ the total decay width and $f_V$ the decay constant of the vector resonance, which can be obtained by lattice or experimental data~\cite{Merlo:2019anv}. $\tilde{F}_V$ is a coefficient which depends on QCD dynamics, and naively can be estimated as
\begin{equation}
\label{eq:naive_approx}
    \tilde{F}\simeq 2\frac{f_V}{m_V}\left[1+\mathcal{O}(\alpha_S)\right].
\end{equation}

Recently, this hadronic constant was computed from lattice in Ref.~\cite{Colquhoun:2025xlx} for the $J/\Psi\to a\gamma$ channel, which we have implemented in \alpaca\!\!. For $\Upsilon$ resonances we use the naive approximation of Eq.~\eqref{eq:naive_approx} which at scales $\mu\simeq m_\Upsilon$ should be a fair approximation.

For direct production of the quarkonia resonances, it is known that the width of these particles and the uncertainty play an important role in computing the final cross-section~\cite{Eidelman:2016aih}. Namely, the width is typically smaller than the uncertainty (except for $\Upsilon(4S)$ which is the opposite), and hence non-resonant ALP production can be enhanced with respect to the resonant production. To take care of these effects both contributions need to be accounted for, 
\begin{equation}
    \sigma_{\textrm{tot}}(s)\simeq \sigma_\mathrm{NR}(s)+\langle \sigma_{R}(s)\rangle \, ,
\end{equation}
where $\langle \cdot \rangle$ corresponds to smearing the resonant cross section with a Gaussian distribution centred on $s=m_V^2$ with a deviation corresponding to the uncertainty on the beam energy, leading to
\begin{equation}
    \langle \sigma_{\rm R}(s)\rangle=\frac{1}{\sqrt{2\pi}}\int dq \,\frac{\sigma_R(q^2)}{\sigma_W}\exp
\left[-\frac{(q-\sqrt{s})^2}{2\sigma_W^2} \right] \, .
\end{equation}
The resonant cross section of the process $e^+ e^- \to V \to \gamma a$ can be given in terms of the branching ratio in Eq.~\eqref{eq:quarkonia_rad_dec}, the Breit-Wigner distribution and the peak cross section, as
\begin{equation}
\sigma_{\rm R}(s)=\sigma_\mathrm{peak} \frac{m_V^2 \Gamma_V^2}{(s-m_V^2)^2 +m^2_V\Gamma_V^2}\mathcal{B}(V\to \gamma a) \, , 
\label{eq:resonant}   
\end{equation}
where $\sigma_\mathrm{peak}$ can be obtained by using the experimental measurement of the branching ratio $\mathcal{B}(V\to e^+ e^-)$, 
\begin{equation}
    \sigma_\mathrm{peak} =\frac{12\pi \mathcal{B}(V\to e^+e^-)}{m_V^2} \, . 
\end{equation}
The non-resonant production cross-section, corresponding to the ALP emission from a photon mediator, is given by
\begin{equation}
    \sigma_\mathrm{NR}(s) = \frac{\alpha_\mathrm{em}^3}{24\pi^2} \frac{c_{\gamma}^2}{f_a^2}\, 
                        \left(1-\frac{m_a^2}{s}\right)^3\,.
\end{equation}
The different inputs implemented in \alpaca are taken from Refs.~\cite{Merlo:2019anv,DiLuzio:2024jip}, and the various searches are summarised in Table~\ref{tab:exp_quarkonia} in App.~\ref{sec:database_exp}.

\subsubsection{FCNC meson decays}
Other important processes in which ALPs can be produced are meson decays involving flavour-changing neutral currents (FCNC), where $M_1 \equiv (q_1 \overline{q}_{1'}) \to a \, M_2\equiv(q_2\overline{q}_{2'}) $. In particular, we will focus on transitions $q_1\overline{q} \to q_2\overline{q} \, a$ where the initial and final meson share a common quark (``spectator quark''), while the other quark can change its flavour by emitting an ALP through a flavour-violating coupling, which can be either a tree-level coupling, or an effective coupling generated by the running or by EW loops. 

Considering the experimental measurements included in \alpaca\!\!, the relevant transitions are of the type $P\to P \,a$ and $P\to V\,a$. 

Starting with decays of the type $P_1 \to P_2 a$, the decay width reads
\begin{equation}
    \Gamma(P_1\to P_2 a) = \frac{1}{16\pi}|A(P_1\to P_2 a)|^2 \frac{\sqrt{\lambda(m_{P_1}^2, m_{P_2}^2, m_a^2)}}{m_{P_1}^3}\,
\end{equation}
where $\lambda(a,b,c)=a^2+b^2+c^2-2(ab+ac+bc)$ is the K\"all\'en function and $A(P_1\to P_2 a)$ is the amplitude given by
\begin{equation}
    A_{FV}(P_1\to P_2a) = \left(m_{P_1}^2-m_{P_2}^2\right)f_0(m_a^2)\frac{c_{q_2q_1}^V}{2f_a}\,,
\end{equation}
where $f_0(m_a^2)$ is the hadronic matrix element.
It is important to note that the flavour-violating amplitude for the process $K\to\pi a$ must be calculated in the framework of chiral perturbation theory (for the full expressions of the FV amplitude for $K^-\to\pi^- a$ and $\overline{K}^0\to\pi^0 a$, see Refs.~\cite{Bauer:2021wjo,Bauer:2021mvw}).

In the case of $P\to Va$ decays, we can write the amplitude of a decay to a meson $V$ with polarisation $\lambda_V$ as 
\begin{equation}\label{eq:ampl_pol_PVa}
    A_{(\lambda_V)}(P\to V a) = \tilde{A}_P\ p_P\cdot\epsilon^*(p_V,\lambda_V) + \tilde{A}_a\ p_a\cdot\epsilon^*(p_V,\lambda_V)\,.
\end{equation}
Since there are only two independent momenta, $p_P$ and $p_a$, and since $p_V\cdot \epsilon^*(p_V,\lambda_V) = (p_P - p_a)\cdot \epsilon^*(p_V,\lambda_V) =0$, we can define $\tilde{A} = \tilde{A}_P + \tilde{A}_a$. The decay rate, after summing over the final polarisation, becomes
\begin{equation}
    \Gamma(P\to Va) = \frac{1}{64\pi}|\tilde{A}(P\to V a)|^2 \frac{\lambda^{3/2}(m_P^2, m_V^2, m_a^2)}{m_P^3 m_V^2}\,.
\end{equation} 
The flavour--violating amplitude is
\begin{equation}
    \tilde{A}_\mathrm{FV}(P\to V a) = -i m_V A_0(m_a^2)\frac{c_{q_2q_1}^A}{f_a}\,,
\end{equation}
where $A_0(q^2)$ is the pseudo-scalar decay form factor for the $P\rightarrow V$ transition.
Within \alpaca\!\!\!, the form factors are obtained using \texttt{flavio}\cite{flavio}, which also implements the results of the references in Table~\ref{tab:refs_hadronicelements}.

\begin{table}[tb]
    \centering
    \begin{tabular}{c|c}
        Decay & Reference \\\hline
         \rowcolor{lightgray!40}$K\to\pi $ &\cite{FlavourLatticeAveragingGroupFLAG:2021npn}\\
         $B\to\pi$ & \cite{FlavourLatticeAveragingGroupFLAG:2021npn}\\
         \rowcolor{lightgray!40}$B\to K$ & \cite{Bailey:2015dka} \\
         $D\to\pi$ & \cite{Lubicz:2017syv} \\
         \rowcolor{lightgray!40}$D_s \to K$ & \cite{Wang:2008ci} \\
         $D\to\eta$ & \cite{Fajfer:2004mv,Palmer:2013yia}\\
         \rowcolor{lightgray!40}$D\to\eta'$ & \cite{Fajfer:2004mv,Palmer:2013yia}\\
    \end{tabular}\hspace{2cm}
    \begin{tabular}{c|c}
        Decay & Reference \\\hline
        \rowcolor{lightgray!40}$B\to K^*$ & \cite{Horgan:2015vla}\\
        $B\to\rho$ & \cite{Bharucha:2015bzk}\\
        \rowcolor{lightgray!40}$B_s\to\phi$ & \cite{Bharucha:2015bzk}\\
        $D\to\rho$ & \cite{Chang:2019mmh}\\
        \rowcolor{lightgray!40}$D_s\to K^*$ & \cite{Chang:2019mmh}
    \end{tabular}
    
    \caption{\em Hadronic matrix elements $\langle M_2|\overline{q}_2 \Gamma q_1|M_1\rangle$. The left (right) table represents $P \to P$ ($P\to V$) processes.}
    \label{tab:refs_hadronicelements}
\end{table}

The couplings of the ALP will determine the relative size of the flavour transition. In absence of tree-level flavour-violating couplings, the flavour-changing transition in the down sector will be dominated by couplings to the top-quark induced by the one-loop penguin diagram, enhanced by the top-quark mass~\cite{Bauer:2020jbp,Bauer:2021mvw}, and detailed in Sect.~\ref{sec:running_matching}. 
On the other hand, couplings to down-type quarks will generate, at one-loop level, a flavour-violating interaction $c\to u a$, leading to an effective coupling of the form
\begin{equation}
    c_{uc}^{L,\mathrm{eff}}(\mu;p_a^2)=\sum_{q=d,s,b} \frac{G_F}{\sqrt{2}\pi^2}V_{cq}V_{uq}^\ast (c_{q_L}-c_{q_R})m_q^2\left[1+\mathcal{DB}(p_a^2,m_q,m_q)+\log\frac{\mu^2}{m_q^2}\right]\,,
\end{equation}
where the loop function is
\begin{align}
    \mathcal{DB}(p_a^2, m_q, m_q) &=2i \sqrt{1-\frac{4m_q^2}{p_a^2}} f\!\left(\frac{4m_q^2}{p_a^2}\right)\,\, \mathrm{with}\,\, 
    f(x) = \left\{\begin{matrix}\mathrm{arcsin}\frac{1}{\sqrt{x}} & \quad & \mathrm{for}\ x\geq1 \\\frac{\pi}{2}+\frac{i}{2}\log\frac{1+\sqrt{1-x}}{1-\sqrt{1-x}} & \quad & \mathrm{for}\ x<1 \end{matrix} \right..
\end{align}
It may be important for some processes to include the tree-level initial and final emission of ALPs, as in Refs.~\cite{Guerrera:2021yss,Guerrera:2022ykl}, which we leave for a future version of \alpaca\!\!. Analogous expressions can be obtained for the effective couplings $c_{ds}^{L,\mathrm{eff}}$, $c_{db}^{L,\mathrm{eff}}$ and $c_{sb}^{L,\mathrm{eff}}$ generated by loops of up and charm quarks. However, they will always be subdominant compared to the matching contribution obtained after integrating the top and gauge bosons.

\subsubsection{LFV lepton decays}
In scenarios where ALPs feature tree-level flavour violating couplings to leptons, see Refs.~\cite{Cornella:2019uxs,Calibbi:2020jvd}, the production channel $\ell_i \to \ell_j a$ with $i\neq j$ might present important constraints. 
The partial decay width for this process is \cite{Bauer:2021mvw}
\begin{equation}
    \Gamma(\ell_i\to\ell_j a) = \frac{|c_{\ell_1\ell_2}^A|^2 m_+^2(m_-^2-m_a^2) + |c_{\ell_1\ell_2}^V|^2m_-^2(m_+^2-m_a^2)}{64\pi m_{\ell_1}^3 f_a^2}\sqrt{\lambda(m_{\ell_1}^2, m_{\ell_2}^2, m_a^2)}\,,
\end{equation}
where $m_\pm = m_{\ell_i} \pm m_{\ell_j}$.

\subsection{ALP processes in the NWA}\label{sec:nwa}
We have included processes where an on-shell ALP can either be a final product or act as a mediator. In the latter case, we calculate the branching ratio of the full process a
\begin{equation}\label{eq:nwa}
    \mathrm{BR}(X\rightarrow M N)= \mathrm{BR}(X\rightarrow M a)\times \mathrm{BR}(a\rightarrow N),
\end{equation}
where $X$ and $M, \,N$ are the initial and final states, respectively. It is important to note that $M$, $N$ are not necessarily one--particle states. \alpaca implements all the decays that can be obtained in the NWA with any of the ALP production channels in Table~\ref{tab:production-channel} and any of the decay channels in Table~\ref{tab:decay-channel}.
In the $B^+\to K^+ \mu^+ \mu^-$, $B^0 \to K^{*0}\mu^+\mu^-$ and $K^+\to \pi^+\mu^+ \mu^-$ decays, we have available information about both prompt and displaced vertex searches. The experimental analysis, in order to be as generic as possible, considers the distance between primary and secondary vertices (i.e. the lifetime of the possible intermediate particle) and the semileptonic branching ratios to be independent  quantities. However, once we introduce a specific New Physics model, an ALP model in our case, both quantities are determined only by the ALP mass and couplings. Consequently, for each $(m_a, \tau_a)$ point in the experimental data we can extract the corresponding value of the couplings, use it to predict the semileptonic branching ratio, and check whether it is excluded by the experiments.

\subsection{Processes mediated by off-shell ALPs}\label{sec:ALP_offshell}

In this section, we will study processes described by tree diagrams where the ALP appears as an internal propagator. As such, it is always off-shell, which allows us to probe arbitrarily small or large ALP masses. The price to pay is that, unlike the NWA processes discussed above, these processes only enjoy a resonant enhancement in a narrow region around $m_a\approx m_{M^0}$, where $M^0$ is the decaying meson. One important remark is that the diagrams contain two ALP couplings, and therefore will be of order $\mathcal{O}(f_a^{-2})$. If the process also has a pure SM amplitude, their interference will give rise to a  $\mathcal{O}(f_a^{-2})$ to the observables. On the other hand, if there is no SM contribution and thus no interference between the pure SM and the ALP amplitudes is generated, as is the case of the LFV decays $M^0 \to \ell_1^+ \ell_2^-$, the branching ratios will be suppressed by $\mathcal{O}(f_a^{-4})$, and for this reason are not included in \alpaca\!\!. The processes mediated by an off-shell ALP implemented in the code are listed in Table~\ref{tab:processes_offshell}.

\begin{table}[h!]
    \centering
    \begin{tabular}{c|c|c|c}
       Quark content  & Radiative & Leptonic & Meson mixing \\\hline
       \rowcolor{lightgray!40} & & $K_S^0\to e^+e^-$, $K_S^0\to \mu^+\mu^-$,  &  \\
       \rowcolor{lightgray!40}\multirow{-2}{*}{$sd$} & \multirow{-2}{*}{$K_S^0\to \gamma\gamma$ }  & $K_L^0\to e^+e^-$, $K_L^0\to \mu^+\mu^-$ & \multirow{-2}{*}{$\Delta m_K$, $|\epsilon_K|$} \\
       $bd$ & $B^0\to\gamma\gamma$ & $B^0\to e^+e^-$, $B^0\to \mu^+\mu^-$, $B^0\to \tau^+\tau^-$ & $\Delta m_{B^0}$, $\mathcal{A}_\mathrm{SL}(B^0)$ \\
       \rowcolor{lightgray!40}$bs$ & $B_s^0\to\gamma\gamma$ & $B_s^0\to e^+e^-$, $B_s^0\to \mu^+\mu^-$, $B_s^0\to \tau^+\tau^-$ & $\Delta m_{B_s^0}$, $\mathcal{A}_\mathrm{SL}(B_s^0)$\\
       $cu$ & $D^0\to\gamma\gamma$ & $D^0\to e^+e^-$, $D^0\to \mu^+\mu^-$ & $x_{D^0}$, $\phi_{12,D^0}$
    \end{tabular}
    \caption{\em Processes mediated by an off-shell ALP implemented in \alpaca\!\!.}
    \label{tab:processes_offshell}
\end{table}

\subsubsection{Radiative and leptonic decays of neutral mesons}

In the processes $M^0\to \ell^+\ell^-$ and $M^0\to\gamma\gamma$, with $M^0 = K_L^0,\, K_S^0,\, D^0,\, B^0,\, B_s^0$, the decay chain $M^0\to a\to \ell^+\ell^-$ with an on-shell ALP is not allowed by 4-momentum conservation, unless $M_{M^0}=m_a$; however the process mediated by an off-shell ALP $M^0\to a^*\to \ell^+\ell^-$ and its interference with the SM diagrams must be considered.

The most general amplitudes for $M\to\gamma\gamma$ and $M\to\ell^+\ell^-$ are obtained by splitting the $CP$-even part $B$ and the $CP$-odd part $C$,~\cite{Cirigliano:2011ny}
\begin{align}
    A(M^0\to\gamma\gamma) &= \epsilon^*_\mu(k_1) \epsilon^\ast_\mu(k_2)[(k_2^\mu k_1^\nu -k_1\cdot k_2\, g^{\mu\nu}) B_{M^0\to\gamma\gamma} + i \varepsilon^{\mu\nu\rho\sigma}{k_1}_\rho {k_2}_\sigma C_{M^0\to\gamma\gamma}]\,,\nonumber\\
    A(M^0\to\ell^+\ell^-) &= \overline{u}(p_{\ell^-})\ [i\, B_{M^0\to\ell^+\ell^-}+C_{M^0\to\ell^+\ell^-}\gamma_5]\ v(p_{\ell^+})\,.
\end{align}
In the radiative decays, $B$ corresponds to the amplitude to parallel photon polarisations ($F_{\mu\nu}F^{\mu\nu}$) and $C$ to perpendicular polarisations ($F_{\mu\nu}\widetilde{F}^{\mu\nu}$). The corresponding decay widths are
\begin{align}
    \Gamma(M^0\to\gamma\gamma) &= \frac{m_{M^0}^3}{64\pi}\left(|B_{M^0\to\gamma\gamma}|^2+|C_{M^0\to\gamma\gamma}|^2\right)\,,\nonumber\\
    \Gamma(M^0\to\ell^+\ell^-) &= \frac{m_{M^0}}{8\pi}\sqrt{1-\frac{4m_\ell^2}{m_{M^0}^2}}\left[\left(1-\frac{4m_\ell^2}{m_{M^0}^2}\right)|B_{M^0\to\ell^+\ell^-}|^2 + |C_{M^0\to\ell^+\ell^-}|^2\right]\,.
\end{align}

The ALP-mediated contribution only enters the $CP$-odd amplitude,
\begin{align}
    C^\mathrm{ALP}_{M^0\to\gamma\gamma} &= \frac{\alpha_\mathrm{em}}{\pi} \frac{c_{q_1q_2}^A c_\gamma }{2 f_a^2}f_{M^0}\frac{m_{M^0}^3}{m_{M^0}^2-m_a^2+im_a\Gamma_a}\,,\nonumber\\
    C^\mathrm{ALP}_{M^0\to\ell^+\ell^-} &= -\frac{c_{q_1q_2}^A c_{\ell\ell}^A}{2f_a^2}f_{M^0} \frac{m_{M^0}^3 m_\ell}{m_{M^0}^2-m_a^2+i m_a \Gamma_a}\,.
\end{align}
Note that the interference with the SM amplitude is of order $f_a^{-2}$, while the quadratic term is suppressed as $f_a^{-4}$. The exception is the region where the ALP is resonant $m_a \approx m_{M^0}$, where the ALP amplitude becomes independent of $f_a$, $C\sim 1/(f_a^2 \Gamma_a) \sim \mathcal{O}(1)$. If $m_a \ll m_{M^0}$, the amplitude is independent of $m_a$, while for $m_a \gg m_{M^0}$, it scales as $C\sim (m_{M_0}/m_a)^2$. Even if we are able to extract limits to the ALP couplings for arbitrary values of $m_a$ -- as opposed to processes with on-shell ALPs, that only are valid inside the kinematic range -- those limits will become weaker for heavy ALPs.

In the pure SM amplitudes, the situation is more complicated.
\begin{itemize}
    \item For $B^0$ and $B_s^0$ mesons, the amplitudes are dominated by short-distance physics~\cite{Bosch:2002bv,Arnan:2017lxi}. The meson oscillation must also be taken into account, as the experiments reports time-averaged decay widths~\cite{DeBruyn:2012wj}.
    \item For $D^0$ mesons, the amplitude is dominated by the long-distance processes $D^0\to VV\to\gamma\gamma(\to \ell^+\ell^-)$, where $V$ are vector mesons~\cite{Burdman:2001tf}.
    \item For kaons, short-distance and long-distance contributions are typically of the same magnitude~\cite{Ecker:1991ru,Hoferichter:2023wiy}. The long-distance effects interfere with ALP-mediated amplitudes only in $K_L^0\to \gamma\gamma$ and $K_S^0\to\ell^+\ell^-$. However, the calculation of the $K_L^0\to\gamma\gamma$ long-distance amplitude involves parameters that have not been determined independently, and consequently can not be used as a probe for New Physics.
\end{itemize}

\subsubsection{Neutral meson mixing}\label{sec:mixing}

The ALP enters the oscillation of neutral mesons $K_L^0-K_S^0$, $D^0-\overline{D}^0$, $B^0-\overline{B}^0$ and $B_s^0-\overline{B}_s^0$ through the exchange of the ALP in the $s$-channel and $t$-channel. For heavy ALPs above the mass scale of the meson, the momentum carried by the exchanged ALP can be safely neglected compared to its mass, and the propagators in the $s$-channel $\mathcal{P}_s$ and in the $t$-channel $\mathcal{P}_t$ can be approximated as
\begin{equation}
    \mathcal{P}_s \approx \mathcal{P}_t \approx -\frac{1}{m_a^2}\,.
\end{equation}

In the case of a light ALP, the computation of the $s$-channel is straightforward, since the momentum carried by the ALP is just the momentum of the meson,
\begin{equation}
    \mathcal{P}_s = \frac{1}{(p_{q_1}+p_{q_2})^2 - m_a^2}=\frac{1}{m_{M^0}^2-m_a^2}\,.
\end{equation}
On the other hand, the computation of the $t$-channel is problematic, since $p_a=p_{q_1}-p_{q_2}$ involves the difference of momenta between the quark $q_1$ in the initial meson and the quark $q_2$ in the final meson. For the case of $B^0$ and $B_s^0$ it is possible to apply Heavy Quark Effective Theory (HQET), which gives the prescription for the $t$-channel propagator~\cite{Bauer:2021mvw}
\begin{equation}
    \mathcal{P}_t =\frac{1}{(p_{q_1}-p_{q_2})^2-m_a^2} = \frac{1}{(m_b-\bar{\Lambda
    })^2-m_a^2}\left(1+\mathcal{O}(\Lambda_\mathrm{QCD}^2)\right)\,,
\end{equation}
where $\bar{\Lambda} = m_{B_q^0} - m_b$, and the expression above is only valid for $m_a^2 \ll (m_b-\bar{\Lambda})^2-\Lambda_\mathrm{QCD}^2 \approx (3\,\mathrm{GeV})^2$. The applicability of HQET to lighter mesons is a bit more dubious, and as a conservative approach we will refrain from studying these oscillations for light ALPs, and will postpone it to a future work. 

In the kaon sector, the oscillation parameters are~\cite{Buras:1998raa,Bagger:1997gg}
\begin{align}
    \Delta m_K &= \frac{1}{m_{K^0}}\, \mathrm{Re}\langle K^0|\mathcal{H}_{\mathrm{eff},sd}|\overline{K}^0\rangle\,,\nonumber\\|
    \epsilon_K| &= \frac{1}{2\sqrt{2}}\left|\frac{\mathrm{Im}\langle K^0|\mathcal{H}_{\mathrm{eff},sd}|\overline{K}^0\rangle}{\,\mathrm{Re}\langle K^0|\mathcal{H}_{\mathrm{eff},sd}|\overline{K}^0\rangle}\right|\,,
\end{align}
in the $B_d^0$ and $B^0_s$ sectors,~\cite{Buras:1998raa,Albrecht:2024oyn,Miro:2024fid}
\begin{align}
    \Delta m_{B_q} &= \frac{1}{m_{B_q^0}}|\langle B_q^0|\mathcal{H}_{\mathrm{eff},bq}|\overline{B}_q^0\rangle|\,, \\
    \mathcal{A}_\mathrm{SL}(B_q) &= 2 m_{B_q^0}\left|\frac{\Gamma_{12}}{\langle B_q^0|\mathcal{H}_{\mathrm{eff},bq}|\overline{B}_q^0\rangle} \right|\sin \arg  \frac{-\langle B_q^0|\mathcal{H}_{\mathrm{eff},bq}|\overline{B}_q^0\rangle}{2m_{B_q^0}\Gamma_{12}}\,,
\end{align}
and in the $D^0$ sector,
\begin{align}
    x_{D^0} &= \frac{\Delta m_{D}}{\Gamma_{D^0}} = \frac{1}{m_{D^0} \Gamma_{D^0}}|\langle D^0|\mathcal{H}_{\mathrm{eff},cu}|\overline{D}^0\rangle|\,, \\
    \phi_{12,D^0}&=\arg \frac{\langle D^0|\mathcal{H}_{\mathrm{eff},cu}|\overline{D}^0\rangle}{2m_{D^0}\Gamma_{12}}\,,
\end{align}
where $\Gamma_{12}$ is the off-diagonal part of transition matrix elements from  physical intermediate states, which we assume to be SM-like.

The effective Hamiltonian governing the mixing is
\begin{equation}\label{eq:effH_mixing}
    \mathcal{H}_{\mathrm{eff},q_1q_2} = \sum_{i=1}^5 C_i^{q_1q_2}(m_{q_1}) \mathcal{O}_i^{q_1q_2} +  \sum_{i=1}^3 \tilde{C}_i^{q_1q_2}(m_{q_1}) \tilde{\mathcal{O}}_i^{q_1q_2}\,,
\end{equation}
with the effective operators
\begin{gather}
    \mathcal{O}_1^{q_1q_2} = \overline{q}_2^\alpha \gamma_\mu P_L q_1^\alpha\,\overline{q}_2^\beta \gamma^\mu P_L q_1^\beta\,,\quad\quad \mathcal{O}_2^{q_1q_2} = \overline{q}_2^\alpha  P_L q_1^\alpha\,\overline{q}_2^\beta  P_L q_1^\beta\,,\quad\quad \mathcal{O}_3^{q_1q_2} = \overline{q}_2^\alpha  P_L q_1^\beta\,\overline{q}_2^\beta  P_L q_1^\alpha\,,\nn\\
    \mathcal{O}_4^{q_1q_2} = \overline{q}_2^\alpha  P_L q_1^\alpha\,\overline{q}_2^\beta  P_R q_1^\beta\,,\quad\quad \mathcal{O}_5^{q_1q_2} = \overline{q}_2^\alpha  P_L q_1^\beta\,\overline{q}_2^\beta  P_R q_1^\alpha\,,
\end{gather}
where $\alpha$ and $\beta$ are colour indices, and the operators $\tilde{\mathcal{O}}_i^{q_1q_2}$ are obtained by exchanging $P_L$ and $P_R$. The SM only enters in $C_1^{q_1q_2}$, while the other coefficients have only ALP contributions. The general expressions for the $B^0_q$ coefficients can be found in Ref.~\cite{Bauer:2021mvw} (with the notation $A_+ = \mathcal{P}_s$, $A_- = \mathcal{P}_t$) in the approximation $m_b \gg m_d, m_s$. For the other mesons, as explained previously, we only consider the limit $m_a \gg m_{M^0}$, and we will retain also the terms that depend on the mass of the lighter quark,
\begin{align}
    C_2^{q_1q_2}(m_a) &= \frac{(c_{q_2q_1}^R m_1 - c_{q_2q_1}^L m_2)^2}{2 m_a^2 f_a^2}\,,\nonumber\\
    \tilde{C}_2^{q_1q_2}(m_a) &= \frac{(c_{q_2q_1}^L m_1 - c_{q_2q_1}^R m_2)^2}{2 m_a^2 f_a^2}\,,\\
    C_4^{q_1q_2}(m_a) &= \frac{(c_{q_2q_1}^R m_1 - c_{q_2q_1}^L m_2)(c_{q_2q_1}^L m_1 - c_{q_2q_1}^R m_2)}{m_a^2 f_a^2}\,.\nn
\end{align}
The evolution of the operators down to the scale $m_{q_1}$ is reported in Ref.~\cite{Bagger:1997gg}.
Finally, the evaluation of the hadronic matrix element of the effective operators is given in terms of the bag parameters $B_{M^0}^{(i)}$,
\begin{equation}
    \langle M^0|\mathcal{O}_i^{q_1q_2}|\overline{M}^0\rangle = f_{M^0}^2 m_{M^0}^2\, \eta_{i}^{q_1q_2}(m_{q_1})\, B_{M^0}^{(i)}(m_{q_1})\,,
\end{equation}
and $\langle M^0|\tilde{\mathcal{O}}_i^{q_1q_2}|\overline{M}^0\rangle = \langle M^0|\mathcal{O}_i^{q_1q_2}|\overline{M}^0\rangle$ due to parity invariance of QCD. The normalisation factors can be obtained from Ref.~\cite{Bagger:1997gg}. In \alpaca the bag parameters for $B^0$ and $B_s^0$ are implemented from Ref.~\cite{Dowdall:2019bea}, while the ones for kaons are extracted from Ref.~\cite{Boyle:2024gge}, and for $D^0$ from Ref.~\cite{Bazavov:2017weg}. The SM contributions to the matrix elements are obtained from \texttt{flavio}~\cite{Straub:2018kue}.

\subsection{Meson decay widths}
Finally, the possibility of decays with an ALP in the final state might also significantly alter the decay width, and equivalently the lifetime, of the various mesons. The modification of the decay width of the meson $M$ is given by
\begin{equation}
    \delta\Gamma_M^\mathrm{alp} = \sum_Y \Gamma(M\to Y a)\,,
\end{equation}
where the decay widths $\Gamma(M\to Y a)$ are the ones collected in Table~\ref{tab:production-channel}.

\subsection{Implementation}
\subsubsection{Particle dictionary}
Before we delve into the details of how to work with the processes implemented in \alpaca\!\!, it is instructive to learn how to call a particle within the program. In Table~\ref{tab:particle_codes}, a summary dictionary of the accepted names for the relevant particles is shown. As can be seen, the program is designed in such a way that several names are acceptable for a specific particle.
\begin{table}[!ht]
    \centering
    \begin{tabular}{c|c}
        Particle & Codes \\\hline
        \rowcolor{lightgray!40}$a$ & \texttt{a}, \texttt{alp}, \texttt{ALP}\\
        $\gamma$ & \texttt{A}, \texttt{gamma}, \texttt{photon} \\
        \rowcolor{lightgray!40}$g$ & \texttt{g}, \texttt{gluon}\\  
        $e^\pm$ & \texttt{electron}, \texttt{positron}, \texttt{e}, \texttt{e-}, \texttt{e+}\\
        \rowcolor{lightgray!40}$\mu^\pm$ & \texttt{muon}, \texttt{mu}, \texttt{mu-}, \texttt{mu+}\\
        $\tau^\pm$ & \texttt{tau}, \texttt{ta}, \texttt{ta-}, \texttt{ta+}\\
        \rowcolor{lightgray!40}$c$, $\bar{c}$ & \texttt{charm}, \texttt{c}, \texttt{cbar}\\
        $b$, $\bar{b}$ & \texttt{bottom}, \texttt{b}, \texttt{bbar}\\
        \rowcolor{lightgray!40}Hadrons & \texttt{hadrons}\\
        Pions & \texttt{pion}, \texttt{pi} \\
        \rowcolor{lightgray!40}$\pi^+$ & \texttt{pion+}, \texttt{pi+}\\
        $\pi^-$ & \texttt{pion-}, \texttt{pi-}\\
        \rowcolor{lightgray!40}$\pi^0$ & \texttt{pion0}, \texttt{pi0}\\
        $\eta$ & \texttt{eta} \\
        \rowcolor{lightgray!40}$\eta'$ & \verb|eta'|, \texttt{etap}, \texttt{eta\_prime} \\
        $\omega$ & \texttt{omega} \\
        \rowcolor{lightgray!40}$\rho^0$ & \texttt{rho0} \\
        $\rho^+$ & \texttt{rho+}\\
        \rowcolor{lightgray!40}$\rho^-$ & \texttt{rho-}\\
        $\phi$ & \texttt{phi}
    \end{tabular}\hspace{1cm}
    \begin{tabular}{c|c}
        Particle & Codes \\\hline
        \rowcolor{lightgray!40}$K^+$ & \texttt{K+}\\
        $K^-$ & \texttt{K-}\\
        \rowcolor{lightgray!40}$K^0$ & \texttt{K0}\\
        $K_L^0$ & \texttt{KL}, \texttt{K0L}\\
        \rowcolor{lightgray!40}$K_S^0$ & \texttt{KS}, \texttt{K0S}\\
        $K^{*+}$ & \texttt{K*+}\\
        \rowcolor{lightgray!40}$K^{*-}$ & \texttt{K*-}\\
        $K^{*0}$ & \texttt{K*0}\\
        \rowcolor{lightgray!40}$D^+$ & \texttt{D+}\\
        $D^-$ & \texttt{D-}\\
        \rowcolor{lightgray!40}$D^0$ & \texttt{D0}\\
        $D_s^+$ & \texttt{Ds+}\\
        \rowcolor{lightgray!40}$D_s^-$ & \texttt{Ds-}\\
        $J/\psi$ & \texttt{J/psi} \\
        \rowcolor{lightgray!40}$B^+$ & \texttt{B+}\\
        $B^-$ & \texttt{B-}\\
        \rowcolor{lightgray!40}$B^0$ & \texttt{B0}, \texttt{Bd0}\\
        $B_s^0$ & \texttt{Bs}, \texttt{Bs0}\\
        \rowcolor{lightgray!40}$\Upsilon(1S)$ & \texttt{Upsilon(1S)}, \texttt{Y(1S)}\\
        $\Upsilon(2S)$ & \texttt{Upsilon(2S)}, \texttt{Y(2S)}\\
        \rowcolor{lightgray!40}$\Upsilon(3S)$ & \texttt{Upsilon(3S)}, \texttt{Y(3S)}\\
        $\Upsilon(4S)$ & \texttt{Upsilon(4S)}, \texttt{Y(4S)}\\
        \rowcolor{lightgray!40}Dark sector & \texttt{dark}
        \end{tabular}
    \caption{\em List of accepted particle names.}
\label{tab:particle_codes}
\end{table}

\subsubsection{Working structure}
\label{sec:structure_process}
Working with the different transitions present in the program is relatively simple. The structure is the following:
\begin{center}
    \verb|"initial -> final1 final2 final3 ... finalN"|
\end{center}
The initial and final states are separated by the arrow \texttt{->}. Particles are separated by spaces.
The order in which the particles are placed is irrelevant, as the program internally rearranges them to match the process. For quarks and leptons, particle and antiparticle are considered equivalent since, for the processes considered, they always come in pairs.

There are three types of observables implemented in \alpaca\!\!: 
\begin{itemize}
    \item \texttt{alpaca.decay\_width}\index{\texttt{decay\_width}} for the decay widths of the ALP described in Section~\ref{sec:ALPdecays}. The result is in GeV.
    \item \texttt{alpaca.branching\_ratio}\index{\texttt{branching\_ratio}} for the branching ratios of the ALP described in Section~\ref{sec:ALPdecays} and of the various meson decays described in Sections~\ref{sec:ALPproduction}, \ref{sec:nwa} and \ref{sec:ALP_offshell}.
    \item \texttt{alpaca.cross\_section}\index{\texttt{cross\_section}} for the non-resonant cross sections in Sections~\ref{sec:ALPproduction} and \ref{sec:nwa}. The result is in pb.
\end{itemize}

All three observables use the same arguments:
\begin{itemize}
    \item \texttt{transition}: An \texttt{str} indicating the process, in the syntax described above. In the case of \texttt{alpaca.decay\_width}, the transition \texttt{"a"} produces the total decay width of the ALP, and a transition containing only the name of a meson (e.g. \texttt{"B+"} for the $B^+$ meson) produces the ALP contribution to the decay width of said meson.
    \item \texttt{ma}: The mass of the ALP, in GeV.
    \item \texttt{couplings}: The \texttt{ALPcouplings} object containing the couplings of the ALP.
    \item \texttt{s} (\texttt{cross\_section} only): The center-of mass energy $s$ of the process, in $\mathrm{GeV^2}$.
    \item \texttt{fa}: The decay constant of the ALP $f_a$, in GeV.
    \item \texttt{br\_dark}: Optional parameter providing the branching ratio $\mathrm{BR}(a\to\chi\bar{\chi})$, as explained in Section~\ref{sec:darksector}. If it is not provided, it is set to 0.
    \item Optional arguments for the numerical integration, using \texttt{vegas}, of the phase space in $a\to\pi^+\pi^-\pi^0$, $a\to3\pi^0$, $a\to\eta^{(\prime)}\pi^+\pi^-$, $a\to\eta^{(\prime)}\pi^0\pi^0$ and $a\to\gamma\pi^+\pi^-$ decays:
    \begin{itemize}
        \item \texttt{nitn}: Number of estimates of the integral.
        \item \texttt{neval}: Maximum number of samples of the integrand per estimate.
        \item \texttt{nitn\_adapt}: Number of estimates of the integral in the adaptation phase (they are discarded).
        \item \texttt{neval\_adapt}: Maximum number of samples of the integrand per estimate (they are discarded).
    \end{itemize}
    \item Optional arguments for the running of \texttt{ALPcouplings} from its initial scale down to the physical scale of each observable: \texttt{integrator}, \texttt{beta}, \verb|match_tildecouplings|, \verb|scipy_method|, \verb|scipy_atol| and \verb|scipy_rtol|, described in Section~\ref{sec:running_impl}.
\end{itemize}

In the example below, one can see how to calculate the branching ratio for an ALP of mass $1.2$ GeV for the processes $a\to\eta\pi^+\pi^-$, $B^0\rightarrow K^{\ast 0}a$ and $B^+\rightarrow K^+ e^+e^-$. In this case, the couplings are taken to be universal for LH quarks at the UV scale, $\Lambda=4\pi\TeV$, and run and matched to the scale of the mass of the $B$ meson.

\begin{minted}{python}
import alpaca
mB = 5.279

transitions = [
    'a -> eta pi+ pi-',
    'B0 -> K*0 a',
    'B+ -> K+ e e',
]

couplings = alpaca.ALPcouplings(
{'cqL':1.0}, 
scale = 4*np.pi*1000, 
basis='derivative_above').match_run(
scale_out=mB,
basis='VA_below')

for tr in transitions:
    br = alpaca.branching_ratio(
        transition = tr,
        ma = 1.2,
        couplings = couplings,
        fa = 1000.0
    )
    print(f"BR({tr}) = {br:.5e}")
\end{minted}
\vspace{-8mm}
\begin{minted}{pwsh-session}
BR(a -> eta pi+ pi-) = 3.05992e-01
BR(B0 -> K*0 a) = 1.13747e-01
BR(B+ -> K+ e e) = 7.13931e-08
\end{minted}

The functions \texttt{decay\_rate}, \texttt{branching\_ratio} and \texttt{cross\_section} are vectorised, meaning that the arguments \texttt{ma}, \texttt{couplings}, \texttt{fa} and \texttt{br\_dark} can be \texttt{numpy.array}s of arbitrary shape, and the result will be a \texttt{numpy.array} of the same shape. This can be useful to perform grid scans in a convenient and optimised manner, as illustrated by the following example:

\begin{minted}{python}
import alpaca
import numpy as np

# Create ranges of ma and fa, logarithmically spaced
ma = np.logspace(-1, 1, 30)
fa = np.logspace(3, 8, 30)
Lambda = 4*np.pi*fa
# Create a list of ALPcouplings
couplings = [
    alpaca.uvmodels.QED_DFSZ.get_couplings(
        {alpaca.uvmodels.beta: np.pi/4}, l
    ).match_run(10.0, 'VA_below') for l in Lambda
]
# Combine them into grids
x_ma, y_fa = np.meshgrid(ma, fa)
x_ma, y_couplings = np.meshgrid(ma, couplings)

dw_photons = alpaca.decay_width(
    'a -> gamma gamma',
    ma = x_ma,
    couplings = y_couplings,
    fa = y_fa,
    integrator = 'leadinglog'
)
\end{minted}

We first create ranges for $m_a \in [10^{-1},\,10]\,\mathrm{GeV}$, $f_a\in[10^3,\,10^8]\,\mathrm{GeV}$ and $\Lambda=4\pi f_a$, with 30 points logarithmically spaced. Then, an array is created containing the ALP couplings corresponding to a QED-DFSZ model with $\tan\beta =1$ defined at the scales $\Lambda$ which are run down to $\mu=10\,\mathrm{GeV}$ using numerical integration. These arrays are combined into \texttt{numpy} \texttt{meshgrid}s representing the two-dimensional grids that contain the points in parameter space. And finally, we calculate the decay width of ALPs into photons at the points in parameter space, by using the leading log approximation to run from $\mu=10\,\mathrm{GeV}$ down to the physical scale (running effects below the electroweak scale are generally small, and therefore the error by using this approximation is not too dramatic while the computation is faster). The variable \texttt{dw\_photons} contains a $30\times30$ array of decay widths.

Meson mixing observables are implemented in the function \texttt{alpaca.meson\_mixing}\index{\texttt{meson\_mixing}}. The first argument refers to the observable, with the available options being \texttt{delta\_mK0} and \texttt{epsK} for kaon oscillations, \texttt{x\_D0} and \texttt{phi12\_D0} for $D^0$ oscillations, and \texttt{delta\_mB0}, \texttt{ASL\_B0} and \texttt{delta\_mBs}, \texttt{ASL\_Bs} for $B_q^0$ oscillations. The rest of arguments are the same as in the function \texttt{alpaca.branching\_ratio}, except for \texttt{br\_dark}, that plays no role in neutral meson oscillations.

Finally, it is possible to obtain the partial decay widths for all decay channels of the ALP with the function \texttt{alp\_channels\_decay\_widths}\index{\texttt{alp\_channels\_decay\_widths}}, while the branching ratios are obtained with \texttt{alp\_channels\_branching\_ratios}\index{\texttt{alp\_channels\_branching\_ratios}}. They take the same arguments for \texttt{ma}, \texttt{couplings}, \texttt{fa}, \texttt{br\_dark} and optional arguments described above. Their output is a \texttt{dict} containing the decay width or branching ratio for each channel. Continuing with the previous example,
\begin{minted}{python}
ma = 1.7
fa = 1e4

dw_alp = alpaca.alp_channels_decay_widths(
    ma = ma,
    couplings = y_couplings[5],
    fa = fa
)

br_alp = alpaca.alp_channels_branching_ratios(
    ma = ma,
    couplings = y_couplings[5],
    fa = fa
)
\end{minted}

A summary of the functions available to calculate the observables are reported in Table~\ref{tab:functionsObservables}.

\begin{table}[h!]
\begin{tabularx}{\textwidth}{ c|X } 
 Function name& Output\\
 \hline
  \rowcolor{lightgray!40}\texttt{decay\_width} & Decay width of the desired transition. \\ 
\texttt{branching\_ratio} & Branching ratio of the desired transition.  \\ 
 \rowcolor{lightgray!40} \texttt{cross\_section} & Cross section of the desired transition.  \\ 
\texttt{meson\_mixing} & Calculation of the desired observable related to meson mixing.\\
  \rowcolor{lightgray!40}\texttt{alp\_channels\_decay\_widths} &Calculates the partial decay width for all possible decay channels of the ALP. \\ 
\texttt{alp\_channels\_branching\_ratios}&Calculates the branching ratio for all possible decay channels of the ALP.\\
\end{tabularx}
 \caption{\em Summary of the functions available to calculate observables. The details on how to implement them can be found in the text.}
\label{tab:functionsObservables}
\end{table}

%% file: exp_signatures.tex
This section is devoted to the details of the internal treatment of experimental searches
contained in \alpaca\!\!. In Sect.~\ref{sec:signatures}, we first explain the three different types of experimental signatures relevant for processes involving an on-shell ALP, namely prompt, displaced, or invisible decays, highlighting the importance of considering the kinematic details of the decay and the characteristics of the detector, as well as the probability of each experimental signature, to correctly categorise the process in a phenomenological analysis. In Table~\ref{tab:experiments_decaytypes} we summarise some experimental details necessary for this analysis. Later, in Sect.~\ref{sec:exp_impl}, we show how to work with experimental measurements within \alpaca\!\!, while in Sect.~\ref{sec:exp_sectors} we focus on working with predetermined groups of measurements or sectors.

\begin{table}[th!]
    \centering
    \begin{tabular}{c|c|c|c|c}
$P_1$ & Experiment & Prompt region & Invisible region & $\beta_{P_1}\gamma_{P_1}$ \\\hline
        \rowcolor{lightgray!40}$K$ & NA62 & $\tau_a < 100\,\mathrm{ps}$ & $\tau_a > 5\,\mathrm{ns}$ & --\\
        $K$ & MicroBooNE & $c\tau_a < 0.3\,\mathrm{m}$ & $c\tau_a > 3\times10^7\,\mathrm{m}$ & --\\
        \rowcolor{lightgray!40}$B$ & BaBar & $r_\mathrm{min} = 2\,\mathrm{cm}$  & $r_\mathrm{max} = 3\,\mathrm{m}$ & $0.42$ \\
        $B$ & Belle & $r_\mathrm{min} = 4\,\mathrm{cm}$  & $r_\mathrm{max} = 1\,\mathrm{m}$ & $0.42$ \\
        \rowcolor{lightgray!40}$B$ & Belle II & $c\tau_a < 1\,\mu\mathrm{m}$  & $c\tau_a > 4\,\mathrm{m}$ & $0.28$ \\
        $B$ & LHCb & $\tau_a < 0.1\,\mathrm{ps}$ & $\tau_a > 1\,\mathrm{ns}$ & --\\
        \rowcolor{lightgray!40}$B$ & CHARM & $\tau_a <250\,\mathrm{ps}$ & $\tau_a>1\,\mu\mathrm{s}$ & -- \\
        $B$ & SHiP & $\tau_a <10\,\mathrm{ps}$ & $\tau_a>1\,\mu\mathrm{s}$ & -- \\
        \rowcolor{lightgray!40}$\Upsilon(1S) $ & Belle & &$r_\mathrm{max} =1\,\mathrm{m}$ & 0.0\tablefootnote{The $\Upsilon(1S)$ at Belle are generated in the three-body decay $\Upsilon(2S)\to\Upsilon(1S) \pi^+\pi^-$, and therefore its boost follows a probability distribution. We have checked that in the ranges of interest $|\vec{p}_a|\gg|\vec{p}_{\Upsilon(1S)}|$, and as a simplifying assumption we set the $\Upsilon(1S)$ approximately at rest.} \\
    \end{tabular}
    \caption{\em Summary of the experiment characteristics depending on the initial meson.}
    \label{tab:experiments_decaytypes}
\end{table}

\subsection{Experimental signatures: prompt, displaced vertex and invisible processes}\label{sec:signatures}

In processes where an on-shell ALP is produced, depending on the decay and the characteristics of the detector, each decay can be classified in three distinct categories, namely prompt, displaced vertex and invisible decays. From the experiment viewpoint, in prompt decays, the detector cannot distinguish between the primary and secondary vertices; while in displaced vertices the detector does have the resolution to characterise both of them. In the case of invisible decays, the particle is identified as missing energy. This is the case of  long-lived ALPs that decay outside the detector, or ALPs that mainly decay into a dark sector or even active neutrinos.

When performing a phenomenological analysis considering experimental results, taking into account the different experimental signatures when calculating the theoretical contribution is of uttermost importance. Therefore, \alpaca directly calculates the probability that each process leaves one of the experimental signatures mentioned above. To do that, it considers the boost function and resolution of each experiment. 

Consider the decay $P_1\to P_2 a$. In the frame where $P_1$ is at rest, its four-momentum is $p_{P_1}^* = (m_{P_1}, 0, 0, 0)^T$. Choosing the $x$ axis as the direction of $P_1$ in the LAB frame, and the $x-y$ plane where $P_2$ and the ALP are contained, we can use conservation of momentum to obtain $p_a^* =(E_a^*, p^* \cos\theta, p^* \sin\theta, 0)^T$ and $p_{P_2}^*=(E_{P_2}^*, -p^* \cos\theta, -p^* \sin\theta, 0)^T$, being $\theta$ the angle between the momenta of $a$ and $P_2$. Next, by requiring $(p_a^*)^2 = m_a^2$, $(p_{P_2}^*)^2 = m_{P_2}^2$ and $m_{P_1} = E_a^* + E_{P_2}^*$, we can solve these equations to obtain
\begin{equation}
    E_a^* = \frac{m_{P_1}^2 + m_a^2 - m_{P_2}^2}{2 m_{P_1}}\,,
\end{equation}
\begin{equation}
    E_{P_2}^* = \frac{m_{P_1}^2 + m_{P_2}^2 - m_a^2}{2 m_{P_1}}\,,
\end{equation}
\begin{equation}
    p^* = \frac{\sqrt{m_{P_1}^4+m_a^4+m_{P_2}^4-2 m_{P_1}^2 m_a^2 - 2 m_{P_1}^2 m_{P_2}^2 - 2m_a^2 m_{P_2}^2}}{2 m_{P_1}} \equiv \frac{\sqrt{\lambda(m_{P_1}^2, m_a^2, m_{P_2}^2) }}{2m_{P_1}}\,.
\end{equation}
where $\lambda(x,y,z) = x^2 + y^2 + z^2 -2xy-2xz-2yz $ is the Källén function. 
We move from the $P_1$ rest frame to the LAB frame by a Lorentz boost,
\begin{equation}
    \begin{pmatrix}
        E^\mathrm{LAB} \\ p_x^\mathrm{LAB} \\ p_y^\mathrm{LAB} \\ p_z^\mathrm{LAB}
    \end{pmatrix} = \begin{pmatrix}
        \gamma_{P_1} & \beta_{P_1} \gamma_{P_1} & 0 & 0 \\ \beta_{P_1} \gamma_{P_1} & \gamma_{P_1} & 0 & 0 \\ 0 & 0 & 1 & 0 \\ 0 & 0 & 0 & 1
    \end{pmatrix}
    \begin{pmatrix}
        E^* \\ p_x^* \\ p_y^* \\ p_z^*
    \end{pmatrix} \,.
\end{equation}
For example, in the specific case of $B\to K a$ in Belle II(Belle), the experiment reports a boost factor of $\beta_B\gamma_B = 0.28(0.42)$ (and $\gamma_{P_1} = \sqrt{1+(\beta_{P_1}\gamma_{P_1})^ 2}$). Table~\ref{tab:experiments_decaytypes} reports various information on the experiments considered in \alpaca\!\!. 

The energy and momentum of the ALP in the LAB frame are
\begin{equation}
    E_a^\mathrm{LAB} = \gamma_{P_1} E_a^*+ \gamma_{P_1}\beta_{P_1} p^*\cos\theta\,,
\end{equation}
\begin{equation}
    |\vec{p}_a^\mathrm{LAB}| = \sqrt{\left(\beta_{P_1}\gamma_{P_1} E_a^*+ \gamma_{P_1} p^*\cos\theta\right)^2 + \left(p^* \sin\theta\right)^2}\,.
\end{equation}
The boost needed to go from the rest frame of the ALP to the LAB frame can be directly obtained from the expressions above,
\begin{equation}
    \gamma_a = \frac{E_a^\mathrm{LAB}}{m_a}\,,\qquad \beta_a\gamma_a = \frac{|\vec{p}_a^\mathrm{LAB}|}{m_a}\,.
\end{equation}

The probability density function of finding an ALP that decays after traveling a distance $r$ in the LAB frame is \cite{Ferber:2022rsf,Bruggisser:2023npd,Bauer:2021mvw}
\begin{equation}
    P(r) = \int_0^{\pi/2} \sin\theta d\theta\frac{\exp\left(-\frac{r }{c \tau_a \beta_a \gamma_a}\right)}{c \tau_a \beta_a \gamma_a}\,,
\end{equation}
and integrating the pdf, the probability of finding an ALP decaying between $r_1$ and $r_2$ is simply
\begin{equation}
    P(r_1,r_2) = \int_0^{\theta/2}\sin\theta d\theta\exp\left(-\frac{r_1}{c \tau_a \beta_a \gamma_a}\right)-\int_0^{\pi/2}\sin\theta d\theta\exp\left(-\frac{r_2}{c \tau_a \beta_a \gamma_a}\right)\,.
\end{equation}
If the detector is able to resolve secondary vertices located between $r_\mathrm{min}$ and $r_\textrm{max}$ of the collision point, then the probability of a prompt decay is $P_\textrm{prompt} = P(0, r_\textrm{min})$, for a displaced-vertex decay $P_\textrm{d.v.} = P(r_\textrm{min}, r_\textrm{max})$, and for a decay outside the detector $P_\textrm{out} = P(r_\textrm{max}, \infty)$. The explicit expressions for the prompt and invisible decays have relatively compact expressions,
\begin{align}
    P_\textrm{prompt}&= 1-\int_0^{\pi/2}\sin\theta d\theta\exp\left(-\frac{r_\textrm{min}}{c\tau_a \, \beta_a \gamma_a}\right)\, , \\  P_\textrm{out}&= \int_0^{\pi/2}\sin \theta d \theta\exp\left(-\frac{r_\textrm{max}}{c\tau_a \, \beta_a \gamma_a}\right)\,.
\end{align}

In \alpaca all cases considered above are taken into account internally. For each transition, \alpaca calculates the probability of the process occurring promptly, having a displaced vertex or invisibly. For that, the characteristics of each experiment (see Table~\ref{tab:experiments_decaytypes} for a summary of them) that measure the transition are considered. Once we have the probability distribution, in the calculation of the $\chi^2$-distribution we confront the theoretical prediction, $\mathrm{BR}^\text{theo}\times P_\text{decay type}$, with the experimental measurement. Depending on the decay type, there are some subtleties involved, which we describe below:
\begin{itemize}
    \item If we are dealing with \textbf{prompt decays}, $P_1\to P_2 a(\to P_3)$ (where $P_3$ denotes all the particles in the final state of the ALP decay), the quantity $P_\mathrm{prompt} \times \mathrm{BR}(P_1\to P_2 a)\times\mathrm{BR}(a\to P_3)$ will be confronted in the $\chi^2$ with the experimental determination of $\mathrm{BR}(P_1\to P_2 P_3)$ for those points where $P_\mathrm{prompt}$ is above certain threshold $P_\star$, which is user-defined (the default value is $P_\star=0$). 
    \item In the case of \textbf{displaced vertices}, we follow the approach consistent with the tool provided by LHCb to obtain the bounds on $\mathrm{BR}(B^0 \to K^{*0}a) \times \mathrm{BR}(a\to \mu^+\mu^-)$ in Ref.~\cite{LHCb:2015nkv}. In this case, experimental collaborations report bounds on $\mathrm{BR}(P_1\to P_2 a)\times\mathrm{BR}(a\to P_3)$ as functions of $m_a$ and $c\tau_a$. Hence, we can directly compare $\mathrm{BR}(P_1\to P_2 a)\times\mathrm{BR}(a\to P_3)$ to the bounds on $\mathrm{BR}(P_1\to P_2 a)\times\mathrm{BR}(a\to P_3)$ in the region of the parameter space where $(c\tau_a)_\mathrm{min} < c\tau_a < (c\tau_a)_\mathrm{max}$. If $c\tau_a < (c\tau_a)_\mathrm{min}$, the ALP is promptly decaying and we assume the bounds are equal to those at $(c\tau_a)_\mathrm{min}$ and that $P_\mathrm{prompt}+P_\mathrm{d.v.} = 1$. If $c\tau_a > (c\tau_a)_\mathrm{max}$, the ALP is longer-lived $(c\tau_a)_\mathrm{max}$ so the bound gets suppressed by being multiplied by $P_\mathrm{d.v.}$, again considering only the region where $P_\mathrm{d.v.} > P_\star$. 
    \item Finally, for \textbf{invisible decays}, we compare $P_\mathrm{out} \times\mathrm{BR}(P_1\to P_2 a)$ with the experimental $\mathrm{BR}(P_1\to P_2 + \mathrm{inv.})$ for the points in parameter space where $P_\mathrm{inv.} > P_\star$, if the ALP only decays into SM particles. However, if the ALP can also decay into a dark sector, then our prediction is given by $\mathrm{BR}(P_1\to P_2 a)\times [P_\mathrm{out}\times \mathrm{BR}(a\to \mathrm{SM}) + \mathrm{BR}(a\to\chi\bar{\chi})]$, and the value inside the square brackets has to be larger than $P_\star$.
\end{itemize}

\begin{figure}[t!]
    \centering
    \includegraphics[width=0.6\linewidth]{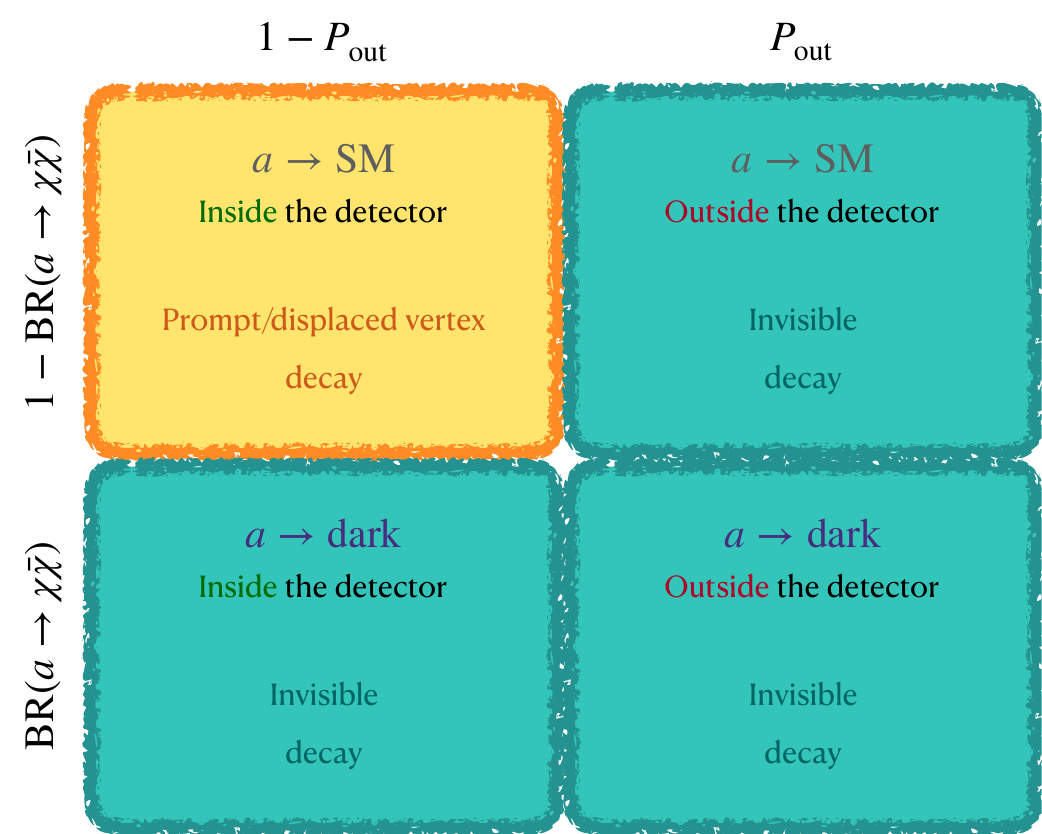}
    \caption{\em Probability for a process corresponding to an invisible decay (teal) or prompt or displaced vertex decay (orange), depending on the probability of decays outside the detector and branching ratio into an invisible sector.}
    \label{fig:diagram_invdecay}
\end{figure}

The diagram in Fig.~\ref{fig:diagram_invdecay} helps identifying the cases when the ALP can be interpreted as missing energy, that is invisible, in a given experiment. We show in teal the case when the ALP is sufficiently long-lived to escape the detector with or without the existence of an ALP decay channel into the dark sector, and when it actually decays within the detector due to a large branching ratio into the dark sector. In orange, instead, the prompt decay and displaced vertices scenarios.

%% file: exp.tex
\subsection{Implementation}\label{sec:exp_impl}
\alpaca is equipped with the latest experimental measurements 
that can be easily retrieved with the function \texttt{alpaca.experimental\_data.get\_measurements}\index{\texttt{experimental\_data}!\texttt{.get\_measurements}}. For example, the following code retrieves all the experimental measurements corresponding to $B^+ \to K^+ \mu^+ \mu^-$:
\begin{minted}{python}
    from alpaca.experimental_data import get_measurements

    exp_BKmumu = get_measurements('B+ -> K+ mu mu')
\end{minted}
By default, \texttt{get\_measurements} excludes projections for future measurements, that instead can be included with the argument \texttt{exclude\_projections = False}. The list of all observables and projections available in \alpaca are listed in the tables of App.~\ref{sec:database_exp}.

The output of \texttt{get\_measurements} is a \texttt{dict}, whose keys identify the experiment and the values are objects of type \texttt{alpaca.experimental\_data.MeasurementBase}\index{\texttt{experimental\_data}!\texttt{.MeasurementBase}}. The following methods are defined for all measurements:
\begin{itemize}
    \item \texttt{get\_central(ma, ctau)}\index{\texttt{experimental\_data}!\texttt{.MeasurementBase}!\texttt{.get\_central}}: Reports the central value for the process with an ALP of mass \texttt{ma} and proper lifetime \texttt{ctau}. The lifetime parameter can be omitted for measurements that are not displaced vertex. In the case of upper limit bounds, this value corresponds with the bound itself.
    
    \item \texttt{get\_sigma\_left(ma, ctau)}\index{\texttt{experimental\_data}!\texttt{.MeasurementBase}!\texttt{.get\_sigma\_left}}: Reports the left one-sided uncertainty for the process with an ALP of mass \texttt{ma} and proper lifetime \texttt{ctau}. The lifetime parameter can be omitted for measurements that are not displaced vertex. In the case of upper limit bounds, this value is zero.
    
    \item \texttt{get\_sigma\_right(ma, ctau)}\index{\texttt{experimental\_data}!\texttt{.MeasurementBase}!\texttt{.get\_sigma\_right}}: Reports the right one-sided uncertainty for the process with an ALP of mass \texttt{ma} and proper lifetime \texttt{ctau}. The lifetime parameter can be omitted for measurements that are not displaced vertex. In the case of upper limit bounds, this value corresponds with Eq.~\eqref{eq:sigmaright}.

    \item \texttt{get\_values(ma, ctau)}\index{\texttt{experimental\_data}!\texttt{.MeasurementBase}!\texttt{.get\_values}}: Reports the central value, left and right one-sided uncertainty for the process with an ALP of mass \texttt{ma} and proper lifetime \texttt{ctau}. The lifetime parameter can be omitted for measurements that are not displaced vertex.
    
    \item \texttt{decay\_type}\index{\texttt{experimental\_data}!\texttt{.MeasurementBase}!\texttt{.decay\_type}}: Returns the experimental signature of the experiment, as described in Section~\ref{sec:signatures}. The possible values are \texttt{"prompt"}, \texttt{"displaced"}, \texttt{"invisible"} or \texttt{"flat"}, the later corresponding to processes without an on-shell ALP.
    
    \item \texttt{decay\_probability(ctau, ma, br\_dark)}\index{\texttt{experimental\_data}!\texttt{.MeasurementBase}!\texttt{.decay\_probability}}: Given an ALP of mass \texttt{ma}, proper lifetime \texttt{ctau}, and optionally branching ratio into the dark sector \texttt{br\_dark}, computes the probability of the process corresponding to the experimental signature \texttt{decay\_type}, taking into account the characteristics of the detector used for the measurement.
\end{itemize}
In all the methods, the ALP mass \texttt{ma} is expressed in GeV and the proper lifetime \texttt{ctau} in cm. 
 
We illustrate these capabilities with the detection of an ALP of mass $m_a=1.2\,\mathrm{GeV}$ and proper lifetime $c\tau_a=100\,\mathrm{cm}$ in $B^+\to K^+\mu^+\mu^-$ at LHCb:

\begin{minted}{python}
from alpaca.experimental_data import get_measurements

exp_BKmumu = get_measurements('B+ -> K+ mu mu')
lhcb_BKmumu = exp_BKmumu['LHCb']

central = lhcb_BKmumu.get_central(ma=1.2, ctau=100.0)
sigma_l = lhcb_BKmumu.get_sigma_left(ma=1.2, ctau=100.0)
sigma_r = lhcb_BKmumu.get_sigma_right(ma=1.2, ctau=100.0)
sign = lhcb_BKmumu.decay_type
prob = lhcb_BKmumu.decay_probability(ma=1.2, ctau=100.0)

print(f"Bound = {central} + {sigma_r} - {sigma_l}")
print(f"{sign} decay with probability {prob}")
    
\end{minted}
\vspace{-8mm}
\begin{minted}{pwsh-session}
Bound = [2.54599857e-08] + [1.06916471e-09] - 0.0
displaced decay with probability [0.47188142]
\end{minted}
\subsection{Experimental sectors}
\label{sec:exp_sectors}
Experimental measurements are grouped in \alpaca in sectors, which are implemented using the class \texttt{alpaca.sectors.Sector}\index{\texttt{sectors}!\texttt{.Sector}}. A large variety of sectors are predefined in \texttt{alpaca.sectors.default\_sectors}\index{\texttt{sectors}!\texttt{.default\_sectors}}, and are summarised in Table~\ref{tab:sectors}.

\begin{table}[h!]
    \centering
    \begin{tabular}{c|c}
    Sector & Observables\\\hline
       \rowcolor{lightgray!40}\texttt{BKinv}  & $B^0 \to K^0  +\mathrm{inv}$, $B^+ \to K^+  +\mathrm{inv}$, $B^+ \to K^{*+} +\mathrm{inv}$  \\
        \texttt{Kpiinv} & $K_L^0\to \pi^0 +\mathrm{inv}$, $K^+ \to \pi^+ +\mathrm{inv}$ \\
        \rowcolor{lightgray!40}& $K^+\to \pi^+ +\mathrm{inv}$, $K^0_L\to \pi^0 +\mathrm{inv}$, \\
        \rowcolor{lightgray!40}&$K^+\to \pi^+\gamma \gamma$, $K^+\to \pi^+ e^+e^-$, $K^+\to \pi^+\mu^+\mu^-$, \\
        \rowcolor{lightgray!40}\multirow{-3}{*}{\texttt{sda\_lfu}}&$K^0_S\to\gamma\gamma$, $K^0_L\to e^+e^-$, $K^0_S\to e^+e^-$, $K^0_L\to \mu^+\mu^-$, $K^0_S\to \mu^+\mu^-$\\
         \texttt{sda\_lfv} & $K^+\to \pi^+\mu e$, $K^0_L\to \pi^0\mu e$ \\
        \rowcolor{lightgray!40}& $B^+\to \pi^++\mathrm{inv}$, $B^0\to \pi^0 +\mathrm{inv}$, $B^+\to \rho^++\mathrm{inv}$, $B^0\to \rho^0 +\mathrm{inv}$, \\
        \rowcolor{lightgray!40}&$B^+\to \pi^+ e^+e^-$, $B^0\to \pi^0 e^+e^-$, $B^+\to \pi^+ \mu^+\mu^-$, $B^0\to \pi^0 \mu^+\mu^-$, \\
        \rowcolor{lightgray!40}\multirow{-3}{*}{\texttt{bda\_lfu} }&$B^0\to \gamma \gamma$, $B^0\to e^+e^-$, $B^0\to \mu^+\mu^-$, $B^0\to \tau^+\tau^-$ \\
         \texttt{bda\_lfv} & $B^+\to \pi^+ \tau e$, $B^+\to \pi^+ \tau \mu$\\
          \rowcolor{lightgray!40}& $B^+\to K^+ +\mathrm{inv}$, $B^0\to K^0 +\mathrm{inv}$, $B^+\to K^{+\ast} +\mathrm{inv}$, $B_s\to \phi +\mathrm{inv}$, \\
         \rowcolor{lightgray!40}&$B^+\to K^+ \gamma \gamma$, $B^+\to K^+ e^+e^-$, $B^+\to K^+ \mu^+\mu^-$,  $B^+\to K^+ \tau^+\tau^-$\\
         \rowcolor{lightgray!40}&$B^0\to K^{0\ast} e^+e^-$, $B^0\to K^{0\ast} \mu^+\mu^-$, $B^0\to K^{0\ast} \tau^+\tau^-$,\\
         \rowcolor{lightgray!40}&$B^+\to K^+ \pi^0\pi^+\pi^-$, $B^0\to K^0\pi^0\pi^+\pi^-$, $B^+\to K^+ \eta\, \pi^+\pi^-$, \\
         \rowcolor{lightgray!40}\multirow{-5}{*}{\texttt{bsa\_lfu}}&$B_s\to \gamma \gamma$, $B_s\to e^+e^-$, $B_s\to \mu^+\mu^-$, $B_s\to \tau^+\tau^-$\\
          & $B^+\to K^+ \mu e$, $B^0\to K^0 \mu e$, $B^+\to K^{\ast+} \mu e$, $B^0\to K^{\ast0} \mu e$, \\
         \multirow{-2}{*}{\texttt{bsa\_lfv}}&$B^+\to K^+ \tau e$, $B^+\to K^+ \tau \mu$,  $B^0\to K^{\ast 0} \tau \mu$\\
         \rowcolor{lightgray!40} & $D^0\to \pi^0+\mathrm{inv}$, $D^0\to \pi^0 e^+e^-$, $D^0\to \eta\, e^+e^-$, $D^0\to \rho^0 e^+e^-$, \\
         \rowcolor{lightgray!40}&$D^+\to \pi^+ e^+e^-$, $D^+_s\to K^+ e^+e^-$,\\
         \rowcolor{lightgray!40}&$D^0\to \pi^0 \mu^+\mu^-$, $D^0\to \eta \mu^+\mu^-$, $D^0\to \rho^0 \mu^+\mu^-$,\\
         \rowcolor{lightgray!40}&$D^+\to \pi^+ \mu^+\mu^-$, $D^+\to \rho^+ \mu^+\mu^-$, $D^+_s\to K^+ \mu^+\mu^-$, \\
         \rowcolor{lightgray!40}\multirow{-5}{*}{\texttt{cua\_lfu}}&$D^0\to\gamma\gamma$, $D^0\to e^+e^-$, $D^0\to \mu^+\mu^-$\\
         \texttt{cua\_lfv} & $D^0\to \pi^0 \mu e$, $D^0\to \eta \mu e$, $D^0\to \rho^0 \mu e$, $D^+_s\to K^+ \mu e$, $D^+_s\to K^{\ast +} \mu e$ \\
        \rowcolor{lightgray!40} & $e^+e^-\to \gamma \gamma\gamma$, $J/\psi\to \gamma +\mathrm{inv} $, $\Upsilon(1S)\to \gamma +\mathrm{inv} $, $\Upsilon(3S)\to \gamma +\mathrm{inv} $, \\
        \rowcolor{lightgray!40}&$J/\psi\to \gamma\gamma\gamma$, $J/\psi\to \gamma \mu^+\mu^-$, $\Upsilon(1S)\to \gamma \mu^+\mu^-$, $\Upsilon(3S)\to \gamma \mu^+\mu^-$, \\
        \rowcolor{lightgray!40} \multirow{-3}{*}{\texttt{quarkonia\_lfu}}&$\Upsilon(1S)\to \gamma \tau^+\tau^-$, $\Upsilon(3S)\to \gamma \tau^+\tau^-$, $\Upsilon(1S)\to \gamma c\overline{c}$, $\Upsilon(3S)\to \gamma +\mathrm{hadrons}$\\
         \texttt{quarkonia\_lfv} & $\Upsilon(1S)\to \gamma \mu e$, $\Upsilon(1S)\to \gamma \tau e$, $\Upsilon(1S)\to \gamma \mu \tau$ \\
         \rowcolor{lightgray!40}\texttt{meson\_mixing} & $\Delta m_{K^0}$, $\epsilon_K$, $x_{D^0}$, $\phi^{12}_{D_0}$, $\Delta m_{B^0}$, $\mathcal{A}_\mathrm{SL}(B^0)$, $\Delta m_{B_s^0}$, $\mathcal{A}_\mathrm{SL}(B_s^0)$\\
         \texttt{meson\_dw} & $\Gamma_{B^+}$, $\Gamma_{B^0}$, $\Gamma_{D^+}$, $\Gamma_{D^0}$, $\Gamma_{D_s^+}$, $\Gamma_{K^+}$, $\Gamma_{K^0_S}$, $\Gamma_{K^0_L}$\\
         \rowcolor{lightgray!40} & $\mu \to e+\mathrm{inv}$, $\mu \to e\gamma\gamma$, $\mu \to e\,e\,e$, \\
         \rowcolor{lightgray!40}&$\tau \to e+\mathrm{inv}$, $\tau \to e\gamma\gamma$, $\tau \to e\,e\,e$, $\tau \to e\,\mu\,\mu$,\\
         \rowcolor{lightgray!40} \multirow{-3}{*}{\texttt{lepton\_lfv}}&$\tau \to \mu+\mathrm{inv}$, $\tau \to \mu\gamma\gamma$, $\tau \to \mu\,e\,e$, $\tau \to \mu\,\mu\,\mu$\\
         \texttt{all}& All transitions\\
         \hline
    \end{tabular}
    \caption{\em Sectors implemented in \texttt{alpaca.sectors.default\_sectors}.}
    \label{tab:sectors}
\end{table}

Additionally, users can define their own sectors, by providing the following information:
\begin{itemize}
    \item \texttt{name}: text string that identifies the sector.
    \item \texttt{tex}: text string for the \TeX\  of the sector, used for plot labels. \TeX\  must be inserted between \$ symbols, and special characters such as \textbackslash~ must be properly escaped (for example, by using a \texttt{r-string}).
    \item \texttt{observables}: list containing all observables in the sector, for which all experiments are included.
    \item \texttt{obs\_measurements}: dictionary of observables and lists of experiments included in the sector.
    \item Optionally, a \texttt{description} of the sector.
    \item Additionally, the user can customise the plotting style of the sector via the optional arguments \texttt{color} (a \texttt{str} with the hex code of the color), \texttt{lw} (the line width, in points), and \texttt{ls} (the line style, as defined by \texttt{matplotlib}).
\end{itemize}

As an example, the following code defines a sector including only Belle II measurements:
\begin{minted}{python}
import alpaca

sector_BelleII = alpaca.sectors.Sector(
    name = 'Belle II',
    tex = r'$\mathrm{Belle II}$ bounds',
    observables = [
        'B+ -> K+ e e',
        'B0 -> K*0 e e',
        ('e e -> gamma gamma gamma', 10.58**2),
        'delta_mB0',
        'tau -> e a',
        'tau -> mu a',
    ],
    obs_measurements = {
        'B+ -> K+ mu mu': ['Belle II'],
        'B+ -> K+ a': ['Belle II'],
        'tau -> e e e e': ['Belle II'],
        'tau -> e mu mu mu': ['Belle II'],
        'tau -> mu mu mu': ['Belle II'],
        'tau -> mu e e': ['Belle II'],
    },
    description = 'Belle II measurements implemented in ALPaca.'
)
\end{minted}

The method \texttt{save()}\index{\texttt{sectors}!\texttt{.Sector}!\texttt{.save}} of the \texttt{Sector} instance is used to save it as a \texttt{JSON} file. Conversely, \texttt{sectors.Sector.load()}\index{\texttt{sectors}!\texttt{.Sector}!\texttt{.load}} reads a sector from \texttt{JSON} file. Additionally, the function \texttt{sectors.initialize\_sectors()}\index{\texttt{sectors}!\texttt{.initialize\_sectors}} loads all the sectors stored in a folder into a dictionary.

Several sectors can be merged together by using \texttt{sectors.combine\_sectors}\index{\texttt{sectors}!\texttt{.combine\_sectors}}. For example, we can merge the $B\to K+\mathrm{inv}$ and $K\to \pi +\mathrm{inv}$ as follows,

\begin{minted}{python}
import alpaca

sector_invisible = alpaca.sectors.combine_sectors(
    [
        alpaca.sectors.default_sectors['BKinv'],
        alpaca.sectors.default_sectors['Kpiinv']
    ],
    name = 'MesonsInv',
    tex = r'$M_1 \to M_2 + \mathrm{inv}$',
    description = 'B -> K inv and K -> pi inv observables.'
)
\end{minted}

Finally, it is possible to check if a given observable is included in the sector using the method \texttt{contains\_observable}\index{\texttt{sectors}!\texttt{.Sector}!\texttt{.contains\_observable}}. 

A summary of the methods available within \texttt{alpaca.sectors} is given in Table~\ref{Tab:methodsSectors}.

\begin{table}[h!]
\begin{tabularx}{\textwidth}{ c|X } 
 Method name& Output\\
 \hline
  \rowcolor{lightgray!40}\texttt{Sector.save()} & Save \texttt{Sector} as \texttt{JSON} file. \\ 
\texttt{Sector.load()} & Load \texttt{Sector} from \texttt{JSON} file.  \\ 
 \rowcolor{lightgray!40} \texttt{initialize\_sectors} & Load all sectors stored in a folder.  \\ 
\texttt{combine\_sectors} & Merge several sectors.\\
  \rowcolor{lightgray!40}\texttt{contains\_observable} &Check if desired observable is included in the sector. \\ 
 \hline
\end{tabularx}
 \caption{\em Summary of the methods available within \texttt{alpaca.sectors}.}
\label{Tab:methodsSectors}
\end{table}

%% file: stats.tex
The question that we aim to answer is which points of the parameter space are compatible with the experimental results, and to what degree. 

According to the Neyman-Pearson lemma, under certain regularity conditions, the likelihood ratio is the most powerful test for hypothesis testing. In a model with parameters $\theta$, the likelihood function $L(\theta)$ is obtained by evaluating the probability distribution of the data $p(x|\theta)$ at the observed data $x_\mathrm{obs}$, $L(\theta) = p(x=x_\mathrm{obs}|\theta)$. Then, the likelihood ratio test establishes that, given the hypothesis $H_0\!: \theta=\theta_0$ and $H_1\!: \theta=\theta_1$, it is possible to reject $H_0$ with probability $\alpha$ if the test statistic~\cite{Cousins:2018tiz}
\begin{equation}
    \lambda=\frac{L(\theta_0)}{L(\theta_1)}
\end{equation}
satisfies $\lambda < \lambda_\alpha$, where $\lambda_\alpha$ is such that $P(\lambda<\lambda_\alpha|H_0) = \alpha$.  The test is the most powerful in the sense that it minimises the probability $\beta$ of type II error (failing to reject a false hypothesis) for a given probability $\alpha$ of type I error (rejecting a true hypothesis).

If $x$ is a vector of $n$ statistically independent variables, the likelihood function factorises as
\begin{equation}
    L(\theta) = \prod_{i=1}^n \ell_i(\theta)\,.
\end{equation}
In the case of Gaussian random variables, commonly reported as $x_i = \mu_i\pm \sigma_i$, the individual likelihood function takes the well-known expression
\begin{equation}
    \ell_i(\theta) = \frac{1}{\sqrt{2\pi\sigma_i^2}}\,\mathrm{exp}\!\!\left(-\frac{(\tilde{x}_i(\theta) - \mu_i)^2}{2\sigma_i^2}\right)\,,
\end{equation}
where $\tilde{x}_i(\theta)$ is the prediction of the variable $x_i$ assuming the model parameters $\theta$. There are cases in which experimental collaborations only report upper limit bounds as $x_i <u_i\,(c_i\ \mathrm{C.L.})$. For these type of bounds we assume a uniform distribution below $u_i$ and a half-normal distribution above $u_i$, that is,
\begin{equation}
    \ell_i(\theta) = \left\{ \begin{matrix}
        \frac{c_i}{u_i} & \quad & \tilde{x}_i(\theta) \leq u_i\,,\\
        \frac{c_i}{u_i}\,\mathrm{exp}\!\!\left(-\frac{(\tilde{x}_i(\theta) - u_i)^2}{2\sigma_i^2}\right) & \quad & \tilde{x}_i(\theta) > u_i\,,
    \end{matrix} \right.
\end{equation}
where 
\begin{equation}\label{eq:sigmaright}
\sigma_i = \sqrt{\frac{2}{\pi}} u_i \frac{1-c_i}{c_i}
\end{equation}
is chosen so that the likelihood is continuous and properly normalised.

Up to this point, the discussion is completely general for any choices of $H_0\!: \theta=\theta_0$ and $H_1\!: \theta=\theta_1$. In our case, $H_0$ will correspond to the hypothesis of an ALP with parameters $\theta_0 = \{m_a, f_a, c_{ij}(\Lambda)\}$, each choice of $\theta_0$ is a disjoint hypothesis and therefore there are no degrees of freedom. Since we want to confront $H_0$ directly to experimental data, and not to a competing model (e.g. SM) or to a best-fit determination of $\theta$, we will take $H_1$ as the \textit{saturated model}: for Gaussian distributed variables we choose $\tilde{x}_i(\theta_1) = \mu_i$, and for upper limit bounds, $\tilde{x}_i(\theta_1) = u_i$. Then it is easy to check that the logarithm of the test statistic takes the simple form
\begin{equation}
    -2\log \lambda = \sum_{x_i \in X_g} \frac{(\tilde{x}_i(\theta_0) - \mu_i)^2}{2\sigma_i^2} + \sum_{x_i \in X_{UL+}} \frac{(\tilde{x}_i(\theta_0) - u_i)^2}{2\sigma_i^2}\,,
\end{equation}
where $X_g$ is the set of Gaussian random variables, and $X_{UL+}$ is the set of variables in which the model exceeds the upper bound. Note that the variables in which the model is compatible with the upper bound do not contribute to the test statistic, as the likelihood in $H_0$ and $H_1$ cancel out. The quantity $-2\log\lambda$ is itself a random variable that follows a $\chi^2$ distribution with $n_g + n_{UL+}$ degrees of freedom. $H_0$ will be rejected then if
\begin{equation}
    -2\log\lambda > -2\log\lambda_\alpha\,,\qquad\qquad \int_{-2\log\lambda_\alpha}^\infty f_{\chi^2}(t; n_g + n_{UL+}) dt = \alpha\,,
\end{equation}
where $f_{\chi^2}(t; k)$ is the pdf of the $\chi^2$ with $k$ degrees of freedom.

Alternatively, instead of setting a fixed value of $\alpha$, we can obtain the minimum value of $\alpha$ that would exclude $H_0$, which is known as the $p$-value,
\begin{equation}\label{eq:pvalue}
    p = \int_{-2\log\lambda}^\infty f_{\chi^2}(t; n_g + n_{UL+}) dt\,.
\end{equation}
The $p$-value is typically translated into the significance $Z$, in units of $\sigma$,
\begin{equation}\label{eq:signif_sigma}
    Z = \Phi^{-1}(1-p/2) = \sqrt{2}\mathrm{erf}^{-1}(1-p)\,,
\end{equation}
where $\Phi^{-1}$ is the inverse of the cumulative distribution function of the Gaussian distribution, and $\mathrm{erf}^{-1}$ the inverse error function.

\subsection{Implementation}

The full statistical analysis is implemented in \alpaca in the function \texttt{statistics.get\_chi2}\index{\texttt{statistics}!.\texttt{get\_chi2}}. It takes the following arguments:
\begin{itemize}
    \item \texttt{transitions}: list of sectors included in the total $\chi^2$. Additionally, individual observables can also be included using the syntax described in Section~\ref{sec:structure_process}, which will define a sector for them. The calculation avoids the double-counting of observables included in multiple sectors.
    \item \texttt{ma}: \texttt{float} or container corresponding to the value(s) of the mass of the ALP, in GeV.
    \item\texttt{couplings}: Object of type \texttt{ALPcouplings} or container with the couplings of the ALP to SM particles.
    \item\texttt{fa}: \texttt{float} or container corresponding to the ALP scale $f_a$, in GeV.
    \item\texttt{min\_probability}: Cut-off probability $P_\star$ for the given signature, by default $P_\star = 0.0$. Parameter points that produce \texttt{decay\_probability} (defined in Section~\ref{sec:exp_impl}) below the cut-off $P_\star$ are not included in the $\chi^2$.
    \item Optionally, \texttt{br\_dark}: \texttt{float} or container corresponding to the branching ratio of the ALP into the dark sector.
    \item Additionally, the function can also use the optional arguments for the running of the ALP couplings described in Section~\ref{sec:running_impl}.
\end{itemize}

At each parameter point, the function evolves and matches the ALP couplings to the physical scale, calculates the probability of the experimental signature and the observables, and finally compares them with the experimental distribution, as explained previously. 

The result of the $\chi^2$ analysis of each sector is stored in a variable belonging to the class \texttt{alpaca.statistics.ChiSquared}\index{\texttt{statistics}!\texttt{.ChiSquared}}. All the \texttt{ChiSquared} variables resulting from \texttt{get\_chi2} are contained in a \texttt{alpaca.statistics.ChiSquaredList}\index{\texttt{statistics}!\texttt{.ChiSquaredList}} object, which is a subclass of the \texttt{list} class. It is straightforward to merge the elements of a \texttt{ChiSquaredList} into a single \texttt{ChiSquared} object, and to select or remove specific observables and/or measurements.

The elements contained inside a \texttt{ChiSquaredList} and their respective indices can be visualised by using the standard \texttt{print()} function. Additionally, evaluating the \texttt{ChiSquaredList} object in a \texttt{jupyter} cell will display the same information as a table.\index{\texttt{statistics}!\texttt{.ChiSquaredList}!\texttt{print()}}

Since \texttt{ChiSquaredList} is a subclass of \texttt{list}, individual \texttt{ChiSquared} elements can be accessed with the usual index notation, and new \texttt{ChiSquaredList} objects can be created from lists of them. The elements of a list can be merged into a single \texttt{ChiSquared} object using the method \texttt{combine()}\index{\texttt{statistics}!\texttt{.ChiSquaredList}!\texttt{.combine}}, which takes the name, \TeX\ and optionally description of the combined sector.

\begin{minted}{python}
import alpaca
from alpaca.statistics import get_chi2, ChiSquaredList
import numpy as np

#Create grid of ma, fa and couplings
ma = np.logspace(-2, np.log10(4), 20)
fa = np.logspace(3, 8, 20)
x_ma, y_fa = np.meshgrid(ma, fa)
couplings = [alpaca.ALPcouplings({'cuR': np.diag([0,0,1])},
    4*np.pi*f, 'derivative_above').match_run(4, 'VA_below') for f in fa]
x_ma, y_couplings = np.meshgrid(ma, couplings)

transitions_BKll = [
    'B+ -> K+ mu mu',
    'B+ -> K+ e e',
    'B0 -> K*0 mu mu',
    'B0 -> K*0 e e'
]

chi2_BKll = get_chi2(
    transitions_BKll,
    x_ma,
    y_couplings,
    y_fa,
    integrator='no_rge'   # The running below 4 GeV is negligible
)

#Select only B -> K mu mu observables
chi2_BKmumu = ChiSquaredList([chi2_BKll[0], chi2_BKll[2]])

#Obtain combined chi2
chi2_comb = chi2_BKmumu.combine(
    name = 'B->Kmumu',
    tex = r'$B \to K \mu^+ \mu^-$',
    description = 'Combined chi2 to B -> K mu mu observables'
)
\end{minted}

The rest of the methods in this section have been implemented in both \texttt{ChiSquared} and \texttt{ChiSquaredList} objects and summarised in Table~\ref{Tab:MethodsChiSquare}:

\begin{table}[ht!]
\begin{tabularx}{\textwidth}{ c|X } 
 Method name& Output\\
 \hline
  \rowcolor{lightgray!40}\texttt{combine} & Combine different \texttt{ChiSquared} from \texttt{ChiSquaredList} (only for \texttt{ChiSquaredList} objects). \\ 
\texttt{get\_observables} & List of observables included.  \\ 
  \rowcolor{lightgray!40}\texttt{get\_measurements} & List of measurements included.  \\ 
\texttt{split\_observables} & List of \texttt{ChiSquared} objects, each containing measurement of observable. \\ 
  \rowcolor{lightgray!40}\texttt{split\_measurements} &List of \texttt{ChiSquared} objects, each containing  only one measurement.  \\ 
\texttt{exclude\_observables}&\texttt{ChiSquared} or \texttt{ChiSquaredList} without specified observables.\\
  \rowcolor{lightgray!40}\texttt{exclude\_measurements}&\texttt{ChiSquared} or \texttt{ChiSquaredList} without specified measurements.\\
\texttt{extract\_observables}&\texttt{ChiSquared} or \texttt{ChiSquaredList} with specified observables.\\
  \rowcolor{lightgray!40}\texttt{extract\_measurements}&\texttt{ChiSquared} or \texttt{ChiSquaredList} with specified measurements.\\
\texttt{constraining\_observables}&Extract observables which give the most constraining exclusion bounds.\\
\rowcolor{lightgray!40}\texttt{constraining\_measurements}&Extract measurements which give the most constraining exclusion bounds.\\
\texttt{significance}&Exclusion significance (in units of $\sigma$) at each point.\\
\rowcolor{lightgray!40}\texttt{contour}&Extract the contours of constant significance.\\
\texttt{contour\_to\_csv}&Save the contours of constant significance to \texttt{CSV} file.\\
\rowcolor{lightgray!40}\texttt{slicing}&Select from \texttt{chi2} regions of the parameter space.\\
\end{tabularx}
 \caption{\em Summary of the methods available for \texttt{ChiSquared} and \texttt{ChiSquaredList} objects.}
\label{Tab:MethodsChiSquare}
\end{table}

The methods \texttt{get\_observables()}\index{\texttt{statistics}!\texttt{.ChiSquared}!\texttt{.get\_observables}}\index{\texttt{statistics}!\texttt{.ChiSquaredList}!\texttt{.get\_observables}} and \texttt{get\_measurements()}\index{\texttt{statistics}!\texttt{.ChiSquared}!\texttt{.get\_measurements}}\index{\texttt{statistics}!\texttt{.ChiSquaredList}!\texttt{.get\_measurements}} produce a list with the observables or measurements included in the \texttt{ChiSquared} and \texttt{ChiSquaredList} objects.

The method \texttt{split\_observables()}\index{\texttt{statistics}!\texttt{.ChiSquared}!\texttt{.split\_observables}}\index{\texttt{statistics}!\texttt{.ChiSquaredList}!\texttt{.split\_observables}} returns a list of \texttt{ChiSquared} objects, each containing all the measurements of the corresponding observable. Analogously, \texttt{split\_measurements()}\index{\texttt{statistics}!\texttt{.ChiSquared}!\texttt{.split\_measurements}}\index{\texttt{statistics}!\texttt{.ChiSquaredList}!\texttt{.split\_measurements}} returns a list of \texttt{ChiSquared} objects, each containing only one measurement. Following with the previous example, it is possible to see the difference in the generated outputs.
\begin{minted}{python}
chi2_BKll.split_observables()
\end{minted}
\begin{tcolorbox}[
  colback=black,        
  colframe=black,       
  sharp corners=southwest,
  boxrule=0.5pt
]
\color{lightgray} 
\begin{tabular}{|c|l|}
\hline
\textbf{Index} & \textbf{Sector} \\
\hline
0 & \(B^+ \to K^+ e e\) \\
1 & \(B^+ \to K^+ \mu \mu\) \\
2 & \(B^0 \to K^{*0} e e\) \\
3 & \(B^0 \to K^{*0} \mu \mu\) \\
\hline
\end{tabular}
\end{tcolorbox}

\begin{minted}{python}
chi2_BKll.split_measurements()
\end{minted}
\begin{tcolorbox}[colback=black, colframe=black,  sharp corners=southwest, boxrule=0.5pt
]
\color{lightgray} 
\begin{tabular}{|c|l|}
\hline
\textbf{Index} & \textbf{Sector} \\
\hline
0 & \( B^0 \to K^{*0} \mu \mu \ \mathrm{(LHCb)} \) \\
1 & \( B^+ \to K^+ \mu \mu \ \mathrm{(CHARM)} \) \\
2 & \( B^0 \to K^{*0} e e \ \mathrm{(Belle\ II)} \) \\
3 & \( B^+ \to K^+ \mu \mu \ \mathrm{(Belle\ II)} \) \\
4 & \( B^+ \to K^+ \mu \mu \ \mathrm{(LHCb)} \) \\
5 & \( B^+ \to K^+ e e \ \mathrm{(Belle\ II)} \) \\
6 & \( B^0 \to K^{*0} \mu \mu \ \mathrm{(Belle\ II)} \) \\
\hline
\end{tabular}
\end{tcolorbox}

The method \texttt{exclude\_observables}\index{\texttt{statistics}!\texttt{.ChiSquared}!\texttt{.exclude\_observables}}\index{\texttt{statistics}!\texttt{.ChiSquaredList}!\texttt{.exclude\_observables}} returns a new \texttt{ChiSquared} or \texttt{ChiSquaredList} in which the observables specified (using the format in Section~\ref{sec:structure_process}) have been removed. Analogously, \texttt{exclude\_measurements}\index{\texttt{statistics}!\texttt{.ChiSquared}!\texttt{.exclude\_measurements}}\index{\texttt{statistics}!\texttt{.ChiSquaredList}!\texttt{.exclude\_measurements}} returns a new \texttt{ChiSquared} or \texttt{ChiSquaredList} which does not contain any of the observables specified. Each excluded measurement is indicated as a \texttt{tuple}, with the first element corresponding to the observable and the second to the experiment. In the case of \texttt{ChiSquared} object, if all the observables or measurements have been excluded, the result is \texttt{None}, while in the case of \texttt{ChiSquaredList} objects, any element that has been completely excluded is removed from the list.

Conversely, the method \texttt{extract\_observables}\index{\texttt{statistics}!\texttt{.ChiSquared}!\texttt{.extract\_observables}}\index{\texttt{statistics}!\texttt{.ChiSquaredList}!\texttt{.extract\_observables}} returns a new \texttt{ChiSquared} or \texttt{ChiSquaredList} restricting to only the specified observables. Similarly for the method \texttt{extract\_measurements}\index{\texttt{statistics}!\texttt{.ChiSquared}!\texttt{.extract\_measurements}}\index{\texttt{statistics}!\texttt{.ChiSquaredList}!\texttt{.extract\_measurements}}.

\begin{figure}[t!]
    \centering
    \includegraphics[width=0.5\linewidth]{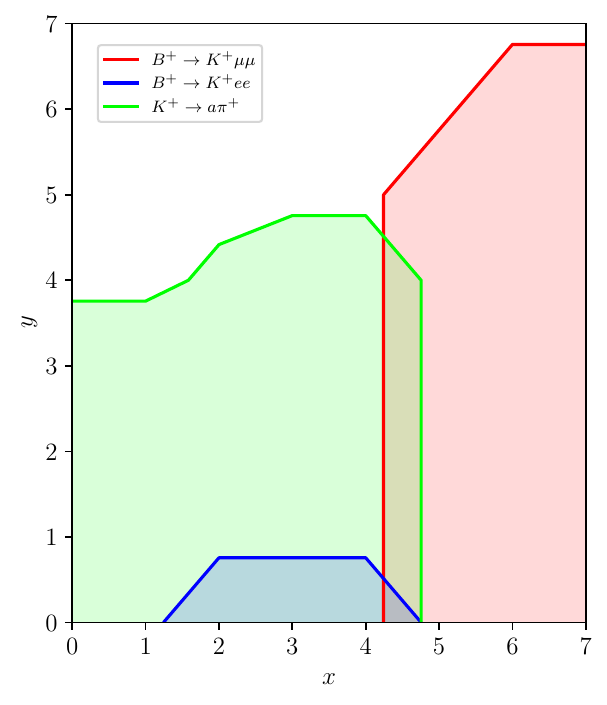}
    \caption{\em Example of the \texttt{constraining\_observables} method. The lines correspond to the $2\,\sigma$ contour for each observable, and the shaded regions for significances greater than $2\,\sigma$. Using the \texttt{"y-inverted"} criterion (default), for the values of $x=0, 1, 2, 3, 4$, starting from $y=7$ and going downwards, the first observable that becomes excluded is $K^+\to \pi^+a$. For $x=5,6,7$, the first observable that becomes excluded is $B^+\to K^+\mu\mu$. Therefore, the result of \texttt{constraining\_observables("y-inverted")} would include only these two observables.}
    \label{fig:constraining}
\end{figure}

The method \texttt{constraining\_observables}\index{\texttt{statistics}!\texttt{.ChiSquared}!\texttt{.constraining\_observables}}\index{\texttt{statistics}!\texttt{.ChiSquaredList}!\texttt{.constraining\_observables}} extracts from the \texttt{ChiSquared} or \texttt{ChiSquaredList} the observables that provide the most constraining exclusion bounds. There are three criteria to decide the most constraining observables:
\begin{itemize}
    \item \texttt{"y-inverted"} (default): Only if the parameter space is a two-dimensional grid. An observable is extracted if, for any value of the coordinate $x$, it provides the maximum significance at the largest $y$ coordinate, as long as the significance is at least $2\,\sigma$. If the maximum significance is below $2\,\sigma$, the $y$ direction is traversed in decreasing order, until one observable above $2\,\sigma$ is found. An example of this criterion is shown in Fig.~\ref{fig:constraining}.
    \item \texttt{"y"}: Only if the parameter space is a two-dimensional grid. Similar to the previous criterion, but for each value of $x$, the search starts at the lowest value of $y$ and proceeds in increasing order.
    \item \texttt{"grid"}: An observable is extracted if it provides the largest significance at any point of the grid.
\end{itemize}
The analogous method for measurements, \texttt{constraining\_measurements}\index{\texttt{statistics}!\texttt{.ChiSquared}!\texttt{.constraining\_measurements}}\index{\texttt{statistics}!\texttt{.ChiSquaredList}!\texttt{.constraining\_measurements}}, also exists.

The method \texttt{significance()}\index{\texttt{statistics}!\texttt{.ChiSquared}!\texttt{.significance}}\index{\texttt{statistics}!\texttt{.ChiSquaredList}!\texttt{.significance}} provides the exclusion significance, in units of $\sigma$, at each point in parameter space. Note that due to numerical precision, the minimum $p$-value is $10^{-16}$, which corresponds to a maximum exclusion significance of $Z = 8.3\,\sigma$.

In the following example, we use the functions in the \texttt{statistics} module to compute the exclusion significance for $B \to K^{(*)} \nu \bar{\nu}$ and $K \to \pi \nu \bar{\nu}$ for a top-philic ALP with mass between $0.01\,\mathrm{GeV}$ and $4\,\mathrm{GeV}$ and $f_a$ between $10^3\,\mathrm{GeV}$ and $10^8\,\mathrm{GeV}$:

\begin{minted}{python}
import alpaca
import numpy as np

#Create grid of ma, fa and couplings
ma = np.logspace(-2, np.log10(4), 20)
fa = np.logspace(3, 8, 20)
x_ma, y_fa = np.meshgrid(ma, fa)
couplings = [alpaca.ALPcouplings({'cuR': np.diag([0,0,1])},
    4*np.pi*f, 'derivative_above').match_run(4, 'VA_below') for f in fa]
x_ma, y_couplings = np.meshgrid(ma, couplings)

transitions_inv = [
    alpaca.sectors.default_sector['BKinv'],
    alpaca.sectors.default_sector['Kpiinv']
]

chi2_inv = alpaca.statistics.get_chi2(
    transitions_inv,
    x_ma,
    y_couplings,
    y_fa,
    integrator='no_rge'   # The running below 4 GeV is negligible
)

# Calculate significance for B->Ka, K->pi a and total
sigmas_BKinv = chi2_inv[0].significance()
sigmas_Kpiinv = chi2_inv[1].significance()
sigmas_total = chi2_inv.significance()
\end{minted}

If the parameter space is two dimensional, the method \texttt{contour}\index{\texttt{statistics}!\texttt{.ChiSquared}!\texttt{.contour}}\index{\texttt{statistics}!\texttt{.ChiSquaredList}!\texttt{.contour}} extracts the contours of constant significance, using as arguments the $x$ and $y$ coordinates in parameter space, and the desired significance (if not provided, $2\,\sigma$ is assumed by default). If the contour lines have disjoint segments, they will be separated by \texttt{NaN} values, in a way that can be directly used by \texttt{matplotlib}. The contours can be saved to a \texttt{CSV} file using the method \texttt{contour\_to\_csv}\index{\texttt{statistics}!\texttt{.ChiSquared}!\texttt{.contour\_to\_csv}}\index{\texttt{statistics}!\texttt{.ChiSquaredList}!\texttt{.contour\_to\_csv}}, which also requires the file name and optionally the labels for the columns of the file.

In the following example, we plot and save the contours for the $\chi^2$ of $B \to K^{(*)}\nu\bar{\nu}$ observables previously obtained:

\begin{minted}{python}
import matplotlib.pyplot as plt

for sigma in [1, 2, 3]:
    x_contour, y_contour = chi2_inv[0].contour(ma, fa, sigma)
    plt.loglog(x_contour, y_contour, color='r', lw = 4-sigma)
    chi2[0].contour_to_csv(ma, fa, f'BKnunu_{sigma}sigma.csv',
                            sigma, 'ma', 'fa')
plt.xlabel(r'$m_a$ [GeV]')
plt.ylabel(r'$f_a$ [GeV]')
\end{minted}

Finally, the method \texttt{slicing}\index{\texttt{statistics}!\texttt{.ChiSquared}!\texttt{.slicing}}\index{\texttt{statistics}!\texttt{.ChiSquaredList}!\texttt{.slicing}} allows to select regions of the parameter space. It takes as many arguments as dimensions the grid has: the $i$-th argument can be either \texttt{int} (to select one index of the $i$-th dimension),  \texttt{slice(start,stop,step)} (to select a range of indices of the $i$-th dimension) or \texttt{None} (to select the whole $i$-th dimension). In the case of \texttt{ChiSquared} objects, slicing is also possible using \texttt{[]} and the same syntax as \texttt{numpy} arrays.

Continuing with the previous examples,
\begin{minted}{python}
# we select the first 15 values of ma, and the third value of fa
x_ma_slice = x_ma[2,:15]
fa_slice = y_fa[2,:15]
chi2_slice = chi2.slicing(2, slice(0,15))
chi2BKa_slice = chi2[0].slicing(2, slice(0,15))
# or equivalently
chi2BKa_slice = chi2[0][2,:15]
\end{minted}

%% file: plotting.tex
\alpaca allows to produce some plots that are of interest for phenomenological studies, using \texttt{matplotlib}~\cite{Hunter:2007} as backend.

The first type of plot implemented is a representation of all the partial decay widths or branching ratios,  \texttt{plotting.mpl.alp\_channels\_plot}\index{\texttt{plotting.mpl}!\texttt{.alp\_channels\_plot}}. Its arguments are:
\begin{itemize}
    \item \texttt{x}: array of coordinates for the $x$ variable.
    \item \texttt{channels}: \texttt{dict} containing the partial decay widths or branching ratios. The keys are the corresponding transitions, using the syntax in Section~\ref{sec:structure_process}, and the values are arrays with the information to be plotted. The output of \texttt{alp\_channels\_decay\_widths} or \texttt{alp\_channels\_branching\_ratios} can be used directly.
    \item\texttt{xlabel}, \texttt{ylabel}: Labels for both axes.
    \item \texttt{ymin}: optional value for the minimum of the $y$ axis. If provided, the decay channels that are below this value will not be displayed, and if omitted, all channels will be shown.
    \item \texttt{title}: optional, title of the plot.
    \item \texttt{ax}: \texttt{matplotlib.Axes} where to draw the plot. If not provided, a new \texttt{figure} is created.
\end{itemize}
The output of the functon is a \texttt{matplotlib} axes object that can be further modified. Note that plot elements added after \texttt{alp\_channels\_plot} will not be added to the legend.

\begin{minted}{python}
import alpaca
from alpaca.plotting.mpl import alp_channels_plot
import numpy as np
import matplotlib.pyplot as plt

#Create grid of ma, fa and couplings
ma = np.logspace(-2, np.log10(4), 30)
fa = 10000
coupling = alpaca.ALPcouplings({'cuR': np.diag([0,0,1])},
    4*np.pi*fa, 'derivative_above').match_run(4, 'VA_below')
brs = alpaca.alp_channels_branching_ratios(ma,
        coupling, fa, integrator='no_rge')
dws = alpaca.alp_channels_decay_widths(ma,
        coupling, fa, integrator='no_rge')

# Plot the branching ratios
fig = plt.figure(figsize=(25.4/2.54, 16/2.54), dpi=400)
ax = fig.gca()
ax = alp_channels_plot(ma, brs, r'$m_a$ [GeV]', r'BR', 1e-4, ax=ax)
fig.savefig('br.pdf', bbox_inches='tight')

# Plot the partial decay widths
fig = plt.figure(figsize=(25.4/2.54, 16/2.54), dpi=400)
ax = fig.gca()
ax = alp_channels_plot(ma, dws, r'$m_a$ [GeV]', r'$\Gamma$ [GeV]', ax=ax)
fig.savefig('dw.pdf', bbox_inches='tight')
\end{minted}

Another interesting application of \alpaca is the production of plots showing the excluded regions in the ALP parameter space, implemented by \texttt{plotting.mpl.exclusionplot}\index{\texttt{plotting.mpl}!\texttt{.exclusionplot}}. The arguments of \texttt{exclusionplot} are:
\begin{itemize}
    \item \texttt{x}: Array of coordinates for the $x$ variable.
    \item \texttt{y}: Array of coordinates for the $y$ variable.
    \item \texttt{chi2}: List of $\chi^2$ distributions calculated beforehand with the desired couplings, in the form of a \texttt{ChiSquaredList} object.
    \item\texttt{xlabel}, \texttt{ylabel}: Labels for both axes.
    \item \texttt{title}: Label for the title, 
    \item \texttt{ax}: Optionally, a \texttt{matplotlib} axes object where to draw the plot. If not provided, a new \texttt{figure} is created.
    \item \texttt{global\_chi2}: Optionally, a \texttt{ChiSquared} object representing the global $\chi^2$, if it does not coincide with the combination of all \texttt{chi2}. If set to \texttt{False}, no global $\chi^2$ is plotted.
\end{itemize}

Each sector is represented by the $2\,\sigma$ contour level. If no point of the parameter space is excluded at least by $2\,\sigma$, the corresponding sector will not be included in the legend. However, its contribution to the $\chi^2$ is indeed included. The combined significance for all sectors is depicted with a color scale ranging from $0\,\sigma$ to $5\,\sigma$.

The output is a \texttt{matplotlib} axes object that can be further modified, for example changing to a linear scale (logarithmic scale is used by default).

Below we show an example of how to do an exclusion plot.
\begin{minted}{python}
import alpaca
from alpaca.plotting.mpl import exclusionplot
import numpy as np
import matplotlib.pyplot as plt

#Create grid of ma, fa and couplings
ma = np.logspace(-2, np.log10(4), 30)
fa = np.logspace(3, 8, 30)
x_ma, y_fa = np.meshgrid(ma, fa)
couplings = [alpaca.ALPcouplings({'cuR': np.diag([0,0,1])},
    4*np.pi*f, 'derivative_above').match_run(4, 'VA_below') for f in fa]
x_ma, y_couplings = np.meshgrid(ma, couplings)

#Select channels to study
transitions = [
    alpaca.sectors.default_sectors['BKinv'],
    alpaca.sectors.default_sectors['Kpiinv']
]

#Compute chi2
chi2 = alpaca.statistics.get_chi2(
    transitions,
    x_ma,
    y_couplings,
    y_fa,
    integrator='no_rge'   # The running below 4 GeV is negligible
)

#Plot 
fig, ax=plt.subplots(nrows=1, ncols=1, figsize=(25.4/2.54, 16/2.54),\
dpi=400)
ax = exclusionplot(
    x_ma, 1/y_fa, chi2,
    r'$m_a$ [GeV]', r'$1/f_a$ [GeV$^{-1}$]', r'Top-philic',
    ax=ax
)
ax.set_xscale('linear')
plt.savefig('plot_BK_Kpi.pdf', bbox_inches='tight')
\end{minted}

\subsection{Customisation of the plots}

For a complete overview of plot styling and customisation, we refer the reader to the \texttt{matplotlib} manual~\cite{Hunter:2007}. Here we will only show some of the basic functionalities that can be of use on a regular basis.

Using the statistical analysis of the previous example, we will produce a side-by-side plot of the $K\to\pi \nu\bar{\nu}$ and $B\to K \nu\bar{\nu}$ observables, modifying the legends and colorbars and plotting additional plot lines in one subplot. 
\begin{minted}{python}
chi2_BKa = chi2[0].split_observables()

mB, mK, mpi = 5.28, 0.49, 0.139 # GeV

# Create subplots
fig, axs = plt.subplots(nrows=1, ncols=2, figsize=(25.4/2.54, 16/2.54),\
dpi=400)

# Plot a line in the 1st subplot
# BEFORE creating the exclusionplot
axs[0].axvline(mK - mpi, c='k', ls=':', label = 'Kinematic limit')
exclusionplot(
    x_ma, 1/y_fa, chi2[1],
    r'$m_a$ [GeV]', r'$1/f_a$ [GeV$^{-1}]$',
    r'$K\to \pi+\mathrm{inv}$ with top-philic ALP',
    ax=axs[0], # This goes to 1st subplot
)
# Remove the legend of the 1st subplot
legend0 = axs[0].get_legend()
legend0.remove()

# Plot a line in the 2nd subplot
# BEFORE creating the exclusionplot
axs[1].axvline(mB - mK, c='k', ls=':', label = 'Kinematic limit')
exclusionplot(
    x_ma, 1/y_fa, chi2_BKa,
    r'$m_a$ [GeV]', r'$1/f_a$ [GeV$^{-1}]$',
    r'$B\to K^{(*)} +\mathrm{inv}$ with top-philic ALP',
    ax=axs[1] # This goes to 2nd subplot
)
# Get legend of 2nd subplot
legend1 = axs[1].get_legend()
# We will reposition the lower left corner of the legend
legend1.set_loc('lower left')
# Reference point at x=0.1*plot_width
# and y=0.3*plot_height
# from lower left corner of the plot
legend1.set_bbox_to_anchor((0.1,0.3))
# Remove padding between legend and reference point
legend1.borderaxespad = 0

# Removing the colorbar of the first plot
# Each colorbar has its own axis,
# typically at the end of fig.axes
fig.axes[2].remove()
# Optimise blank spaces
fig.tight_layout()
\end{minted}

Items plotted in the same axis as an \texttt{exclusionplot} will only be shown in the legend if they are created \textit{before} the \texttt{exclusionplot}.

%% file: references.tex
\alpaca is built upon a large number of bibliographical references for the theoretical framework, experimental measurements and coding implementation. Luckily, it also keeps track of which references are used in each function and declaration. After completing the calculations, the user can obtain a \texttt{.bib} file containing only the relevant references\index{\texttt{biblio}!\texttt{.citations}!\texttt{.generate\_bibtex}}:

\begin{minted}{python}
import alpaca
from alpaca.biblio import citations

# Some ALPaca calculations

citations.generate_bibtex("bibliography.bib")
\end{minted}
This command requires an internet connection, as it uses the iNSPIRE-HEP REST API~\cite{Moskovic:2021zjs} to generate the bibliography.

It is also possible to generate a bibliography file for a block of code, omitting any references used only outside of it, with the context manager \texttt{biblio.citations\_context}\index{\texttt{biblio}!\texttt{.citations\_context}} inside a \texttt{with} block: 
\begin{minted}{python}
import alpaca
from alpaca.biblio import citations, citations_context

with citationscontext(merge=True):
    # Some ALPaca calculations
    citations.generate_bibtex("bibliography_block.bib")
\end{minted}

If using the option \texttt{merge=True} (default option), the references used inside the block will also be added to the general bibliography.

Note that some results are cached to improve the efficiency. As a consequence, if a calculation (including intermediate steps) is re-done inside a context manager, its references will not be added to the bibliography.

Finally, when performing an statistical analysis with \texttt{ChiSquared} or \texttt{ChiSquaredList}, \alpaca can generate a bibliographical report of the measurements used to constraint the analysis:\index{\texttt{statistics}!\texttt{.ChiSquared}!\texttt{.citation\_report}}\index{\texttt{statistics}!\texttt{.ChiSquaredList}!\texttt{.citation\_report}}
\begin{minted}{python}
my_chi2 = alpaca.statistics.get_chi2(
    # Options for get_chi2
)
my_chi2.citation_report("my_chi2")
\end{minted}
This method generates a \texttt{.tex} file and a \texttt{.bib} file that contain all the measurements used in the analysis and their corresponding references.

%% file: database_experiments.tex
\label{sec:database_exp}
\subsection*{FCNC meson decays with on-shell ALP}
\begin{table}[H]
    \centering
    \begin{tabular}{c|c}
        Process ($s\to d a$)& Experiment \\\hline
        \rowcolor{lightgray!40}$K^+\to\pi^+ + \mathrm{inv}$ & E949~\cite{BNL-E949:2009dza}\\
        $K^+\to\pi^+ + \mathrm{inv}$ & NA62~\cite{NA62:2020pwi,NA62:2025upx}\\
        \rowcolor{lightgray!40}$K_L^0\to \pi^0+\mathrm{inv}$ & KOTO~\cite{KOTO:2018dsc}\\
        $K^+\to\pi^+\gamma\gamma$ & NA62~\cite{NA62:2023olg,NA62:2025upx} \\
        \rowcolor{lightgray!40}$K^+\to\pi^+ e^+e^-$  & MicroBooNE~\cite{MicroBooNE:2021sov} \\
        $K^+\to\pi^+\mu^+\mu^-$ & NA48/2~\cite{NA482:2016sfh}\\
        \rowcolor{lightgray!40}$K^+\to\pi^+\mu^+\mu^-$ & NA62~\cite{NA62:2022qes,NA62:2025upx}\\
        $K^+\to \pi^+ \mu^\pm e^\mp$ & NA48~\cite{NA62:2021zxl}\\
        \rowcolor{lightgray!40}$K_L^0\to\pi^0\mu^\pm e^\mp$ & KTeV~\cite{KTeV:2007cvy}\\
        \multicolumn{2}{c}{}\\
        \multicolumn{2}{c}{}\\       
        Process ($b\to d a$) & Experiment \\\hline
        \rowcolor{lightgray!40}$B^+\to\pi^++\mathrm{inv}$ & Belle~\cite{Belle:2017oht}\\
        $B^0\to\pi^0+\mathrm{inv}$ & Belle~\cite{Belle:2017oht}\\
        \rowcolor{lightgray!40}$B^+\to\rho^++\mathrm{inv}$ & Belle~\cite{Belle:2017oht}\\
        $B^0\to\rho^0+\mathrm{inv}$ & Belle~\cite{Belle:2017oht}\\
        \rowcolor{lightgray!40}$B^+\to \pi^+ e^+e^-$ & Belle~\cite{Belle:2008tjs}\\
        $B^+\to\pi^+ e^+ e^-$ & BaBar~\cite{BaBar:2013qaj}\\
        \rowcolor{lightgray!40}$B^0\to \pi^0 e^+e^-$ & Belle~\cite{Belle:2008tjs}\\
        $B^0\to\pi^0 e^+ e^-$ & BaBar~\cite{BaBar:2013qaj}\\
        \rowcolor{lightgray!40}$B^+\to \pi^+ \mu^+\mu^-$ & Belle~\cite{Belle:2008tjs}\\
        $B^+\to\pi^+ \mu^+ \mu^-$ & BaBar~\cite{BaBar:2013qaj}\\
        \rowcolor{lightgray!40}$B^0\to \pi^0 \mu^+\mu^-$ & Belle~\cite{Belle:2008tjs}\\
        $B^0\to\pi^0 \mu^+ \mu^-$ & BaBar~\cite{BaBar:2013qaj}\\
        \rowcolor{lightgray!40}$B^+\to \pi^+ \tau^\pm e^\mp$ & BaBar~\cite{BaBar:2012azg}\\
        $B^+\to \pi^+ \tau^\pm \mu^\mp$ & BaBar~\cite{BaBar:2012azg}
        
    \end{tabular}\quad  \begin{tabular}{c|c}
        Process ($c\to ua$) & Experiment \\\hline
        \rowcolor{lightgray!40}$D^0\to\pi^0+\mathrm{inv}$ & BESIII~\cite{BESIII:2021slf}\\
        $D^0\to\pi^0e^+e^-$ & BESIII~\cite{BESIII:2018hqu}\\
        \rowcolor{lightgray!40}$D^0\to\eta e^+e^-$ & BESIII~\cite{BESIII:2018hqu}\\
        $D^0\to\rho^0e^+e^-$ & E791~\cite{E791:2000jkj}\\
        \rowcolor{lightgray!40}$D^+\to\pi^+e^+e^-$ & LHCb~\cite{LHCb:2020car}\\
        $D_s^+\to K^+e^+e^-$ & LHCb~\cite{LHCb:2020car}\\
        \rowcolor{lightgray!40}$D^0\to\pi^0\mu^+\mu^-$ & E653~\cite{E653:1995rpz}\\
        $D^0\to\eta \mu^+\mu^-$ & CLEO II~\cite{CLEO:1996jxx}\\
        \rowcolor{lightgray!40}$D^0\to\rho^0\mu^+\mu^-$ & E791~\cite{E791:2000jkj}\\
        $D^+\to\pi^+\mu^+\mu^-$ & LHCb~\cite{LHCb:2020car}\\
        \rowcolor{lightgray!40}$D^+\to\pi^+\mu^+\mu^-$ & LHCb HL-LHC~\cite{LHCb:2018roe}${}^\dagger$\\
        $D^+\to\rho^+\mu^+\mu^-$ & E653~\cite{E653:1995rpz}\\
        \rowcolor{lightgray!40}$D_s^+\to K^+\mu^+\mu^-$ & LHCb~\cite{LHCb:2020car}\\
        $D^0 \to \pi^0 \mu^\pm e^\mp$ & BaBar~\cite{BaBar:2020faa}\\
        \rowcolor{lightgray!40}$D^+ \to \pi^+ \mu^\pm e^\mp$ &  LHCb~\cite{LHCb:2020car}\\
        $D^0 \to \eta \mu^\pm e^\mp$ & BaBar~\cite{BaBar:2020faa}\\
        \rowcolor{lightgray!40}$D^0 \to \rho^0 \mu^\pm e^\mp$ & BaBar~\cite{BaBar:2020faa}\\
        $D_s^+ \to K^+ \mu^\pm e^\mp$ &  LHCb~\cite{LHCb:2020car}\\
        \rowcolor{lightgray!40}$D_s^+ \to K^{*+} \mu^\pm e^\mp$ &  E653~\cite{E653:1995rpz}\\
        \multicolumn{2}{c}{}\\
        \multicolumn{2}{c}{}\\ 
    \end{tabular} 
    \caption{\em FCNC decays into on-shell ALPs. ${}^\circ$: Theoretical re-cast of experimental data. \linebreak ${}^\dagger$: Projection of future measurement.}
    \label{tab:exp_FCNCon}
\end{table}

\begin{table}[H]
\centering
\begin{tabular}{c|c}
Process ($b\to s a$) & Experiment \\\hline
        \rowcolor{lightgray!40}$B^+\to K^+ + \mathrm{inv}$ & Belle II~\cite{Belle-II:2023esi,Altmannshofer:2023hkn}${}^\circ$ \\
        $B^0\to K^0+\mathrm{inv}$ &  Belle~\cite{Belle:2017oht}\\
        \rowcolor{lightgray!40}$B^+\to K^{+}+\mathrm{inv}$ & BaBar~\cite{BaBar:2013npw,Altmannshofer:2023hkn}${}^\circ$\\
        $B^+\to K^{*+}+\mathrm{inv}$ & BaBar~\cite{BaBar:2013npw,Altmannshofer:2023hkn}${}^\circ$\\
        \rowcolor{lightgray!40}$B_s^0\to\phi +\mathrm{inv}$ & DELPHI~\cite{DELPHI:1996ohp} \\
        $B^+\to K^+ \gamma\gamma$ & BaBar~\cite{BaBar:2021ich}\\
        \rowcolor{lightgray!40}$B^+\to K^+e^+e^-$ & Belle II~\cite{Belle-II:2023ueh}\\
        $B^0\to K^{*0}e^+e^-$  & Belle II~\cite{Belle-II:2023ueh}\\
        \rowcolor{lightgray!40}$B^+\to K^+\mu^+\mu^-$ & LHCb~\cite{LHCb:2016awg}\\
        $B^+\to K^+\mu^+\mu^-$ & Belle II~\cite{Belle-II:2023ueh}\\
        \rowcolor{lightgray!40}$B^+\to K^+\mu^+\mu^-$ & CHARM~\cite{CHARM:1985anb,Dobrich:2018jyi}${}^\circ$\\
        $B^+\to K^+\mu^+\mu^-$ & NA62~\cite{Dobrich:2018jyi}${}^\dagger$\\
        \rowcolor{lightgray!40}$B^+\to K^+\mu^+\mu^-$ & SHiP~\cite{Dobrich:2018jyi}${}^\dagger$\\
        $B^0\to K^{*0}\mu^+\mu^-$ & LHCb~\cite{LHCb:2015nkv}\\
        \rowcolor{lightgray!40}$B^0\to K^{*0}\mu^+\mu^-$ & Belle II~\cite{Belle-II:2023ueh}\\
        $B^+\to K^+\tau^+\tau^-$ & BaBar~\cite{BaBar:2016wgb}\\
        \rowcolor{lightgray!40}$B^+\to K^+\tau^+\tau^-$ & Belle II~\cite{Belle-II:2018jsg}${}^\dagger$\\
        $B^0\to K^{*0}\tau^+\tau^-$ & Belle~\cite{Belle:2021ecr}\\
        \rowcolor{lightgray!40}$B^0 \to K^{*0}\tau^+\tau^-$ & Belle II~\cite{Belle-II:2022cgf}${}^\dagger$\\
        $B^+\to K^+ \pi^0\pi^+\pi^-$ & Belle~\cite{Belle:2013nby,Chakraborty:2021wda}${}^\circ$\\
        \rowcolor{lightgray!40}$B^0\to K^{0} \pi^0\pi^+\pi^-$ & Belle~\cite{Belle:2013nby,Chakraborty:2021wda}${}^\circ$\\
        $B^+\to K^+\eta\pi^+\pi^-$ & BaBar~\cite{BaBar:2008rth,Chakraborty:2021wda}${}^\circ$\\
        \rowcolor{lightgray!40}$B^+ \to K^+ \mu^\pm e^\mp$ & BaBar~\cite{BaBar:2006tnv}\\
        $B^0 \to K^0 e^\pm \mu^\mp$ & BaBar~\cite{BaBar:2006tnv}\\
        \rowcolor{lightgray!40}$B^+ \to K^{*+} \mu^\pm e^\mp$ & BaBar~\cite{BaBar:2006tnv}\\
        $B^0 \to K^{*0} \mu^\pm e^\mp$ & BaBar~\cite{BaBar:2006tnv}\\
        \rowcolor{lightgray!40}$B^+\to K^+ \tau^\pm e^\mp$ & BaBar~\cite{BaBar:2012azg}\\
        $B^+\to K^+ \tau^\pm e^\mp$ & Belle II~\cite{Belle-II:2018jsg}${}^\dagger$\\
        \rowcolor{lightgray!40}$B^+\to K^+ \tau^\pm \mu^\mp$ & BaBar~\cite{BaBar:2012azg}\\
        $B^+\to K^+ \tau^\pm \mu^\mp$ & Belle II~\cite{Belle-II:2018jsg}${}^\dagger$\\
        \rowcolor{lightgray!40}$B^0 \to K^{*0}\tau^\pm\mu^\pm$ & LHCb~\cite{LHCb:2022wrs}
    \end{tabular}
    \caption{\em FCNC decays into on-shell ALPs. ${}^\circ$: Theoretical re-cast of experimental data. \linebreak ${}^\dagger$: Projection of future measurement.}
\end{table}

\subsection*{Quarkonia decays with on-shell ALP}
\begin{table}[H]
    \centering
    \begin{tabular}{c|c}
        Process & Experiment \\\hline
        \rowcolor{lightgray!40}$J/\psi\to\gamma +\mathrm{inv}$ & BESIII~\cite{BESIII:2020sdo}\\
        $\Upsilon(1S)\to\gamma+\mathrm{inv}$ & Belle~\cite{Belle:2018pzt}\\
        \rowcolor{lightgray!40}$\Upsilon(3S)\to\gamma+\mathrm{inv}$ & BaBar~\cite{BaBar:2008aby}\\
        $J/\psi\to\gamma\gamma\gamma$ & BESIII~\cite{BESIII:2024hdv}\\
        \rowcolor{lightgray!40}$e^+e^-\to\gamma \gamma\gamma$, $s=(10.58\,\mathrm{GeV})^2$ & Belle II~\cite{Belle-II:2020jti}\\
        $J/\psi\to\gamma\mu^+\mu^-$ & BESIII~\cite{BESIII:2021ges}\\
        \rowcolor{lightgray!40}$\Upsilon(1S)\to \gamma \mu^+\mu^-$ & BaBar~\cite{BaBar:2012wey}\\
        $\Upsilon(1S)\to\gamma\mu^+\mu^-$ & Belle~\cite{Belle:2021rcl}\\
        \rowcolor{lightgray!40}$\Upsilon(3S)\to \gamma\mu^+\mu^-$ & BaBar~\cite{BaBar:2009lbr}\\
        $\Upsilon(1S)\to\gamma\tau^+\tau^-$ & Belle~\cite{Belle:2021rcl}\\
        \rowcolor{lightgray!40}$\Upsilon(3S)\to\gamma\tau^+\tau^-$ & BaBar~\cite{BaBar:2009oxm}\\
        $e^+e^-\to\gamma \tau^+\tau^-$, $s=(10.58\,\mathrm{GeV})^2$ & Belle II~\cite{Alda:2024cxn}${}^\dagger$ \\
        \rowcolor{lightgray!40}$\Upsilon(3S)\to\gamma+\mathrm{hadrons}$ & BaBar~\cite{BaBar:2011kau}\\
        $\Upsilon(1S) \to \gamma c\overline{c}$ & BaBar~\cite{BaBar:2015cce}\\
        \rowcolor{lightgray!40}$\Upsilon(1S)\to\gamma\mu^\pm e^\mp$ & Belle~\cite{Belle:2022cce}\\
        $\Upsilon(1S)\to\gamma\tau^\pm e^\mp$ & Belle~\cite{Belle:2022cce}\\
        \rowcolor{lightgray!40}$\Upsilon(1S)\to\gamma\tau^\pm \mu^\mp$ & Belle~\cite{Belle:2022cce}
    \end{tabular}
    \caption{\em Quarkonia decays into on-shell ALPs.  ${}^\dagger$: Projection of future measurement.}
    \label{tab:exp_quarkonia}
\end{table}

\subsection*{FCNC meson decays and mixing with off-shell ALP}
\begin{table}[H]
    \centering
    \begin{tabular}{c|c}
        Process ($sd$)& Experiment \\\hline
        \rowcolor{lightgray!40}$K_S^0\to\gamma\gamma$ & NA48~\cite{Lai:2002sr}\\
        $K_S^0\to \gamma\gamma$ & KLOE~\cite{KLOE:2007rta}\\
        \rowcolor{lightgray!40}$K_L^0\to e^+e^-$ & E871~\cite{BNLE871:1998bii}\\
        $K_S^0\to e^+e^-$ & KLOE~\cite{KLOE:2008acb}\\
        \rowcolor{lightgray!40}$K_L^0\to\mu^+\mu^-$ & PDG combination~\cite{E871:2000wvm,Akagi:1994bb,E791:1994xxb,ParticleDataGroup:2024cfk} \\
        $K_S^0\to\mu^+\mu^-$ & LHCb~\cite{LHCb:2020ycd}\\
        \rowcolor{lightgray!40}$\Delta m_K$ & PDG~\cite{ParticleDataGroup:2024cfk} \\
        $|\epsilon_K|$ & PDG~\cite{ParticleDataGroup:2024cfk}\\
        \multicolumn{2}{c}{}\\
        \multicolumn{2}{c}{}\\Process ($bd$) & Experiment \\\hline
        \rowcolor{lightgray!40}$B^0\to\gamma\gamma$ & BaBar~\cite{BaBar:2010qny}\\
        $B^0\to e^+e^-$ & LHCb~\cite{LHCb:2020pcv}\\
        \rowcolor{lightgray!40}$B^0\to\mu^+\mu^-$ & LHCb~\cite{LHCb:2021awg,LHCb:2021vsc}\\
        $B^0\to\mu^+\mu^-$ & CMS~\cite{CMS:2022mgd}\\
        \rowcolor{lightgray!40}$B^0 \to \mu^+\mu^-$ & ATLAS HL-LHC~\cite{ATLAS:2025eaw}${}^\dagger$\\
        $B^0 \to \mu^+\mu^-$ & CMS HL-LHC~\cite{Collaboration:2928094}${}^\dagger$\\
        \rowcolor{lightgray!40}$B^0 \to\mu^+\mu^-$ & LHCb HL-LHC~\cite{LHCb:2018roe}${}^\dagger$\\
        $B^0\to\tau^+\tau^-$ & LHCb~\cite{LHCb:2017myy}\\
        \rowcolor{lightgray!40}$B^0\to\tau^+\tau^-$ & Belle II~\cite{Belle-II:2018jsg}${}^\dagger$\\
        $\Delta m_{B^0}$ & Belle II~\cite{Belle-II:2023bps} \\
        \rowcolor{lightgray!40} $\mathcal{A}_\mathrm{SL}(B^0)$ & HFLAV fit~\cite{HeavyFlavorAveragingGroupHFLAV:2024ctg}
    \end{tabular}\quad\begin{tabular}{c|c}
        Process ($cu$) & Experiment\\\hline
        \rowcolor{lightgray!40}$D^0 \to \gamma\gamma$ & Belle~\cite{Belle:2015pzk}\\
        $D^0\to\gamma\gamma$ & Belle II~\cite{Belle-II:2022cgf}${}^\dagger$\\
        \rowcolor{lightgray!40}$D^0\to e^+e^-$ & Belle~\cite{Belle:2010ouj}\\
        $D^0\to\mu^+\mu^-$ & LHCb~\cite{LHCb:2022jaa}\\
        \rowcolor{lightgray!40}$D^0 \to \mu^+\mu^-$ & CMS~\cite{CMS:2025fmx}\\
        $x_{D^0}$ & HFLAV fit~\cite{HeavyFlavorAveragingGroupHFLAV:2024ctg}\\
        \rowcolor{lightgray!40}$\phi_{12,D^0}$ & HFLAV fit~\cite{HeavyFlavorAveragingGroupHFLAV:2024ctg}\\
        \multicolumn{2}{c}{}\\
        \multicolumn{2}{c}{}\\
        Process ($bs$) & Experiment \\\hline
        \rowcolor{lightgray!40}$B_s^0\to\gamma\gamma$ & Belle~\cite{Belle:2014sac}\\
        $B_s^0\to e^+e^-$ & LHCb~\cite{LHCb:2020pcv}\\
        \rowcolor{lightgray!40}$B_s^0\to\mu^+\mu^-$ & LHCb~\cite{LHCb:2021awg,LHCb:2021vsc}\\
        $B_s^0\to\mu^+\mu^-$ & CMS~\cite{CMS:2022mgd}\\
        \rowcolor{lightgray!40}$B_s^0 \to \mu^+\mu^-$ & ATLAS HL-LHC~\cite{ATLAS:2025eaw}${}^\dagger$\\
        $B_s^0 \to \mu^+\mu^-$ & CMS HL-LHC~\cite{Collaboration:2928094}${}^\dagger$\\
        \rowcolor{lightgray!40}$B_s^0 \to\mu^+\mu^-$ & LHCb HL-LHC~\cite{LHCb:2018roe}${}^\dagger$\\
        $B_s^0\to\tau^+\tau^-$ & LHCb~\cite{LHCb:2017myy}\\
        \rowcolor{lightgray!40}$B_s^0\to\tau^+\tau^-$ & Belle II~\cite{Belle-II:2018jsg}${}^\dagger$\\
        $B^0\to\tau^+\tau^-$ & LHCb HL-LHC~\cite{LHCb:2018roe}${}^\dagger$\\
        \rowcolor{lightgray!40}$\Delta m_{B_s^0}$ & LHCb~\cite{LHCb:2023sim}\\$\mathcal{A}_\mathrm{SL}(B_s^0)$ & HFLAV fit~\cite{HeavyFlavorAveragingGroupHFLAV:2024ctg}
    \end{tabular}
    \caption{\em FCNC meson decays and mixing with off-shell ALPs.}
    \label{tab:exp_FCNCoff}
\end{table}

\subsection*{LFV lepton decays}
\begin{table}[H]
    \centering
    \begin{tabular}{c|c}
        Process & Experiment \\\hline
        \rowcolor{lightgray!40}$\mu^- \to e^- + \mathrm{inv}$ & TWIST~\cite{TWIST:2014ymv} \\
        $\mu^- \to e^- \gamma\gamma$ & Crystal Box~\cite{Bolton:1988af}\\
        \rowcolor{lightgray!40}$\mu^- \to e^- e^+ e^-$ & SINDRUM~\cite{SINDRUM:1987nra} \\
        $\tau^- \to e^- + \mathrm{inv}$ & Belle II~\cite{Belle-II:2022heu} \\
        \rowcolor{lightgray!40}$\tau^- \to e^- \gamma\gamma$ & BaBar~\cite{Bryman:2021ilc} \\
        $\tau^- \to e^- e^+ e^-$ & Belle~\cite{Hayasaka:2010np} \\
        \rowcolor{lightgray!40}$\tau^- \to e^- e^+ e^-$ & Belle II~\cite{Belle-II:2022cgf}${}^\dagger$\\
        $\tau^- \to e^-\mu^+\mu^-$ & Belle~\cite{Hayasaka:2010np} \\
        \rowcolor{lightgray!40}$\tau^- \to e^-\mu^+\mu^-$ & Belle II~\cite{Belle-II:2022cgf}${}^\dagger$\\
        $\tau^- \to \mu^- +\mathrm{inv}$ & Belle II~\cite{Belle-II:2022heu} \\
        \rowcolor{lightgray!40}$\tau^-\to \mu^-\gamma\gamma$ & BaBar~\cite{Bryman:2021ilc} \\
        $\tau^- \to \mu^- e^+ e^-$ & Belle~\cite{Hayasaka:2010np} \\
        \rowcolor{lightgray!40}$\tau^- \to \mu^- e^+ e^-$ & Belle II~\cite{Belle-II:2022cgf}${}^\dagger$\\
        $\tau^- \to \mu^-\mu^+\mu^-$ & Belle~\cite{Hayasaka:2010np} \\
        \rowcolor{lightgray!40}$\tau^- \to \mu^-\mu^+\mu^-$ & Belle II~\cite{Belle-II:2022cgf}${}^\dagger$\\
    \end{tabular}
    \caption{\em LFV lepton decays with on-shell ALPs.}
    \label{tab:exp_LFVlepton}
\end{table}

\subsection*{Meson decay widths}
\begin{table}[H]
    \centering
    \begin{tabular}{c|c}
        Process & Experiment \\\hline
        \rowcolor{lightgray!40}$B^\pm$ & PDG~\cite{ParticleDataGroup:2024cfk} \\
        $B^0$ & PDG~\cite{ParticleDataGroup:2024cfk} \\
        \rowcolor{lightgray!40}$D^\pm$ & PDG~\cite{ParticleDataGroup:2024cfk} \\
        $D^0$ & PDG~\cite{ParticleDataGroup:2024cfk} \\
        \rowcolor{lightgray!40}$D_s^\pm$ & PDG~\cite{ParticleDataGroup:2024cfk} \\
        $K^\pm$ & PDG~\cite{ParticleDataGroup:2024cfk} \\
        \rowcolor{lightgray!40}$K_L^0$ & PDG~\cite{ParticleDataGroup:2024cfk} \\
        $K_S^0$ & PDG~\cite{ParticleDataGroup:2024cfk} \\
    \end{tabular}
    \caption{\em Meson decay widths.}
    \label{tab:exp_mesonDWs}
\end{table}